\documentclass[twocolumn,twocolappendix]{aastex631}

\usepackage{amsmath}	% Advanced maths commands
\usepackage{xcolor}

\shorttitle{JWST Spectroscopy of $z>9$ Galaxies}
\shortauthors{Tang et al.}
\begin{document}

\title{The JWST Spectroscopic Properties of Galaxies at $z=9-14$}

\author[0000-0001-5940-338X]{Mengtao Tang}
\affiliation{Steward Observatory, University of Arizona, 933 N Cherry Ave, Tucson, AZ 85721, USA}
\email{tangmtasua@arizona.edu}

\author{Daniel P. Stark}
\affiliation{Department of Astronomy, University of California, Berkeley, Berkeley, CA 94720, USA}

\author{Charlotte A. Mason}
\affiliation{Cosmic Dawn Center (DAWN)}
\affiliation{Niels Bohr Institute, University of Copenhagen, Jagtvej 128, 2200 Copenhagen N, Denmark}

\author{Viola Gelli}
\affiliation{Cosmic Dawn Center (DAWN)}
\affiliation{Niels Bohr Institute, University of Copenhagen, Jagtvej 128, 2200 Copenhagen N, Denmark}

\author{Zuyi Chen}
\affiliation{Steward Observatory, University of Arizona, 933 N Cherry Ave, Tucson, AZ 85721, USA}
\affiliation{Cosmic Dawn Center (DAWN)}
\affiliation{Niels Bohr Institute, University of Copenhagen, Jagtvej 128, 2200 Copenhagen N, Denmark}

\author{Michael W. Topping}
\affiliation{Steward Observatory, University of Arizona, 933 N Cherry Ave, Tucson, AZ 85721, USA}



\begin{abstract}
We characterize the {\it JWST} spectra of $61$ galaxies at $z=9-14$, including $30$ newly-confirmed galaxies. We directly compare the $z>9$ spectroscopic properties against $401$ galaxies at $6<z<9$, with the goal of identifying evolution in the star formation histories and ISM. We measure rest-UV emission line properties and UV continuum slopes, while also investigating the rest-optical emission lines for the subset of galaxies at $9.0<z<9.6$. With these spectra, we constrain the stellar masses, specific star formation rates, dust attenuation, and the average metallicity and abundance pattern of $z>9$ galaxies. Our dataset indicates that the emission lines undergo a marked change at $z>9$, with extremely large C~{\small III}], H$\beta$, and H$\gamma$ EWs becoming $2-3\times$ more common at $z>9$ relative to $6<z<9$. Using the spectra, we infer the distribution of SFRs on short (SFR$_{\rm 3Myr}$) and medium (SFR$_{\rm 3-50Myr}$) timescales, finding that rapid SFR upturns (large SFR$_{\rm 3Myr}$/SFR$_{\rm 3-50Myr}$ ratios) are significantly more likely among $z>9$ galaxies. These results may reflect a larger dispersion in UV luminosity at fixed halo mass and larger baryon accretion rates at $z>9$, although other physical effects may also contribute. We suggest that the shift in star formation conditions explains the prevalence of extreme nebular spectra that have been detected at $z>9$, with hard ionizing sources and nitrogen-enhancements becoming more typical at the highest redshifts. Finally, we identify five $z>9$ spectroscopically confirmed galaxies with red UV colors ($\beta\gtrsim-1.5$), either revealing a small population with moderate dust attenuation ($\tau_V=0.23-0.35$) or very high density nebular-dominated galaxies with hot stellar populations.
\end{abstract}



%%%%%%%%%%%% INTRODUCTION %%%%%%%%%%%%

\section{Introduction} \label{sec:introduction}

Over the last few years, deep {\it JWST} \citep{Gardner2023,Rigby2023} imaging has provided the first robust constraints on the census of galaxies at $z>9$ (see \citealt{Stark2026} for a review). 
Early {\it JWST}/NIRCam \citep{Rieke2023} imaging campaigns revealed a surprisingly large volume density of luminous $z\gtrsim10$ galaxies \citep[e.g.,][]{Naidu2022,Castellano2022,Adams2023,Finkelstein2023,Harikane2023,Whitler2023a,Casey2024,Hainline2024a,Robertson2024}. 
Most recently, the redshift frontier has been extended to $z\simeq14$ \citep{Carniani2024,Naidu2025}, revealing that the population of extremely luminous galaxies is already in place within $300$~Myr following the Big Bang. 
Current photometric samples now exceed $400$ $z\gtrsim9$ galaxies, allowing the luminosity function to be reliably measured across numerous fields \citep[e.g.,][]{Bouwens2023,Donnan2023,Donnan2024,Harikane2023,Harikane2025,Adams2024,Finkelstein2024,McLeod2024,Whitler2025}. 
The integrated ultraviolet (UV) luminosity density revealed by these surveys has been shown to be significantly greater than what had been predicted by many groups prior to the launch of {\it JWST}.

The physical origin of the excess UV luminosity density remains unclear. 
Some have argued that the evolution can be explained if the dispersion in UV luminosity at fixed halo mass ($\sigma_{\rm UV}$) increases with redshift \citep[e.g.,][]{Mason2023,Mirocha2023,Shen2023,Kravtsov2024}, or increases with decreasing halo mass \citep[e.g.,][]{Katz2023,Sun2023,Gelli2024,Feldmann2025}, perhaps reflecting stronger bursts of star formation and a shift to galaxy formation in lower mass halos at earlier epochs. 
Others have suggested that the star formation efficiency (SFE) may be larger than previously expected at higher redshifts, perhaps a result of redshift evolution \citep[e.g.,][]{Dekel2023,Qin2023,Somerville2025} or larger SFE at moderately low halo masses \citep{Feldmann2025}. 
Additionally, \citet{Ferrara2023} and \citet{Mason2023} have suggested that the slow evolution at the bright end of the luminosity function at $z\gtrsim9$ may be explained if dust attenuation is reduced at earlier times, driven in part by the efficient ejection of dust from luminous galaxies with extremely large specific star formation rates \citep[e.g.,][]{Ziparo2023}.

New observations will be required to determine which of these physical effects are most important in driving the luminosity function evolution at the highest redshifts.
Deep spectroscopy provides one of the most immediate avenues for insight. 
The ratio of the emission line luminosity and underlying continuum emission (the equivalent width, EW) is very sensitive to the recent star formation history, reaching extremely large values in the midst of strong bursts. 
If $\sigma_{\rm UV}$ increases toward higher redshift rapidly at $z>9$, it should leave its imprint on the emission line EW distributions. 
If star formation conditions evolve at $z>9$, we may also expect to see a sudden change in properties of the gas, dust, and ionizing sources, all of which can be probed with {\it JWST} spectroscopy. 

NIRSpec \citep{Jakobsen2022,Boker2023} observations of the luminous galaxy GN-z11 provided our first look at the spectra of $z\gtrsim9$ galaxies \citep{Bunker2023}. 
The spectrum revealed numerous strong emission lines, pointing to a dense population of massive stars formed in a recent burst (see also \citealt{Maiolino2024}). 
The emission line ratios reveal dense ionized gas, hard ionizing agents, and a nitrogen-enhanced abundance pattern, all of which are atypical in spectra at lower redshifts.
Subsequent spectra have revealed that GN-z11 is not anomalous among the $z\gtrsim9$ population. 
The luminous galaxies GHZ2 \citep{Castellano2024} and MoM-z14 \citep{Naidu2025} both exhibit similar emission line spectra, potentially also related to strong bursts of star formation. 
It is conceivable that the emergence of such extreme nebular spectra may hint at a shift in star formation conditions, plausibly linked to the physics that is driving the excess UV luminosity in early galaxies. 

The goal of this paper is to provide statistical context necessary to identify evolution in the spectroscopic properties of galaxies at $z\gtrsim9$. 
Previous works have investigated the evolution in the mean emission line properties of the galaxy population, utilizing composite spectra with $\simeq30-40$ galaxies at $z\gtrsim9$ \citep{Roberts-Borsani2024,Roberts-Borsani2025,Hayes2025}. 
Others have investigated the UV continuum slopes in prism spectra using a sample of $19$ $z\gtrsim9.5$ galaxies \citep{Saxena2024b}. 
Since these studies, spectroscopic samples at $z\gtrsim9$ have grown in number, following a series of campaigns with NIRSpec. 
Here we utilize the latest spectroscopic database available in the public archive to identify whether there is strong evolution in galaxy properties at $z\gtrsim9$. 
We compare the NIRSpec properties of $61$ galaxies at $z>9$ to a sample of $401$ galaxies at $6<z<9$, quantifying evolution in the dust content, gas properties, and stellar populations at the highest redshifts probed by {\it JWST}. 

The organization of this paper is as follows. 
In Section~\ref{sec:sample}, we describe the sample of galaxies at $z>9$ identified from publicly available {\it JWST}/NIRSpec observations. 
We present the UV continuum slopes and emission line properties of galaxies in our $z>9$ sample in Section~\ref{sec:spec_measure}. 
Using the emission line measurements, we characterize the stellar populations and the gas properties $z>9$ galaxies in Section~\ref{sec:zg9_properties}. 
We then describe the evolution in galaxy spectra from $6<z<9$ to $z>9$ in Section~\ref{sec:evolution}. 
In Section~\ref{sec:discussion}, we discuss the evolution in the stellar populations, ionizing sources, and dust content at $z>9$. 
Finally, we summarize our conclusions in Section~\ref{sec:summary}. 
Throughout the paper we adopt a $\Lambda$-dominated, flat universe with $\Omega_{\Lambda}=0.7$, $\Omega_{\rm{M}}=0.3$, and $H_0=70$~km~s$^{-1}$~Mpc$^{-1}$. 
All magnitudes are quoted in the AB system \citep{Oke1983} and all EWs are quoted in the rest frame.

%%%%%%%%%%%% SAMPLE SELECTION %%%%%%%%%%%%

\section{Spectroscopic Sample Selection} \label{sec:sample}

In this section, we construct a sample of galaxies at $z>9$ with publicly-available NIRSpec spectra. 
We introduce the NIRSpec spectra in Section~\ref{sec:data}, and then select the $z>9$ sample in Section~\ref{sec:zg9}. 
In Section~\ref{sec:stack}, we create $z>9$ composite spectra. 
We then describe the photometry measurements in Section~\ref{sec:photometry}, and the emission line and continuum measurements in Section~\ref{sec:spec_measure}. 
Finally, we introduce the photoionization modeling approaches to infer the physical properties of $z>9$ galaxies in Section~\ref{sec:model}. 

\subsection{JWST/NIRSpec Spectra} \label{sec:data}

We use the NIRSpec spectra obtained from the following public observations: 
the JWST Advanced Deep Extragalactic Survey\footnote{\url{https://jades-survey.github.io/}} (JADES, GTO 1180, 1181, PI: D. Eisenstein, GTO 1210, 1286, PI: N. L\"utzgendorf, GTO 1287, PI: K. Isaak, GO 3215, PIs: D. Eisenstein \& R. Maiolino; \citealt{Eisenstein2023,Eisenstein2025,Bunker2024,DEugenio2025}), 
the GLASS-JWST Early Release Science Program\footnote{\url{https://glass.astro.ucla.edu/ers/}} (GLASS, ERS 1324, PI: T. Treu; \citealt{Treu2022}), 
the Cosmic Evolution Early Release Science\footnote{\url{https://ceers.github.io/}} (CEERS, ERS 1345, PI: S. Finkelstein; \citealt{Finkelstein2025}) and a Director’s Discretionary Time program (DDT 2750, PI: P. Arrabal Haro; \citealt{ArrabalHaro2023a,ArrabalHaro2023b}), 
the Ultra-deep NIRCam and NIRSpec Observations Before the Epoch of Reionization\footnote{\url{https://jwst-uncover.github.io/}} (UNCOVER, GO 2561, PI: I. Labb\'e \& R. Bezanson; \citealt{Bezanson2024,Price2025}), 
the Red Unknowns: Bright Infrared Extragalactic Survey\footnote{\url{https://annadeg.github.io/projects/RUBIES/}} (RUBIES, GO 4233, PIs: A. de Graaff \& G. Brammer; \citealt{deGraaff2025}), 
the CANDELS-Area Prism Epoch of Reionization Survey (CAPERS, GO 6368, PI: M. Dickinson; \citealt{Kokorev2025}), GO 1871 (PI: J. Chisholm; \citealt{Chisholm2024}), GO 3073 (PI: M. Castellano; \citealt{Castellano2024,Napolitano2025a}), and GO 4287 (PI: C. Mason; Whitler et al. 2025, in preparation). 
All the NIRSpec observations were performed with the multi-object spectroscopy (MOS) mode using the micro-shutter assembly (MSA; \citealt{Ferruit2022}). 
We refer readers to the above references for detailed descriptions of the NIRSpec observations. 

Most of the spectroscopic observations described above have used the low resolution ($R\sim100$) prism (JADES, CEERS, UNCOVER, RUBIES, CAPERS, and GO 3073). 
Medium resolution ($R=1000$) grating spectroscopy (G140M, G235M, and G395M) have also been obtained in JADES and CEERS observations. 
The GLASS survey has used high resolution ($R=2700$) gratings with G140H, G235H, and G395H. 
The GO 1871 program has used two high resolution grating/filter pairs G235H and G395H. 
The GO 4287 program has used a combination of the high resolution grating/filter pair G140H and the medium resolution grating/filter pair G395M.

The NIRSpec spectra were reduced following the same approaches described in \citet{Topping2025a} using the standard JWST data reduction pipeline\footnote{\url{https://github.com/spacetelescope/jwst}} \citep{Bushouse2024}. 
For each object, we extract the 1D spectrum from the reduced 2D spectra using a boxcar extraction. 
We set the extraction aperture to match the emission line profile or the continuum along the spatial direction, with a typical width of $5$~pixels ($\sim0.5$~arcsec in spatial direction). 
We refer readers to \citet{Topping2025a} and \citet{Tang2024c} for a full description of the analysis of the spectra of the public spectroscopic sample. 

%%%% Figure: redshift vs. M_UV %%%%

\begin{figure}
\includegraphics[width=\linewidth]{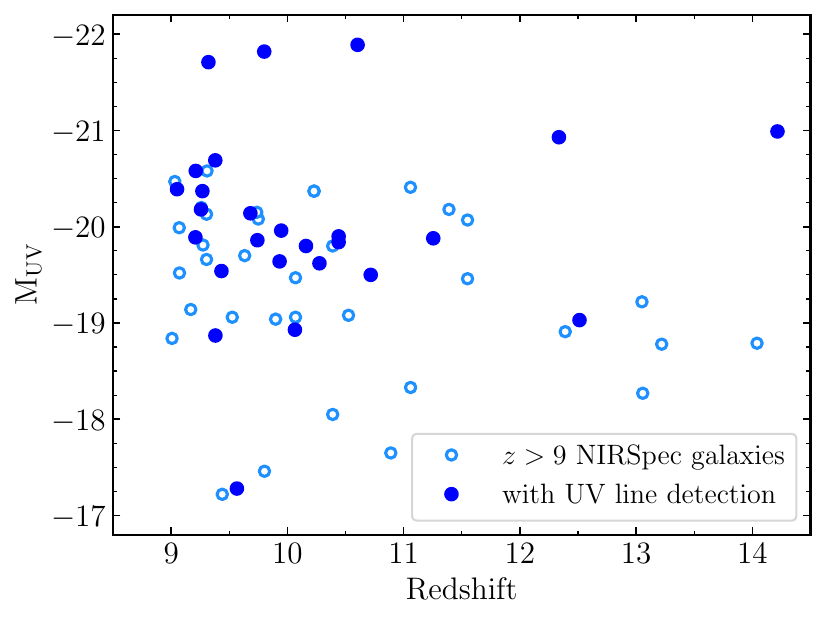}
\caption{Spectroscopic redshifts and absolute UV magnitudes of the $61$ galaxies in our $z>9$ sample. We highlight sources with rest-frame UV emission line (N~{\scriptsize IV}], C~{\scriptsize IV}, He~{\scriptsize II}, O~{\scriptsize III}], N~{\scriptsize III}], C~{\scriptsize III}]) detections with solid blue circles. Those without UV line detections are shown by open dodger blue circles.}
\label{fig:z_muv}
\end{figure}

\subsection{Sample Selection of $z>9$ Galaxies} \label{sec:zg9}

Using the public NIRSpec data sets described in the above subsection, we identify a sample of galaxies with spectroscopic redshifts at $z>9$. 
We determine the spectroscopic redshifts as follows. 
First, we visually inspect the 2D NIRSpec spectra, searching for Ly$\alpha$ break. 
We choose to search for Ly$\alpha$ break instead of emission lines for two reasons: 
1) strong rest-frame optical emission lines (H$\beta$, [O~{\small III}]) are often shifted out of the NIRSpec spectrum at $z>9$, and 2) to ensure we are not biased against sources with weak emission lines (e.g., extremely metal-poor galaxies). 
We identify $61$ galaxies at $z>9$ with Ly$\alpha$ break. 
Next, we search for emission lines in these $61$ galaxies to derive more accurate spectroscopic redshifts. 
We find $34$ galaxies with optical emission line detections ([O~{\small II}], [Ne~{\small III}], H$\gamma$, H$\beta$, or [O~{\small III}]), and we derive their redshifts with the following approach. 
We simultaneously fit the available optical emission lines with Gaussian profiles, and compute the redshifts using the line centers. 
For the remaining $27$ galaxies without emission line detections, we derive their redshifts from Ly$\alpha$ break. 
We verify the Ly$\alpha$ break inferred redshifts of these $27$ galaxies following the approaches described in \citet{Tang2024c}. 
We fit their NIRCam photometry with population synthesis models (see Section~\ref{sec:model}), varying redshift in the range of $0<z<15$. 
The best-fit redshifts are consistent with those derived from Ly$\alpha$ breaks in the NIRSpec spectra within $\Delta z<0.3$. 
Therefore, we adopt the Ly$\alpha$ break inferred redshifts for the subset without emission line detections. 
Overall, we measure the spectroscopic redshifts of the $61$ galaxies in our sample at $9.01<z<14.22$ (Figure~\ref{fig:z_muv}). 

In Table~\ref{tab:zg9} we list the $61$ galaxies in our $z>9$ sample, which are in six independent fields: the Abell 2744, the Cosmological Evolution Survey (COSMOS), the Extended Groth Strip (EGS), the Great Observatories Origins Deep Survey (GOODS) North and South, and the UltraDeep Survey (UDS) fields. 
There are $14$, $3$, $10$, $9$, $17$, and $8$ galaxies in the Abell 2744, COSMOS, EGS, GOODS-N, GOODS-S, and UDS fields, respectively. 
All these $61$ galaxies have low resolution ($R\sim100$) prism spectra, which were taken in the CEERS, JADES, UNCOVER, RUBIES, CAPERS, and GO 3073 programs. 
The exposure time of the $61$ prism spectra at $z>9$ spans from $0.8$ to $46.7$ hours, with a median of $4.4$ hours. 
For a galaxy with $H=27.7$ (the median magnitude of our $z>9$ sources) at $z=10$, this median depth corresponds to a $3\sigma$ rest-frame EW limit of $23$~\AA. 
Medium resolution ($R=1000$) grating spectra have also been obtained in $26$ of the $61$ galaxies at $z>9$, which were all taken in the JADES programs.
For this subset with $R=1000$ spectra, the exposure time spans from $0.9$ to $11.7$ hours for G140M (median $2.2$ hours), $0.9$ to $6.9$ hours for G235M (median $2.6$ hours), and $0.9$ to $46.7$ hours for G395M (median $2.6$ hours). 
The typical EW limits (for an $H=27.7$ galaxy at $z=10$) are $17$, $18$, and $37$~\AA\ in the three configurations, respectively. 

The NIRSpec spectra of $31$ of the $61$ galaxies in our $z>9$ sample have been presented in literature \citep{ArrabalHaro2023a,ArrabalHaro2023b,Bunker2023,Curtis-Lake2023,Wang2023,Carniani2024,Castellano2024,Curti2025,DEugenio2024,Fujimoto2024,Hainline2024b,Heintz2024,Tang2024c,Kokorev2025,Napolitano2025a,Pollock2025,Witstok2025b}. 
We have newly identified $30$ galaxies at $z>9$, and we show their spectra in Figure~\ref{fig:zg9_spec}. 
Our sample of $61$ galaxies is about $2\times$ larger than that used in earlier studies of $z>9$ galaxies with NIRSpec \citep{Roberts-Borsani2024,Heintz2025}. 
We will present the emission line and continuum measurements of our $z>9$ sample in Section~\ref{sec:spec_measure}. 

\subsection{Construction of Composite $z>9$ Spectra} \label{sec:stack}

To investigate the average spectroscopic properties of galaxies in our $z>9$ sample, we create composite spectra by stacking individual spectra.
To maximize the sample size for creating composite spectra, we stack the low resolution prism spectra. 
We construct a stack with the following methods. 
We first shift individual spectra to the rest-frame using the derived spectroscopic redshifts (Section~\ref{sec:zg9}). 
Each spectrum is then interpolated to a common rest-frame wavelength scale ($10$~\AA\ per wavelength bin) and normalized by its luminosity density at rest-frame $1500$~\AA. 
Finally, the spectra are median stacked without weighting to avoid bias towards bright systems or those with deep spectra. 
The composite luminosity density in each wavelength bin is derived from the median of the luminosity densities of individual sources in that bin.
Following the method described in \citet{Roberts-Borsani2024}, the associated uncertainty is calculated as the semi-difference of the 16th and 84th percentiles of the individual luminosity density distribution divided by $\sqrt{N}$, where $N$ is the number of sources in the stack in the given wavelength bin.

\begin{figure*}
\includegraphics[width=0.95\linewidth]{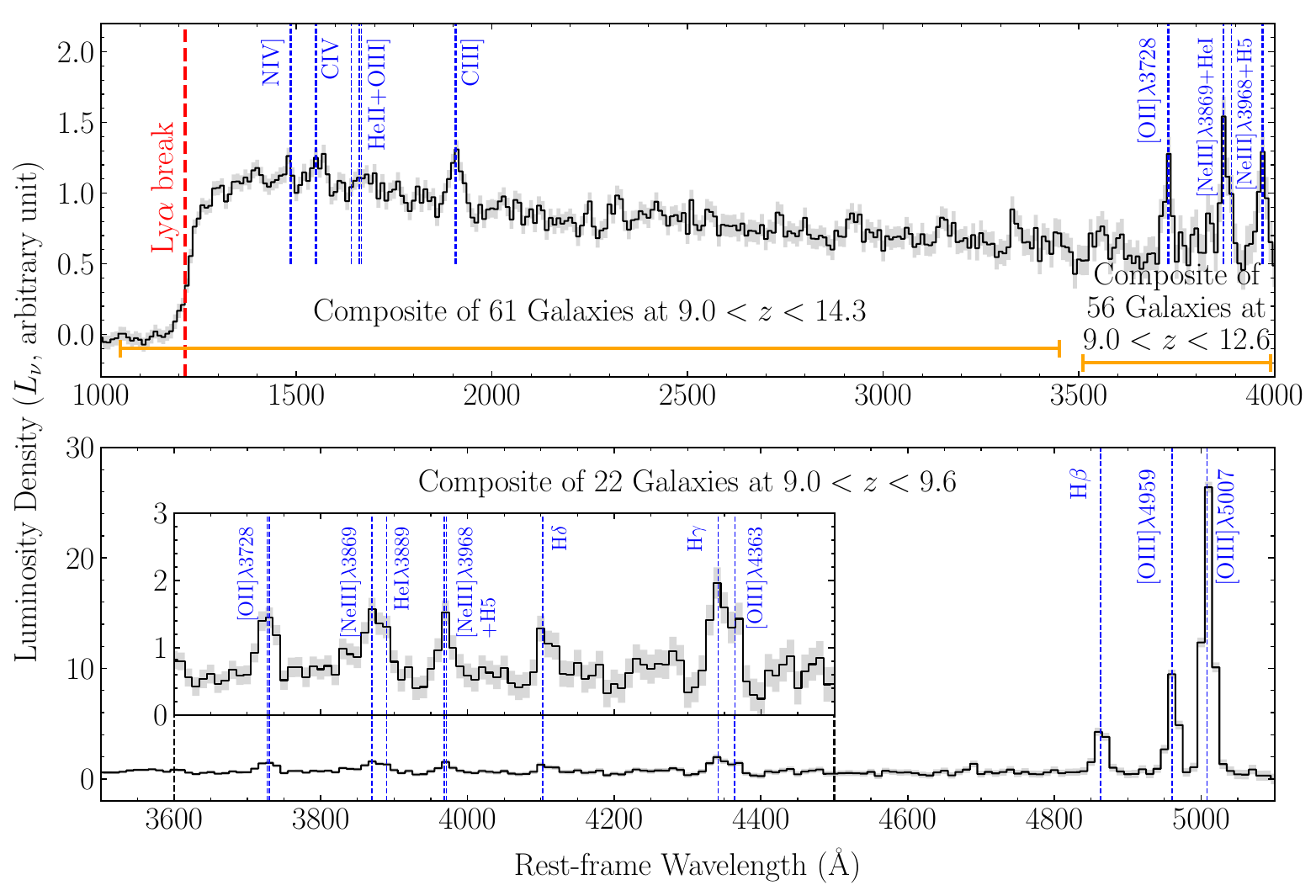}
\caption{Composite prism spectra of galaxies in our $z>9$ sample. The top panel shows the stack of all the $61$ galaxies at in our $z>9$ sample ($9.0<z<14.3$). The spectrum at rest-frame $\sim3500-4000$~\AA\ is composed of $56$ galaxies at $9.0<z<12.6$. The bottom panel shows the composite at rest-frame optical wavelengths obtained by stacking $22$ galaxies at $9.0<z<9.6$. Emission line detections are marked by blue dashed lines. The grey shaded region presents the $1\sigma$ uncertainty of flux density.}
\label{fig:zg9_comp}
\end{figure*}

Here we aim to characterize the rest-frame UV to optical emission lines from the composite spectra. 
We first create a composite spectrum by stacking the prism spectra of all the $61$ galaxies in our sample at $9.0<z<14.3$. 
This allows us to constrain the average EWs of rest-frame UV emission lines (e.g., N~{\small IV}], C~{\small IV}, O~{\small III}], C~{\small III}]) of entire $z>9$ sample, as well as [O~{\small II}] and [Ne~{\small III}] emission lines at redshift up to $z=12.6$ ($56$ galaxies). 
To measure strong rest-frame optical emission lines (e.g., H$\beta$, [O~{\small III}]), we create another composite spectrum by stacking the spectra of galaxies at a narrower redshift window $9.0<z<9.6$, since H$\beta$ and [O~{\small III}] are shifted out of NIRSpec spectra at $z>9.6$. 
The stack at $9.0<z<9.6$ contains $22$ galaxies, and our goal is to constrain the H$\beta$ EW, [O~{\small III}]~$\lambda\lambda4959,5007$/[O~{\small II}]~$\lambda3728$ ratio, and electron temperature and oxygen abundance from [O~{\small III}]~$\lambda4363$ and [O~{\small III}]~$\lambda5007$ measurements. 
In Figure~\ref{fig:zg9_comp} we show the two composite spectra at $9.0<z<14.3$ and $9.0<z<9.6$. 
We will present the emission line measurements of the composite spectra in Section~\ref{sec:spec_measure}. 

%%%% Figure: composite spectra at z>9 %%%%

\subsection{Photometric Measurements} \label{sec:photometry}

We characterize the absolute UV magnitudes and UV slopes of the $61$ galaxies in our $z>9$ sample from their NIRCam photometry. 
Details of the NIRCam photometry used in this work are described in \citet{Tang2024c}. 
In brief, we utilize the reduced NIRCam images \citep{Valentino2023} and photometry catalogs released on the DAWN JWST Archive\footnote{\url{https://dawn-cph.github.io/dja/}} (DJA; \citealt{Valentino2023}). 
The NIRCam images were obtained from CEERS, JADES, UNCOVER, and the Public Release IMaging for Extragalactic Research (PRIMER, GO 1837, PI: J. Dunlop) programs. 
We use imaging taken with six broad-band filters (F115W, F150W, F200W, F277W, F356W, F444W) and two medium-band filters (F335M, F410M) if available. 
For each source, we use the photometry measured in the $0.5$~arcsec diameter circular apertures and corrected to the ``total'' fluxes within elliptical Kron \citep{Kron1980} apertures. 

For each galaxy in the $z>9$ sample, we cross-match its coordinate with the photometry catalog and identify the best-matched source by visually inspecting the image. 
For the $14$ galaxies in the Abell 2744 field, we correct their photometry for gravitational lensing using the magnifications reported in \citet{Price2025}, which are derived from the \citet{Furtak2023} lensing models. 
The median magnification correction factor of these $14$ systems is $1.6$. 
The absolute UV magnitudes (M$_{\rm UV}$) of the $61$ galaxies in our $z>9$ sample range from $-21.9$ to $-17.2$, with a median of M$_{\rm UV}=-19.8$ (Figure~\ref{fig:z_muv}). 

\subsection{Emission Line and Continuum Measurements} \label{sec:spec_measure}

We characterize the emission line fluxes and EWs as well as the UV continuum slopes for the $61$ galaxies in our $z>9$ sample. 
Using the measured spectroscopic redshifts (Section~\ref{sec:zg9}), we search for rest-frame UV emission lines (N~{\small IV}], C~{\small IV}, He~{\small II}, O~{\small III}], N~{\small III}], and C~{\small III}]) and available rest-frame optical emission lines ([O~{\small II}], [Ne~{\small III}], H$\gamma$, [O~{\small III}]~$\lambda4363$, H$\beta$, and [O~{\small III}]~$\lambda4959$, [O~{\small III}]~$\lambda5007$) in both the low resolution prism spectra and the medium resolution grating spectra. 

The line fluxes and EWs are computed from the extracted 1D spectra as follows. 
We first determine the underlying continuum. 
For each emission line we fit the continuum nearby the line with a power law ($f_{\lambda}\propto\lambda^{\beta}$). 
In the grating spectra of a subset of our $z>9$ galaxies, the continua are not detected (S/N $<3$) and we adopt the continua measured from the prism spectra of the same sources.
For the subset of galaxies with continua detected in both prism and grating spectra, we find that the continua measured from both spectra of the same sources are consistent. 
Then we measure the emission line fluxes from the continuum-subtracted spectra. 
For emission lines detected with relatively high S/N ($>5$), we compute the line fluxes by fitting the line profiles with Gaussian functions. 
We use a single Gaussian function for each isolated emission line. 
For emission lines that are close in wavelength (e.g., [Ne~{\small III}]~$\lambda3869$ and He~{\small I}~$\lambda3889$, H$\gamma$ and [O~{\small III}]~$\lambda4363$), we simultaneously fit them with multiple Gaussians. 
For emission lines detected with lower S/N ($<5$), we compute the line fluxes from direct integration. 
Using emission lines with S/N $>5$, we find that Gaussian fit and direct integration result in similar fluxes. 
To compute the uncertainties of the line fluxes, we resample the flux densities of each spectrum $1000$ times by taking the observed value as the mean and the error as standard deviation. 
We then measure the line fluxes from the resampled spectra following the same methods described above and take the standard deviations as the uncertainties. 
We note that for C~{\small IV} emission, the flux (and hence EW) may be underestimated at prism resolution due to the underlying absorption. 
For undetected lines, we derive the $3\sigma$ upper limit of line flux by summing the error spectrum in quadrature over 1 instrument resolution element ($\simeq3000$~km~s$^{-1}$ for low resolution prism and $\simeq300$~km~s$^{-1}$ for medium resolution grating spectrum) around the expected line center. 
The EW (EW limit) of a detected (undetected) emission line is computed by dividing the line flux (flux limit) with the underlying continuum flux density. 

We also search for emission lines in the composite spectra of $z>9$ galaxies constructed in Section~\ref{sec:stack}. 
In the stack of all the $61$ galaxies in our $z>9$ sample ($9.0<z<14.3$, top panel of Figure~\ref{fig:zg9_comp}), we identify a clear C~{\small III}]~$\lambda1908$ emission with S/N of $7$. 
We additionally detect emission line features with S/N of $3-4$ at the expected wavelengths of N~{\small IV}]~$\lambda1485$, C~{\small IV}~$\lambda1549$, and blended He~{\small II}~$\lambda1640$ + O~{\small III}]~$\lambda1663$.
We do not detect other rest-frame UV emission lines with S/N $>3$ in the composite spectrum.
At rest-frame $\sim3500-4000$~\AA, the composite spectrum of the $56$ galaxies at $9.0<z<12.6$ (top panel of Figure~\ref{fig:zg9_comp}) presents [O~{\small II}]~$\lambda3728$, [Ne~{\small III}]~$\lambda3869$, and [Ne~{\small III}]~$\lambda3968$ emission lines. 
For the stack of the $22$ galaxies at $9.0<z<9.6$ (bottom panel of Figure~\ref{fig:zg9_comp}), we identify a suite of rest-frame optical emission lines including [O~{\small II}]~$\lambda3728$, [Ne~{\small III}]~$\lambda3869$, [Ne~{\small III}]~$\lambda3968$, H$\delta$, H$\gamma$, [O~{\small III}]~$\lambda4363$, H$\beta$, [O~{\small III}]~$\lambda4959$, and [O~{\small III}]~$\lambda5007$. 
The fluxes and EWs of emission lines detected in the composite spectra are computed following the same approaches described above. 
We present the emission line EWs and line ratios measured from the composite spectra at $z>9$ in Table~1. 

We measure the UV continuum slopes ($\beta$) from the prism spectra of $z>9$ galaxies. 
Of the $61$ galaxies in our $z>9$ sample, $60$ have prism spectra covering rest-frame UV wavelengths (the rest-frame UV spectrum of CAPERS-UDS-142042 falls in the detector gap, see Figure~\ref{fig:zg9_spec}). 
For each of these $60$ objects, we fit a power law to its prism spectrum in the rest-frame wavelength range $1400-2700$~\AA.
This wavelength range is chosen to be similar to that adopted in the standard \citet{Calzetti1994} approach, minimizing the impact of the Ly$\alpha$ break and the intergalactic medium (IGM) damping wing \citep{Heintz2025,Mason2026}. 
We also mask regions at rest-frame $1440-1590$~\AA, $1620-1680$~\AA, and $1860-1980$~\AA\ to avoid strong emission lines \citep[e.g.,][]{Saxena2024b}. 
To estimate the uncertainty of UV slope measurement of each source, we resample the rest-frame UV continuum flux densities $1000$ times and fit the UV slope for each resampled spectrum with the same approaches described above. 
Then we take the standard deviation of the UV slopes derived from the resampled spectra as the uncertainty. 
For these $60$ galaxies at $z>9$, the 16th-50th-84th percentiles of the UV slopes are $\beta=-2.64$, $-2.33$, and $-2.01$, respectively. 
From the composite spectrum, we measure an average UV slope of $\beta=-2.34\pm0.05$, consistent with the median value of individual sources.
We present the UV slopes and C~{\small III}], C~{\small IV} EWs of our $z>9$ galaxies in Table~\ref{tab:zg9}. 

%%%% Table: composite spectra emission line measurements %%%%

\begin{deluxetable}{cc}
\tablecaption{Emission line EWs, line flux ratios, and UV slope measured from the composite spectra of $z>9$ galaxies.}
\tablehead{\multicolumn{2}{c}{Composite at $9.0<z<14.3$}}
\startdata
\# of Galaxies & $61$ \\
$\beta$ & $-2.34\pm0.05$ \\
N~{\scriptsize IV}] EW & $5.0\pm1.2$~\AA \\
C~{\scriptsize IV} EW & $5.3\pm1.4$~\AA \\
He~{\scriptsize II}+O~{\scriptsize III}] EW & $4.3\pm1.3$~\AA \\
C~{\scriptsize III}] EW & $12.6\pm1.8$~\AA \\
C~{\scriptsize IV}/C~{\scriptsize III}] & $0.77\pm0.24$ \\
\hline
\hline
\multicolumn{2}{c}{Composite at $9.0<z<12.6$} \\
\hline
\# of Galaxies & $56$ \\
{[}Ne~{\small III}]~$\lambda3869$/[O~{\small II}]~$\lambda3728$ & $1.17\pm0.30$ \\
\hline
\hline
\multicolumn{2}{c}{Composite at $9.0<z<9.6$} \\
\hline
\# of Galaxies & $22$ \\
H$\gamma$ EW & $51\pm7$~\AA \\
H$\beta$ EW & $150\pm12$~\AA \\
{[}O~{\scriptsize III}]~$\lambda5007$ EW & $819\pm13$~\AA \\
H$\gamma$/H$\beta$ & $0.460\pm0.071$ \\
{[}O~{\scriptsize III}]~$\lambda5007$/H$\beta$ & $5.2\pm0.4$ \\
([O~{\small III}]+[O~{\small II}])/H$\beta$ & $7.5\pm0.6$ \\
{[}Ne~{\small III}]~$\lambda3869$/[O~{\small II}]~$\lambda3728$ & $0.88\pm0.15$ \\
{[}O~{\small III}]~$\lambda\lambda4959,5007$/[O~{\small II}]~$\lambda3728$ & $13.4\pm1.7$ \\
{[}O~{\scriptsize III}]~$\lambda4363$/H$\gamma$ & $0.42\pm0.12$ \\
{[}O~{\scriptsize III}]~$\lambda4363$/[O~{\scriptsize III}]~$\lambda5007$ & $0.037\pm0.009$ \\
\enddata
\label{tab:zg9_comp}
\end{deluxetable}

\subsection{Photoionization modeling} \label{sec:model}

To explore the stellar population and gas properties of $z>9$ galaxies, we fit the NIRCam spectral energy distributions (SEDs) and emission lines using the Bayesian galaxy SED modeling and interpreting tool BayEsian Analysis of GaLaxy sEds (\texttt{BEAGLE}, version 0.29.2; \citealt{Chevallard2016}). 
The \texttt{BEAGLE} tool utilizes the latest version of the \citet{Bruzual2003} stellar population synthesis models and the \citet{Gutkin2016} photoionization models of star-forming galaxies with the \texttt{CLOUDY} code \citep{Ferland2013}. 

We first fit the rest-frame optical emission line fluxes ([O~{\small II}]~$\lambda3728$, [Ne~{\small III}]~$\lambda3869$, H$\gamma$, [O~{\small III}]~$\lambda4363$, H$\beta$, [O~{\small III}]~$\lambda4959$, [O~{\small III}]~$\lambda5007$) as well as the H$\gamma$ and H$\beta$ EWs measured from the composite spectrum to infer the average interstellar medium (ISM) properties at $z>9$. 
We assume a constant star formation history (CSFH), allowing the galaxy age to vary between $1$~Myr and the age of the Universe at the given redshift with a log-uniform prior. 
In these models, the hydrogen-ionizing spectrum reaches close to a steady state after $10$~Myr of CSFH, and we allow the stellar age to vary to larger values to reproduce the Balmer line EWs. 
We also assume the \citet{Chabrier2003} initial mass function (IMF) with a stellar mass range of $0.1-300\ M_{\odot}$. 
The metallicity is set to vary in the range $-2.2\le\log{(Z/Z_{\odot})}\le0$ ($Z_{\odot}=0.01524$; \citealt{Caffau2011}), and the interstellar metallicity is set to equal to the stellar metallicity. 
We allow the dust-to-metal mass ratio ($\xi_{\rm d}$) to vary within the range $0.1\le\xi_{\rm d}\le0.5$. 
The ionization parameter $U$ is adjusted in the range $-4.0\le\log{U}\le-1.0$. 
We adopt log-uniform priors for metallicity and ionization parameter, and a uniform prior for dust-to-metal mass ratio. 
For dust attenuation, we apply the Small Magellanic Cloud (SMC) extinction curve \citep{Pei1992}, allowing the $V$-band optical depth $\tau_{V}$ to vary between $0.001$ and $5$ with a log-uniform prior. 
We apply the prescription of \citet{Inoue2014} to account for the absorption of IGM. 

To infer the stellar populations and star formation histories (SFHs), we then fit the SEDs and available C~{\small III}], H$\gamma$, H$\beta$, and [O~{\small III}]~$\lambda5007$ EWs of the $61$ individual galaxies in our $z>9$ sample. 
Here we consider two component SFH models (TcSFH), which allow for a more flexible SFH with a wide range of recent variations in star formation rate (SFR). 
We adopt a similar TcSFH modeling procedure as that described in \citet{Endsley2024}. 
The TcSFH models consist of an exponentially delayed SFH component (SFR $\propto t\cdot e^{-t/\tau}$, where $t$ is the time since the onset of star formation) and a CSFH component. 
For the delayed SFH component, we allow the star formation timescale $\tau$ to span the range between $1$~Myr and $30$~Gyr. 
We set $t$ to be between $20$~Myr and the age of the Universe at the given redshift. 
The CSFH component defines the most recent SFH, and the timescale of the recent CSFH component is fitted between $1$~Myr and $20$~Myr. 
The strength of the CSFH component is parameterized in terms of specific star formation rate (sSFR) over this period (relative to the final formed stellar mass) in the range $-14\le \log{({\rm sSFR/yr^{-1}})}\le-6$. 
We adopt log-uniform priors for all the above parameters. 
For gas properties (ionization parameter, metallicity, dust-to-metal ratio) and dust attenuation, we apply the same fitting ranges and priors as for fitting the composite spectrum. 

The assumed SFH can impact the recovered stellar mass \citep[e.g.,][]{Carnall2019,Tacchella2022,Whitler2023b}. 
To explore the range of stellar masses for each object, we also fit the SEDs with two alternative SFHs. 
We consider single CSFH models using \texttt{BEAGLE}, which we expect to provide a lower bound on the stellar masses.
We also fit the SEDs with models that incorporate non-parametric SFHs using \texttt{Prospector} \citep{Leja2019,Johnson2021}, which is based on the Flexible Stellar Population Synthesis code \citep{Conroy2009,Conroy2010} and the nebular emission models of \citet{Byler2017}. 
We adopt the same \texttt{Prospector} model parameters and priors as in \citet{Whitler2023b}, which are also similar to those applied in the \texttt{BEAGLE} models described above. 
The non-parametric SFH models are piecewise constant functions in time. 
We adopt five age bins spanning from the time of observation to the lookback time corresponding to a formation redshift $z_{\rm form}$ (we set $z_{\rm form}=20$): $0-3$~Myr, $3-10$~Myr, $10-30$~Myr, $30-100$~Myr, and $>100$~Myr. 
We use the `continuity' prior in \texttt{Prospector}, which allows a smoothly evolving SFR over time.

From the \texttt{BEAGLE} and \texttt{Prospector} fitting, we derive the median value and the marginalized $68\%$ credible interval from the posterior probability distribution for each fitted parameter. 
We will discuss the photoionization modeling results in Section~\ref{sec:zg9_properties}

%%%%%%%%%%%% SPECTROSCOPIC PROPERTIES %%%%%%%%%%%%

\section{Spectroscopic Properties at $\lowercase{z}>9$} \label{sec:zg9_spectra}

In this section, we present the spectroscopic properties of the galaxies in our $z>9$ sample. 
We discuss the UV continuum slopes in Section~\ref{sec:uv_slope}. 
Then we characterize the C~{\small III}] and O~{\small III}] (Section~\ref{sec:c3_o3}), C~{\small IV} and He~{\small II} (Section~\ref{sec:c4_he2}), N~{\small IV}] and N~{\small III}] (Section~\ref{sec:n4_n3}), Balmer emission lines (Section~\ref{sec:balmer}), [O~{\small II}], [Ne~{\small III}], and [O~{\small III}] in rest-frame optical (Section~\ref{sec:opt_lines}), and the [O~{\small III}]~$\lambda4363$ auroral emission lines (Section~\ref{sec:o3_4363}). 

%%%% Figure: M_UV vs. UV slope & M_UV vs. redshift %%%%

\begin{figure*}
\includegraphics[width=\linewidth]{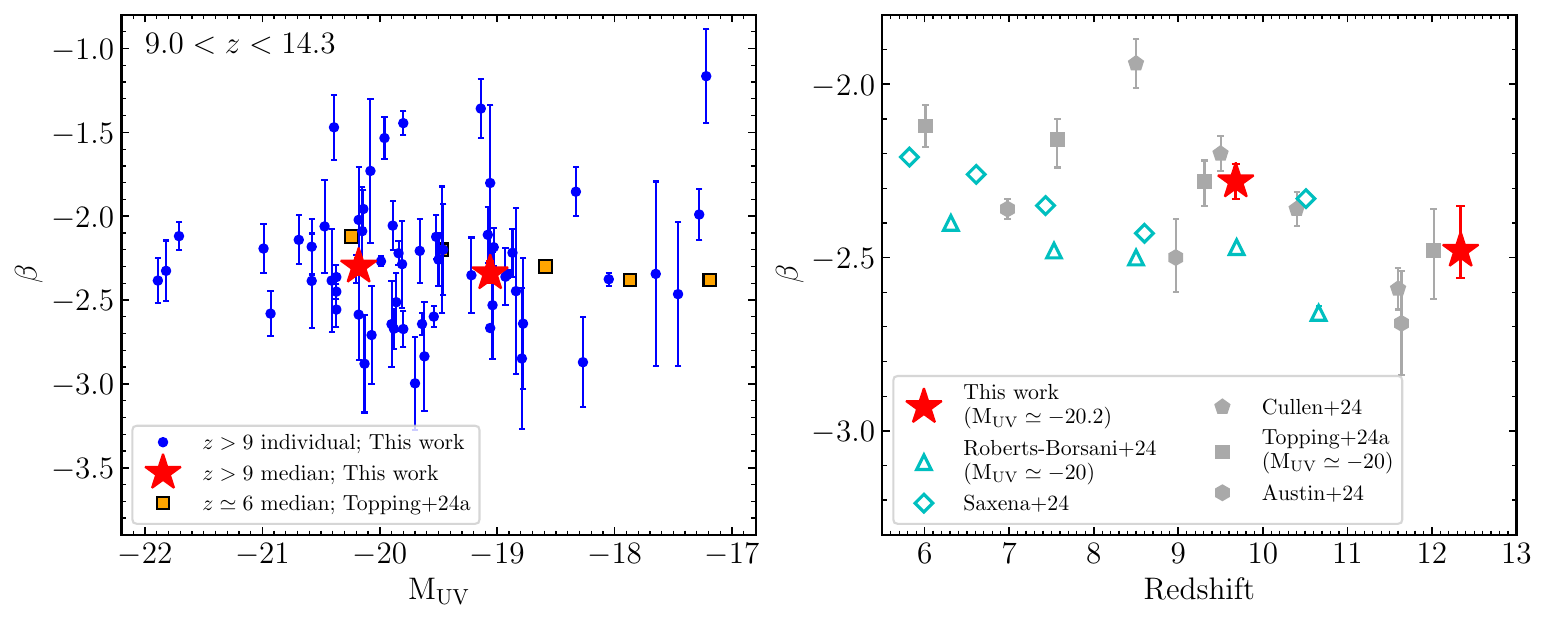}
\caption{Left panel: M$_{\rm UV}$ vs. UV slope of galaxies in our $z>9$ spectroscopic sample (blue circles). UV slopes are derived from NIRSpec prism spectra. The median UV slopes of galaxies with M$_{\rm UV}<-19.8$ (median M$_{\rm UV}<-20.2$) and M$_{\rm UV}>-19.8$ (median M$_{\rm UV}=-19.1$) are shown as red stars. As a comparison, we overplot the $\beta-{\rm M}_{\rm UV}$ relation at $z\simeq6$ \citep{Topping2024b} as orange squares. Right panel: Evolution of UV slopes at $6\lesssim z \lesssim14$. We show the median UV slopes of galaxies in our $z>9$ spectroscopic sample as red stars. As a comparison, we overplot UV slopes of spectroscopic samples in the literature as open cyan symbols (diamonds: \citealt{Saxena2024b}; triangles: \citealt{Roberts-Borsani2024}) and photometrically selected samples as solid grey symbols (pentagons: \citealt{Cullen2024}; squares: \citealt{Topping2024b}; hexagons: \citealt{Austin2025}).}
\label{fig:beta}
\end{figure*}

\subsection{UV continuum slopes} \label{sec:uv_slope}

The UV continuum slopes are sensitive to a range of physical properties. 
The intrinsic slopes are set by the stellar population and ionized gas conditions ($\beta=-2.6$ to $-2.4$) before being reddened by dust in the ISM \citep[e.g.,][]{Cullen2017,Reddy2018,Topping2024b}. 
If $z>9$ galaxies are different from those at $6<z<9$, we may expect to see evolution in the UV colors. 
Early {\it JWST} papers focused on measurement of UV slopes at $z>9$ using NIRCam photometry, revealing blue colors that appear close to the intrinsic values for ionization-bounded H~{\small II} regions ($\beta\simeq-2.6$; e.g., \citealt{Cullen2024,Franco2024,Topping2024b}). 
More recently, spectroscopic samples have grown in number, allowing the UV slopes to be measured with the NIRSpec prism. 
\citet{Saxena2024b} presented an exploration of UV slopes in $295$ spectra at $z>5.5$ including $19$ at $z>9.5$. They found a trend toward bluer slopes at higher redshift over $5.5<z<9.5$. At $z>9.5$, their results indicate that the UV slopes begin to plateau (or even redden), approaching the intrinsic value ($\beta=-2.6$) expected for stellar and nebular continuum. 

We investigate the UV slopes in our spectroscopic sample of $60$ galaxies at $z>9$ (with $41$ at $z>9.5$, $2\times$ more than \citealt{Saxena2024b}). 
In the left panel of Figure~\ref{fig:beta}, we plot the UV slopes as a function of absolute UV magnitude. 
The $z>9$ galaxies are generally blue, with a median UV slope of $\beta=-2.33^{+0.04}_{-0.04}$. 
We divide our sample into two bins of UV luminosity M$_{\rm UV}<-19.8$ (median M$_{\rm UV}=-20.2$) and M$_{\rm UV}\ge-19.8$ (median M$_{\rm UV}=-19.1$), with each group having a same number of systems ($30$ galaxies). 
We find a median UV slope of $\beta=2.30^{+0.05}_{-0.05}$ in UV-luminous systems (M$_{\rm UV}\simeq-20.2$). 
For fainter galaxies (M$_{\rm UV}\simeq-19.1$), the median UV slope is similar to that of the more luminous systems with $\beta=-2.35^{+0.07}_{-0.06}$. 
The absence of a trend with UV slope and absolute magnitude is a departure from what is seen at lower redshifts in the literature, where the UV slopes tend to be bluer in less luminous galaxies \citep[e.g.,][]{Cullen2023,Topping2024b,Austin2025}. 
The photometric UV slopes at $z>9$ also show a shallow $\beta-{\rm M}_{\rm UV}$ relation at M$_{\rm UV}<-19$ (see orange squares in Figure~\ref{fig:beta}).

It is conceivable that the shallower relation between UV slope and absolute magnitude may reflect evolution in the nature of the most luminous galaxies at $z>9$. 
For example, if UV bright galaxies become increasingly dominated by strong bursts at $z>9$ (relative to lower redshifts), the most luminous systems will become weighted toward lower mass systems that have been up-scattered in luminosity. Since less massive galaxies tend to have less attenuation and bluer colors \citep[e.g.,][]{Reddy2010,Finkelstein2012,McLure2018,Shapley2022,Morales2024}, this effect would lead to a flatter slope in the $\beta-{\rm M}_{\rm UV}$ relation at $z>9$. 
It is also possible that we are beginning to probe populations that have yet to build up grains that significantly redden the UV continuum \citep{Narayanan2025} or that the dust that has been formed is ejected \citep{Ferrara2025}.

The UV slopes provide one of our only statistical probes of the nature of galaxies at $z\simeq11-14$. 
Our database includes $14$ sources at $z>11$, which is $\sim2\times$ larger than the spectroscopic samples used in previous prism-based UV slope studies \citep[e.g.,][]{Saxena2024b}. 
Among these $14$ sources, we find an average UV slope of $\beta=-2.37^{+0.05}_{-0.09}$. 
To quantify evolution in UV slopes at $9<z<14$, we consider only galaxies with M$_{\rm UV}<-19.8$. 
At $z=9-11$ we find a median value of $\beta=-2.28^{+0.05}_{-0.05}$, whereas at $z>11$, we find $\beta=-2.48^{+0.13}_{-0.08}$, approaching the intrinsic value expected for galaxies without dust attenuation. 
Hence our results suggest that prism-based UV slopes may become bluer between $z\simeq9$ and $z\simeq14$ (right panel of Figure~\ref{fig:beta}). 
The trend we find is consistent with that seen in most photometric studies \citep{Cullen2024,Topping2024b,Austin2025}. 
While we do not find clear evidence for the plateau or reddening found in \citet{Saxena2024b}, we note that our UV slopes at $z>9$ are broadly consistent with the values reported in that paper. 
Nearly all studies (photometric and prism-based) indicate that average UV slopes approach the intrinsic value of stellar and nebular continuum at $z>9$, likely suggesting minimal reddening from dust. 

With a sample of $60$ $z>9$ galaxies, we can also explore the extremes of the UV color distribution. 
We may expect the reddest slopes to provide signposts of systems with dust attenuation caught during extended off-mode periods of star formation when intrinsic slopes redden \citep[e.g.,][]{Narayanan2025} or extremely dense ionized gas that is ionized by hot stellar populations \citep{Katz2025}. 
We find $5$ $z>9$ galaxies with very red UV slopes ($\beta\gtrsim-1.5$). 
We note that $1$ of the $5$ red galaxies is GHZ9 ($\beta=-1.45$), a potential AGN at $z=10.16$ with a plausible X-ray detection \citep{Kovacs2024,Napolitano2025b}. 
The reddest galaxy is JADES-GS-20064312 ($\beta=-1.17$), the faintest galaxy among the $z>9$ sample (M$_{\rm UV}=-17.2$). 
The spectrum of this source does not reveal H$\beta$ or [O~{\small III}] emission lines (Figure~\ref{fig:zg9_spec}), with $3\sigma$ EW upper limits of $91$ and $93$~\AA, respectively. 
This source appears to be a weak emission line galaxy in addition to having very red UV colors and intrinsically faint UV continuum, all properties that would be expected for a galaxy in a lull of star formation with some dust attenuation. 

The $z>9$ sample also has galaxies with extremely blue UV slopes. 
Galaxies with confident colors indicating $\beta<-2.8$ are difficult to explain unless the nebular continuum has been removed \citep[e.g.,][]{Bouwens2010,Topping2022,Cullen2024,Donnan2025,Dottorini2025,Yanagisawa2025}.
Such blue colors may imply leakage of ionizing photons or a recent downturn in star formation, both of which should reduce the nebular emission contribution to the SED \citep{Topping2024b}. 
There are $5$ galaxies in our sample with $\beta\lesssim-2.8$ (GS-z13-LA, GS-z14-1, CAPERS-COSMOS-109917, JADES-GN-19715, JADES-GN-59720). 
JADES-GN-19715 is the bluest of the subset ($\beta=-3.00$) and has extremely weak rest-frame optical emission lines (H$\beta$ EW is $60$~\AA\ and [O~{\small III}]~$\lambda5007$ EW is $230$~\AA), as is seen in photometric sources with similarly blue sources \citep{Topping2024b}.
To quantify the fraction of galaxies with $\beta<-2.8$ at $z>9$, we follow the approach that \citet{Topping2024b} developed for photometric samples, considering the subset of galaxies with small uncertainties in UV slopes ($\sigma_{\beta}<0.3$). 
This leaves a total sample of $51$ galaxies, including the $5$ sources with $\beta<-2.8$. 
Extremely blue galaxies thus comprise $10^{+6}_{-4}\%$ of our $z>9$ sample. 
We note that this fraction is about $3\times$ larger than that at $z\simeq5-9$ ($3.4\%$; \citealt{Topping2024b}), as may be expected if strong bursts or ionizing photon leakage is more common at $z>9$.

%%%% Figure: UV line EW vs. M_UV %%%%

\begin{figure*}
\includegraphics[width=\linewidth]{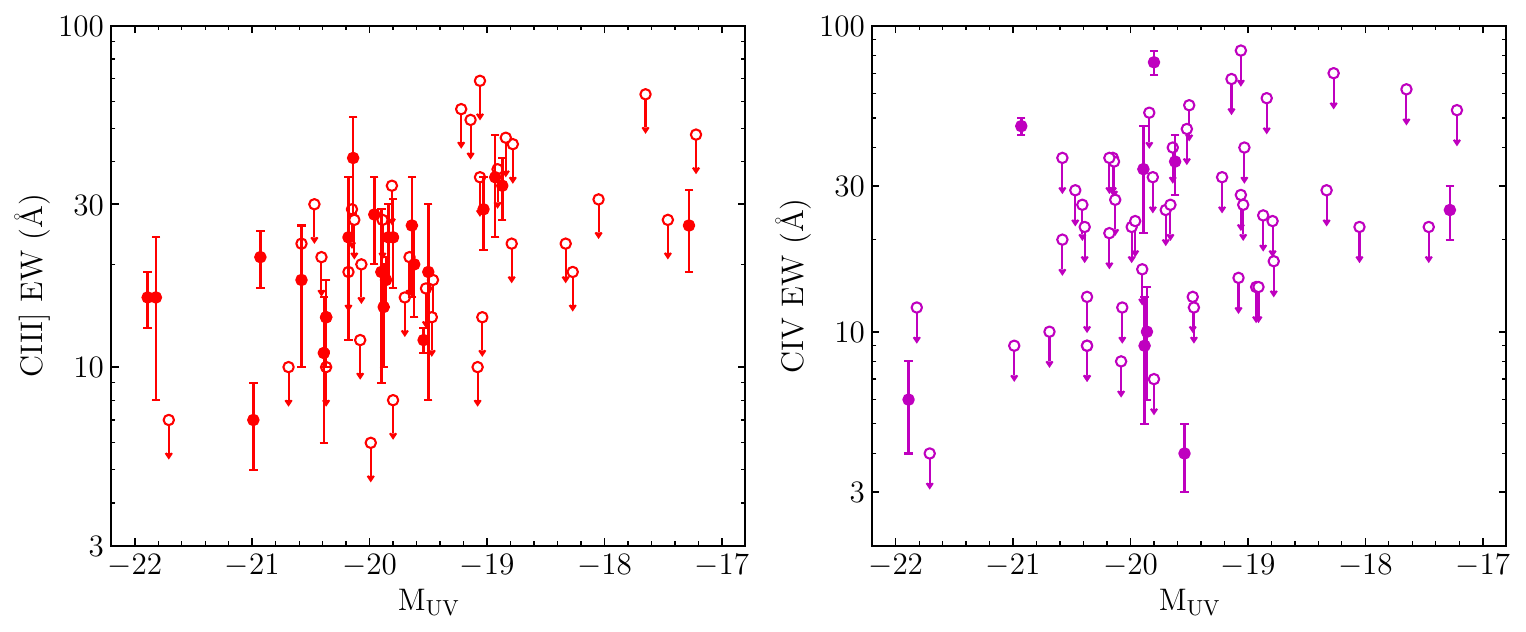}
\caption{C~{\scriptsize III}] (left) and C~{\scriptsize IV} EWs (right) versus M$_{\rm UV}$ of galaxies in our $z>9$ sample. We show detections as solid circles (red: C~{\scriptsize III}]; magenta: C~{\scriptsize IV}) and $3\sigma$ upper limits for non-detections as open circles.}
\label{fig:uvew_muv}
\end{figure*}

\subsection{C~{\small III}] and O~{\small III}] emission lines} \label{sec:c3_o3}

C~{\small III}] is the most commonly detected rest-frame UV emission line among the $z>9$ galaxies. 
Of the $61$ galaxies in our $z>9$ sample, $60$ have prism spectra covering C~{\small III}]. 
In the left panel of Figure~\ref{fig:uvew_muv}, we show the C~{\small III}] EWs of these $60$ galaxies as a function of M$_{\rm UV}$ (see Table~\ref{tab:zg9} for measurements). 
We detect blended C~{\small III}]~$\lambda1908$ emission with S/N $>3$ in the prism spectra in $16$ of these $60$ galaxies. 
We measure C~{\small III}] EWs spanning $7-41$~\AA\ for these $16$ galaxies, with a median of $23$~\AA. 
C~{\small III}] emission in $9$ of the $16$ objects have been reported previously \citep{Bunker2023,Carniani2024,Castellano2024,Curti2025,DEugenio2024,Fujimoto2024,Napolitano2025a}, and our measurements indicate similar C~{\small III}] EWs to those presented in literature. 

We newly identify $7$ galaxies at $z>9$ with C~{\small III}] detections with S/N $>3$ (CAPERS-COSMOS-109917, CAPERS-EGS-25297, CAPERS-EGS-87132, CAPERS-UDS-22431, JADES-GN-55757, JADES-GS-20015720, JADES-GS-20088041; Figure~\ref{fig:zg9_spec}). 
We further note lower S/N C~{\small III}] emission lines (S/N $=2-3$) in other $7$ sources (Table~\ref{tab:zg9}). We consider these as tentative detections requiring deeper spectra for confirmation. 
We do not identify C~{\small III}] in the prism spectra of the remaining $37$ objects, enabling us to place $3\sigma$ upper limits on their C~{\small III}] EWs. 
The median $3\sigma$ C~{\small III}] EW upper limit of these $37$ objects is $23$~\AA. 
We see in Figure ~\ref{fig:uvew_muv} that many of the non-detections are in faint sources, where the EW limits are only conducive to detecting the most extreme line emission. 
At the bright end, we do measure several galaxies with upper limits implying EWs below $5-10$~\AA, indicating that a subset of the population at $z>9$ does have at most moderate strength C~{\small III}] emission. 

C~{\small III}] emission is detected (S/N $=7$) in the composite spectrum of galaxies at $9.0<z<14.3$ in our sample (top panel of Figure~\ref{fig:zg9_comp}), allowing us to infer the average C~{\small III}] EW at $z>9$. 
From the stack we derive an average C~{\small III}] EW of $12.6\pm1.8$~\AA, consistent with the average C~{\small III}] EW ($12.8-13.7$~\AA) measured from the composite spectrum of $30$ galaxies at $z>9$ in \citet{Roberts-Borsani2024}. 
Such large C~{\small III}] EWs are generally found in galaxies with moderately metal poor gas and large sSFRs, as we will show in Section~\ref{sec:zg9_properties}. 

Nineteen galaxies in our sample also have medium resolution grating spectra covering C~{\small III}]. 
In three of the nineteen galaxies, C~{\small III}] is detected in both grating and prism spectra (GN-z11, JADES-GS-55757, GS-z9-0). 
The C~{\small III}] EWs measured from grating and prism spectra are consistent within $1\sigma$ uncertainties. 
Another two galaxies show C~{\small III}] emission in the prism but not the grating (JADES-GS-20015720, GS-z14-0). 
The C~{\small III}] EWs measured from the prism spectra of these two objects are within the $3\sigma$ EW upper limits evaluated from their grating spectra (see also \citealt{Carniani2024}). 

If strong bursts become more common at $z>9$, we should expect to see a larger fraction of the earliest galaxies with large sSFRs and extremely large C~{\small III}] EWs ($>20$~\AA). 
These strong emission lines typically are thought to trace extremely young stellar populations (CSFH age $\lesssim10$~Myr) formed in recent upturns of star formation \citep[e.g.,][]{DEugenio2024,Kumari2024,Topping2024a}.
To test this possibility, we characterize the incidence of galaxies with C~{\small III}] EW $>20$~\AA\ in our $z>9$ sample. 
Here we focus on the subset with deep enough spectra to reach C~{\small III}] EW limits of $20$~\AA\ at $3\sigma$. 
There are $33$ such galaxies in our $z>9$ sample, and $9$ of them have C~{\small III}] detections with EW $>20$~\AA\ with S/N $>3$. 
This results in a strong C~{\small III}] emitter fraction of $9/33=27^{+10}_{-8}\%$ at $z>9$ (uncertainties derived using the statistics for small numbers of events; \citealt{Gehrels1986}). 
We will come back to discuss the evolution of the strong C~{\small III}] emitter fraction in Section~\ref{sec:c3_evolution}.

We also characterize the O~{\small III}]~$\lambda\lambda1661,1666$ doublet with the goal of investigating the carbon-to-oxygen (C/O) ratio of the ionized gas. 
At the resolution of the prism, O~{\small III}]~$\lambda\lambda1661,1666$ tends to be blended with He~{\small II}~$\lambda1640$. 
We detect the O~{\small III}]~$\lambda\lambda1661,1666$ doublet with S/N $>3$ in six individual galaxies (GS-z9-0, GN-z11, GS-z12-0, GHZ2, CAPERS-EGS-25297, CAPERS-EGS-87132). 
The O~{\small III}] detections of four of these six galaxies have been presented previously (GN-z11, \citealt{Bunker2023}; GHZ2, \citealt{Castellano2024}; GS-z12-0, \citealt{DEugenio2024}; GS-z9-0, \citealt{Curti2025}). 
We measure He~{\small II}+O~{\small III}] EWs of $9-20$~\AA\ for these four galaxies, consistent with the values reported in the literature. 
We identify two new detections of the blended He~{\small II}+O~{\small III}] emission lines (CAPERS-EGS-25297, CAPERS-EGS-87132; Figure~\ref{fig:zg9_spec}). 
We derive a He~{\small II}+O~{\small III}] EW of $20\pm6$~\AA\ for CAPERS-EGS-25297 and $20\pm5$~\AA\ for CAPERS-EGS-87132. 

We also detect blended He~{\small II}+O~{\small III}] emission in the composite spectrum of the $61$ galaxies at $9.0<z<14.3$ (top panel of Figure~\ref{fig:zg9_comp}). 
The average He~{\small II}+O~{\small III}] EW measured from the composite is $4.3\pm1.3$~\AA, which is smaller than those of individual detections.
In Section~\ref{sec:CO}, we will use the combination of O~{\small III}] and C~{\small III}] to constrain the C/O ratios in $z>9$ galaxies.

\subsection{C~{\small IV} and He~{\small II} emission lines} \label{sec:c4_he2}

We search for emission lines from high ionization species C~{\small IV} and He~{\small II} in the spectra of $z>9$ galaxies. These lines likely indicate the presence of hard radiation fields, powered by either low metallicity massive stars \citep[e.g.,][]{Stark2015,Mainali2017} or AGNs \citep{Nakajima2018}. 
The C~{\small IV} EWs of galaxies in our $z>9$ sample are shown in the right panel of Figure~\ref{fig:uvew_muv}. 
We detect C~{\small IV}~$\lambda1549$ emission with S/N $>3$ in the prism spectra of $6$ galaxies, with a median EW of $31$~\AA\ (see Table~\ref{tab:zg9}). 
Five of the C~{\small IV} detections have been reported previously in the literature (GN-z11, \citealt{Bunker2023}; GHZ2, \citealt{Castellano2024}; UNCOVER-22223, \citealt{Fujimoto2024}; GS-z9-0, \citealt{Curti2025}; GHZ9, \citealt{Napolitano2025a}). 

The new C~{\small IV} detection is associated with CAPERS-COSMOS-109917 (Figure~\ref{fig:zg9_spec}), with a measured EW of $36\pm8$~\AA. 
We also note the presence of tentative C~{\small IV} detections (S/N $=2-3$) in three galaxies (JADES-GN-17858, JADES-GN-55757, JADES-GS-20015720; Figure~\ref{fig:zg9_spec}). 
However, the vast majority of the galaxies in our sample ($51$ of $60$ objects) do not show C~{\small IV} emission lines. 
Most of the non-detections yield upper limits on C~{\small IV} which can only rule out extremely strong line emission, with a median $3\sigma$ upper limit of $24$~\AA. 
In the composite spectrum, we detect C~{\small IV} emission with S/N of $4$ (top panel of Figure~\ref{fig:zg9_comp}). 
We measure an average C~{\small IV} EW of $5.3\pm1.4$~\AA\ for the $z>9$ galaxies from the stack. 

To constrain how commonly $z>9$ galaxies have hard radiation fields, we quantify the fraction of $z>9$ galaxies with strong C~{\small IV} emission. 
We consider galaxies with deep enough spectra to reach a C~{\small IV} EW limit of $10$~\AA\ at $3\sigma$. 
We identify $13$ such galaxies, with $4$ showing strong C~{\small IV} emission (EW $>10$~\AA) in their spectra. 
This indicates that the fraction of strong C~{\small IV} emitters is $4/13=31^{+18}_{-14}\%$ at $z>9$. 
We will investigate the evolution of this fraction in Section~\ref{sec:c4_evolution}.

The detection of nebular He~{\small II}~$\lambda1640$ emission (ionization potential $=54$~eV) provides another probe of the presence of a hard ionizing spectrum. 
Since He~{\small II} is blended with O~{\small III}] in the prism spectra, we only quantify He~{\small II} emission line strengths in the grating spectra of galaxies in our $z>9$ sample. 
Among the $26$ galaxies with grating spectra, we identify He~{\small II} emission lines in two objects: GN-z11 and GS-z9-0. 
The He~{\small II} emission lines of these two galaxies are relatively narrow (FWHM $=250-330$~km~s$^{-1}$), consistent with a nebular (and not stellar) origin. 
We derive a He~{\small II} EW of $5.7\pm1.8$~\AA\ for GN-z11 and $4.3\pm1.5$~\AA\ for GS-z9-0, comparable to the EWs reported for these two objects in literature \citep{Bunker2023,Curti2025}. 
For the remaining $24$ galaxies without He~{\small II} detection, we place a median $3\sigma$ upper limit on He~{\small II} EW of $14$~\AA. 
Future deep medium or high resolution grating spectroscopic observations are required to place more robust constraints on the He~{\small II} emission in $z>9$ galaxies. 

\subsection{N~{\small IV}] and N~{\small III}] emission lines} \label{sec:n4_n3}

JWST/NIRSpec observations have begun revealing strong N~{\small IV}] and N~{\small III}] emission lines in a handful of $z>5$ galaxies \citep[e.g.,][]{Bunker2023,Castellano2024,Marques-Chaves2024,Topping2025a}, suggesting nitrogen-enriched gas, potentially related to strong bursts of star formation \citep{Topping2024a,Topping2025a} or in other cases AGN \citep[e.g.,][]{Ji2024,Isobe2025}.
With our large $z>9$ sample assembled from the public NIRSpec database, we look for N~{\small IV}] or N~{\small III}] emission lines in galaxies in the very early Universe. 
We identify blended N~{\small IV}]~$\lambda1485$ detections (S/N $>3$) in the prism spectra of $4$ galaxies at $z>9$ (GN-z9p4, GN-z11, UNCOVER-3686, GHZ2), which all have been reported previously 
\citep{Bunker2023,Castellano2024,Fujimoto2024,Schaerer2024}. 
The composite spectrum of galaxies at $9.0<z<14.3$ does show a N~{\small IV}] detection with S/N of $4$ (top panel of Figure~\ref{fig:zg9_comp}). 
We measure an average N~{\small IV}] EW of $5.0\pm1.2$~\AA\ at $z>9$ from the stack. We note that non-detections of N~{\small IV}] are only able to rule out the strongest line emission in most cases. 
The median $3\sigma$ N~{\small IV}] EW upper limit is $24$~\AA, well above the EW of the detected emission lines.

We recover the N~{\small III}]~$\lambda1750$ emission in two galaxies in our sample, GN-z11 \citep{Bunker2023} and GHZ2 \citep{Castellano2024}. 
We measure a N~{\small III}] EW of $14\pm3$~\AA\ for GN-z11 and $13\pm4$~\AA\ for GHZ2, comparable to the EWs reported for these systems in the literature. 
The composite spectrum does not present an N~{\small III}] emission feature, placing a $3\sigma$ upper limit $<4.4$~\AA\ to the average N~{\small III}] EW at $z>9$. 

With the constraints on nitrogen emission lines, we quantify the fraction of galaxies with N~{\small IV}] emission linked to enhanced nitrogen and hard ionizing spectra at $z>9$. 
Because the detected N~{\small IV}] emission lines are relatively weak (EW $=5-11$~\AA), we only consider galaxies with N~{\small IV}] EW limits that reach to $5$~\AA\ at $3\sigma$. 
There are only $6$ such galaxies in our $z>9$ sample.
Although the current statistics are poor, four of these six galaxies present N~{\small IV}] emission, indicating a very large N~{\small IV}] emitter fraction of $67^{+21}_{-28}\%$ at $z>9$. 

%%%% Figure: H-beta EW vs. M_UV %%%%

\begin{figure}
\includegraphics[width=\linewidth]{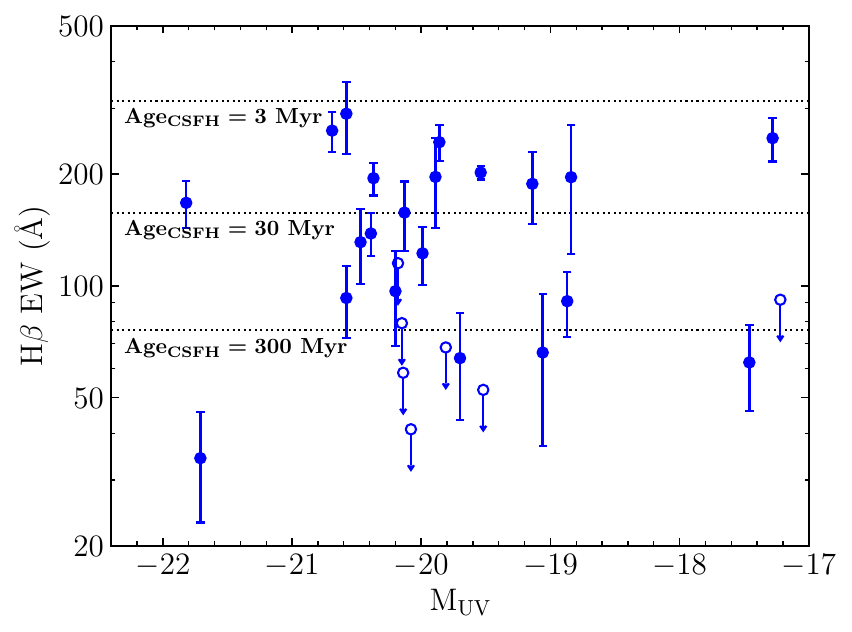}
\caption{H$\beta$ EW versus M$_{\rm UV}$ of galaxies in our $z>9$ sample. We show detections as blue solid circles and $3\sigma$ upper limits for non-detections as open circles. The dotted lines show the corresponding CSFH ages for H$\beta$ EWs. The CSFH ages are derived from \texttt{BEAGLE} models assuming the \citet{Chabrier2003} IMF with stellar masses in the range $0.1-300\ M_{\odot}$, an ionization parameter $\log{U}=-2.5$, and a metallicity $Z=0.1\ Z_{\odot}$ (see Section~\ref{sec:model}).}
\label{fig:hbew_muv}
\end{figure}

\subsection{H$\beta$, H$\gamma$, and Balmer Decrements} \label{sec:balmer}

The hydrogen Balmer emission lines provide constraints on the light-weighted stellar population ages (via H$\beta$ EWs) and the dust attenuation of the H~{\small II} regions (via the Balmer decrements). 
The NIRSpec spectra allow us to measure H$\beta$ and H$\gamma$ emission lines in galaxies at redshifts up to $z\simeq9.8$ and $z\simeq11.1$, respectively. 

Among the $61$ galaxies in our $z>9$ sample, $29$ are at $9.0<z<9.8$ with NIRSpec spectra covering the observed wavelength of H$\beta$. 
We identify H$\beta$ emission lines in the spectra of $20$ galaxies (see Figure~\ref{fig:hbew_muv}). 
The H$\beta$ EWs span a wide range, with 16th-50th-84th percentiles of $91$~\AA, $162$~\AA, and $242$~\AA, respectively. 
For context, stellar population ages of $3$, $30$, and $300$ Myr correspond to H$\beta$ EWs of $300$, $160$, and $80$~\AA, respectively, assuming constant star formation history in the stellar population models described in Section~\ref{sec:model} (Figure~\ref{fig:hbew_muv}).
It is clear that a large fraction of the population appears to have very large H$\beta$ EWs, as might be expected in galaxies caught in strong bursts. 
However, there are also non-detections indicating very weak H$\beta$ ($<75$~\AA), perhaps pointing to systems in a lull following a recent downturn in star formation. 

We also find a large dispersion in the H$\gamma$ EW distribution. 
There are $49$ galaxies at $9.0<z<11.1$ with H$\gamma$ constraints, of which 18 show detections. 
The H$\gamma$ EWs are described by a distribution with 16th-50th-84th percentiles of $39$~\AA, $73$~\AA, and $125$~\AA. 
The average values of the H$\beta$ and H$\gamma$ distributions are consistent with the $9.0<z<9.6$ composite (bottom panel of Figure~\ref{fig:zg9_comp}) which reveals H$\beta$ EW of $150\pm12$~\AA\ and H$\gamma$ EW of $51\pm7$~\AA, similar to the EWs reported in \citet{Roberts-Borsani2024}. 
We will discuss the star formation histories implied by the spectra in Section~\ref{sec:mass_sfh}.

We quantify the fraction of galaxies with extremely large H$\beta$ or H$\gamma$ EWs. 
At $9.0<z<9.8$, we consider systems with H$\beta$ EWs $>240$~\AA, which the EWs are associated with very young stellar populations (CSFH age $\lesssim5$~Myr). 
At higher redshifts where H$\gamma$ is visible ($9.8<z<11.1$), we adopt an equivalent threshold of H$\gamma$ EW greater than $80$~\AA. 
There are $40$ galaxies in our $z>9$ sample with deep enough spectra to reach H$\beta$ EW limit of $240$~\AA\ or H$\gamma$ EW limit of $80$~\AA\ at $3\sigma$. 
Nine of these forty galaxies present H$\beta$ EW $>240$~\AA\ at $9.0<z<9.8$ or H$\gamma$ EW $>80$~\AA\ at $9.8<z<11.1$, indicating a strong Balmer line emitter fraction of $23^{+8}_{-7}\%$. 
We will investigate the evolution of this fraction in Section~\ref{sec:balmer_evolution}.

The H$\gamma$ to H$\beta$ emission line ratios constrain the dust attenuation. 
There are $12$ galaxies in our sample with both H$\gamma$ and H$\beta$ emission detections. 
We derive H$\gamma$/H$\beta$ ratios spanning between $0.436\pm0.087$ and $0.618\pm0.204$. 
Systems with H$\gamma$ non-detections are also consistent with this range. 
Assuming the electron temperature that we infer from the composite spectrum in Section~\ref{sec:opt_lines} ($T_{\rm e}=2.1\times10^4$~K), the intrinsic H$\gamma$/H$\beta$ ratio expected from case B recombination is $0.474$ \citep{Osterbrock2006}. 
We thus find that the H$\gamma$/H$\beta$ ratios are typically consistent with the case B value within uncertainties. 
Assuming the \citet{Cardelli1989} extinction curve, the lowest H$\gamma$/H$\beta$ ratio ($0.436\pm0.087$) suggests an $E(B-V)$ of $0.16^{+0.28}_{-0.16}$. 
The composite spectrum reveals an average H$\gamma$/H$\beta$ ratio ($0.460\pm0.071$) that also indicates a fairly small value of reddening, $E(B-V)$ of $0.06^{+0.20}_{-0.06}$. 
Based on the Balmer line ratios, we find that $z>9$ galaxies have little attenuation toward their H~{\small II} regions, consistent with the conclusions based on the UV continuum slopes.

%%%% Figure: ionization-sensitive line ratio vs. [OIII] EW %%%%

\begin{figure*}
\includegraphics[width=\linewidth]{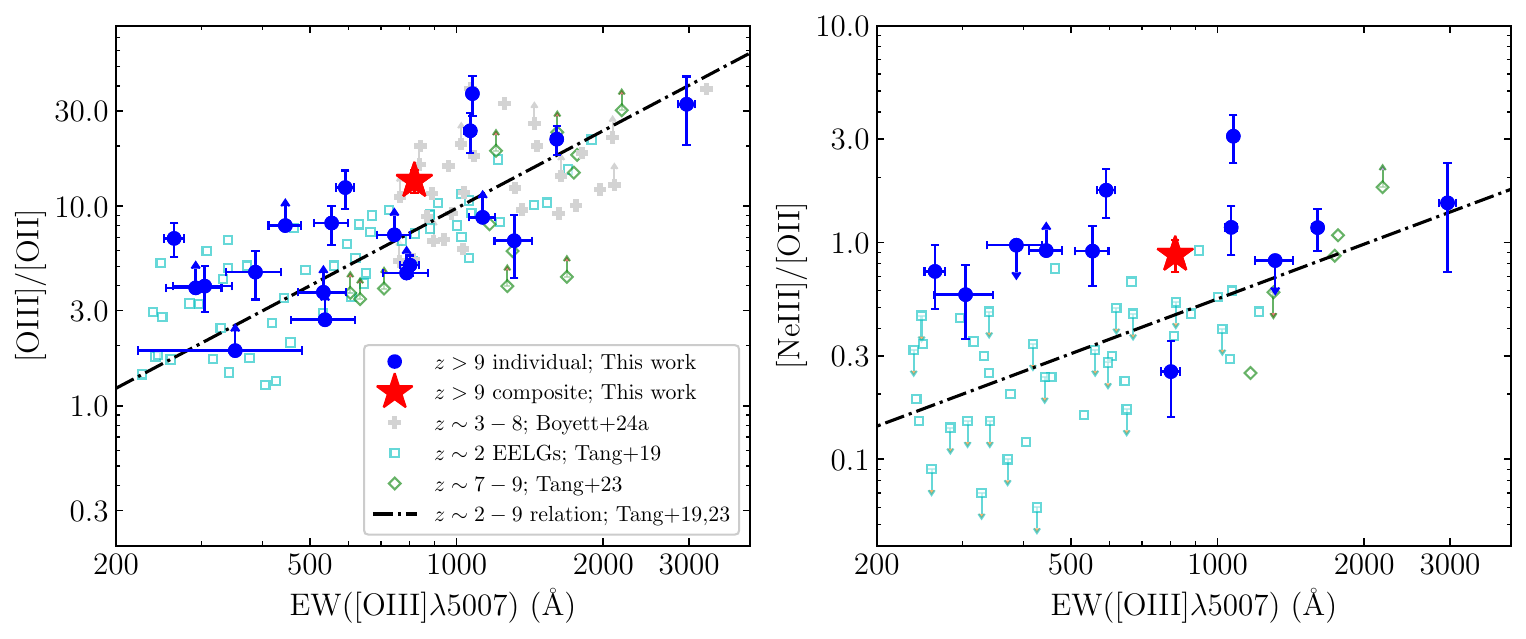}
\caption{Ionization-sensitive line ratios ([O~{\scriptsize III}]/[O~{\scriptsize II}], left; [Ne~{\scriptsize III}]/[O~{\scriptsize II}], right) versus [O~{\scriptsize III}]~$\lambda5007$ EW for $z>9$ galaxies (blue circles). Measurements from the composite spectrum of galaxies at $9.0<z<9.6$ are shown by red stars. For comparison, we overplot the data at $z\sim2-9$ in literature, including extreme emission line galaxies (EELGs) at $z\sim2$ (\citealt{Tang2019}; cyan open squares), {\it JWST} measurements at $z\sim7-9$ (\citealt{Tang2023}; green open diamonds) and at $z\sim3-8$ (\citealt{Boyett2024b}; grey cross). We also show the relationships between line ratios and [O~{\scriptsize III}] EW derived from $z\sim2-9$ dataset \citep{Tang2019,Tang2023} as black lines.}
\label{fig:ion_ew}
\end{figure*}

A subset of the $z>9$ galaxies in our sample have relatively red UV slopes (Section~\ref{sec:uv_slope}), potentially indicating a combination of dust attenuation or redder intrinsic UV slopes. 
We find that these systems have the lowest H$\gamma$/H$\beta$ ratios, potentially indicating moderate (but still relatively low) level of reddening. 
For example, RUBIES-EGS-929993 has $\beta=-1.36$, and its H$\gamma$/H$\beta$ ratio ($<0.44$) suggests an $E(B-V)$ of $>0.1$. 
JADES-GS-20077159 has $\beta=-1.47$ with H$\gamma$/H$\beta=0.436\pm0.087$, indicating an $E(B-V)$ of $0.16^{+0.28}_{-0.16}$. 
The reddest object (JADES-GS-20064312) does not have H$\gamma$ or H$\beta$ detections. 
Deeper spectra are required to better quantify the Balmer decrements of the red systems, as uncertainties currently prevent confident statements on the nebular attenuation.

\subsection{[O~{\small II}], [Ne~{\small III}], and [O~{\small III}] emission lines} \label{sec:opt_lines}

Rest-frame optical emission lines (e.g., [O~{\small II}], [Ne~{\small III}], [O~{\small III}]) provide useful insights into the properties of ionized gas. 
At $z>9$, we are able to access some of the optical lines with NIRSpec in certain redshift windows (e.g., [O~{\small II}] and [Ne~{\small III}] at $z\lesssim12.6$, [O~{\small III}]~$\lambda5007$ at $z\lesssim9.6$). 
We characterize the available rest-frame optical emission lines in the spectra of galaxies in our $z>9$ sample, and we use these measurements to explore the gas conditions and ionizing spectra of $z>9$ systems. 

[O~{\small III}]~$\lambda5007$ is the most luminous emission line in the rest-frame optical. 
We identify [O~{\small III}]~$\lambda5007$ detections in $20$ of the $22$ galaxies at $9.0<z<9.6$ that have coverage of [O~{\small III}]. 
Similar to the H$\beta$ and H$\gamma$ EWs, the [O~{\small III}]~$\lambda5007$ EWs of these $20$ galaxies span a wide range, with 16th-50th-84th percentiles of $301$, $745$, and $1395$~\AA, respectively. 
From the composite spectrum of the $22$ galaxies at $9.0<z<9.6$ in our sample (bottom panel of Figure~\ref{fig:zg9_comp}), we derive an average [O~{\small III}]~$\lambda5007$ EW of $819\pm13$~\AA, broadly consistent with the median [O~{\small III}]~$\lambda5007$ EW measured from individual detections. 
The large dispersion in [O~{\small III}] EWs may point to a range of star formation histories and gas conditions. 
We will discuss this further in Section~\ref{sec:zg9_properties}.

Ionization-sensitive emission line ratios ([O~{\small III}]/[O~{\small II}], hereafter O32, and [Ne~{\small III}]/[O~{\small II}], hereafter Ne3O2) have been found to closely correlate with each other \citep[e.g.,][]{Izotov2021,Schaerer2022,Tang2023,Boyett2024b} and both scale with rest-frame optical emission line EWs in star-forming galaxies over cosmic time ($z\sim0-9$; e.g., \citealt{Tang2019,Tang2023,Sanders2020,Izotov2021,Boyett2024b}). 
This may reflect a variety of physical effects (i.e., ionization parameter, metallicity, density) that correlate with the emission line EWs. 
Here we investigate whether the $z>9$ sample follows similar trends. 
We compute the O32 and Ne3O2 ratios of galaxies with measurements on [O~{\small II}], [Ne~{\small III}], and [O~{\small III}], investigating the ionization state of the ISM in these systems. 
To determine the sample average, we will also consider the composite spectrum and systems where we can only place limits on the line ratios.

We detect [O~{\small II}]~$\lambda3728$ emission in the spectra of $26$ galaxies and [Ne~{\small III}]~$\lambda3869$ emission in $19$ galaxies at $9.0<z<12.6$. 
When computing O32 ratios (and limits on O32), we correct for dust using the $E(B-V)$ inferred from H$\gamma$/H$\beta$ ratios and assuming the \citet{Cardelli1989} extinction curve. 
For those without H$\gamma$/H$\beta$ measurements, we use $E(B-V)=0.06$ inferred from the average H$\gamma$/H$\beta$ ratio (Section~\ref{sec:balmer}). 
In Figure~\ref{fig:ion_ew}, we show O32 (left panel) and Ne3O2 vs. [O~{\small III}]~$\lambda5007$ EW (right panel) for $z>9$ galaxies (blue circles). 
For comparison, we also show the correlations between ionization-sensitive line ratios and [O~{\small III}] EW that have been found at lower redshift\footnote{We fit the relationship between O32 and [O~{\scriptsize III}]~$\lambda5007$ EW for galaxies in \citet{Tang2019} and \citet{Tang2023} using `Orthogonal Distance Regression', a linear regression method which allows us to account for errors along both the x- and y-axis \citep{Chevallard2018}. We find $\log{\rm (O32)}=1.29\cdot\log_{10}{({\rm EW}_{{\rm [OIII]}\lambda5007}/{\rm \AA})}-2.88$. Similarly, we derive the relationship between Ne3O2 and [O~{\scriptsize III}]~$\lambda5007$ EW as: $\log{\rm (Ne3O2)}=0.84\cdot\log_{10}{({\rm EW}_{{\rm [OIII]}\lambda5007}/{\rm \AA})}-2.78$.} (black dash-dotted lines). 

There are two key points to take away from the O32 and Ne3O2 data. 
First, we find that $z>9$ galaxies with larger [O~{\small III}] EWs tend to have larger O32 and Ne3O2 ratios, similar to the general trend seen at lower redshift. 
Second, many of the $z>9$ galaxies have O32 and Ne3O2 ratios above the ionization vs. [O~{\small III}] EW sequence derived at lower redshift. 
Among the $11$ galaxies at $z>9$ with robust O32 measurements (i.e., instead of lower limits), $8$ of them ($73\%$) have O32 larger than those inferred from the O32 $-$ [O~{\small III}] EW sequence at lower-$z$ (left panel of Figure~\ref{fig:ion_ew}). 
We find similar result for Ne3O2 ratios, that $7$ of the $9$ $z>9$ galaxies ($78\%$) with robust Ne3O2 measurements have Ne3O2 larger than the Ne3O2 $-$ [O~{\small III}] EW sequence at lower-$z$ (right panel of Figure~\ref{fig:ion_ew}).
This may suggest that $z>9$ galaxies have even more extreme ionization conditions in their ISM comparing to galaxies at lower redshifts with similar [O~{\small III}] EWs. 
Additionally, if the ionized gas densities are very high in these systems (e.g., $\gtrsim10^4$~cm$^{-3}$, above the critical densities of the $^2$D$^0$ level of the O~{\small II} ion), the low-ionization [O~{\small II}] lines will be suppressed by collisional de-excitation, increasing the O32 and Ne3O2 ratios. 
Collisional de-excitation of [O~{\small II}] has been shown to be important in the strongest bursts \citep[e.g.,][]{Topping2024a}. 
If very high gas densities are also seen in more typical $z>9$ sources (i.e., not just the highest [O~{\small III}] EWs), we may expect to see elevated O32 and Ne3O2 ratios. 

The picture is similar when considering the composite spectrum. 
In the stack of the $22$ galaxies at $9.0<z<9.6$ (bottom panel of Figure~\ref{fig:zg9_comp}), we derive a large average O32 ratio of $13.4\pm1.7$. 
This is about $3-10\times$ larger than the typical O32 seen at $z\simeq1-5$ (O32 $\simeq1-5$; e.g., \citealt{Steidel2016,Sanders2016,Shapley2023}), consistent with line ratios expected for a highly ionized ISM. 
We also derive a large average Ne3O2 ratio ($0.88\pm0.15$) from the stack. 
Both the average O32 and Ne3O2 ratios are above the ionization $-$ EW sequence derived at lower redshift at $>3\sigma$ significance (see red stars in Figure~\ref{fig:ion_ew}), consistent with the picture revealed by individual systems in our sample. 

\subsection{[O~{\small III}]~$\lambda4363$ emission line} \label{sec:o3_4363}

The [O~{\small III}]~$\lambda4363$ auroral emission line provides the potential for constraints on the electron temperature and the gas-phase oxygen abundance. 
There are three galaxies with [O~{\small III}]~$\lambda4363$ detections, with two previously reported in the literature, GS-z9-0 \citep{Curti2025} and GN-z9p4 \citep{Schaerer2024}. 
We newly-identify [O~{\small III}]~$\lambda4363$ in CAPERS-EGS-25297 (Figure~\ref{fig:zg9_spec}). 
We find large [O~{\small III}]~$\lambda4363$/H$\gamma$ ratios ($0.33-0.44$) for these three objects comparing to other $z=2-9$ systems with [O~{\small III}]~$\lambda4363$ detections \citep[e.g.,][]{Nakajima2023,Sanders2024}, suggesting high electron temperatures in the ISM. 
We also identify [O~{\small III}]~$\lambda4363$ auroral line in the composite spectrum of the $9.0<z<9.6$ galaxies in our sample (bottom panel of Figure~\ref{fig:zg9_comp}), allowing us to estimate the average temperature and oxygen abundance of $z>9$ galaxies. 
The [O~{\small III}]~$\lambda4363$/H$\gamma$ ratio measured from the composite is relatively large of $0.42\pm0.12$.
We will use these measurements to calculate the temperature and oxygen abundance of the H~{\small II} regions in $z>9$ galaxies in Section~\ref{sec:OH}.

%%%%%%%%%%%% Physical Properties at z > 9 %%%%%%%%%%%%

\section{Physical Properties at $\lowercase{z>9}$} \label{sec:zg9_properties}

Using the emission line measurements and SEDs of $z>9$ galaxies in our sample, we characterize the stellar masses and SFHs (Section~\ref{sec:mass_sfh}), the gas-phase oxygen abundances (Section~\ref{sec:OH}), the carbon-to-oxygen ratios (Section~\ref{sec:CO}), and the nitrogen-to-oxygen ratios (Section~\ref{sec:NO}). 

%%%% Figure: stellar mass vs. M_UV %%%%

\begin{figure*}
\includegraphics[width=\linewidth]{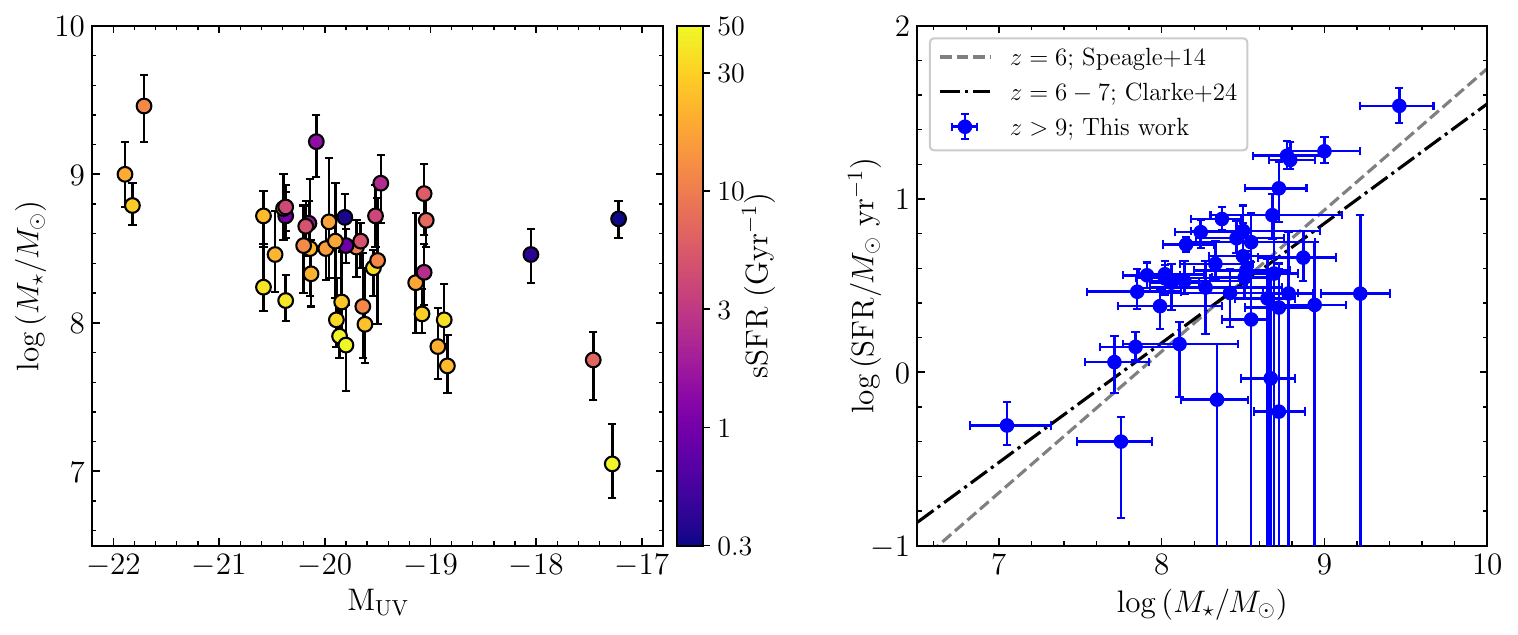}
\caption{Left panel: Stellar mass versus M$_{\rm UV}$ for galaxies at $9.0<z<10.8$ in our sample. The stellar masses are derived from \texttt{BEAGLE} models assuming a two component SFH (Section~\ref{sec:model}). Colors show the sSFRs (averaged over the most recent $10$~Myr) derived from \texttt{BEAGLE}. Right panel: SFR (averaged over the most recent $10$~Myr) versus $M_{\star}$ for galaxies at $9.0<z<10.8$ (blue circles). For comparison, we overplot the star forming main sequence derived from {\it JWST} observations in \citet{Clarke2024} at $z=6-7$ (black dash-dotted line) and derived by \citet{Speagle2014} computed at $z=6$ (grey dashed line).}
\label{fig:mass_muv_sfr}
\end{figure*}

\subsection{Stellar masses and star formation histories} \label{sec:mass_sfh}

We first characterize the stellar populations (stellar masses, SFRs, sSFRs) of the $z>9$ galaxies in our sample. 
Because the rest-frame optical continuum is essential for constraining the stellar mass, we focus on the subset of $45$ galaxies at $9.0<z<10.8$. 
At these redshifts, there is at least one NIRCam filter that covers the continuum redward the Balmer break. 
While line contamination is still a source of error, the emission lines in the rest-frame optical filters at these redshifts tend to be relatively weaker features ([O~{\small II}], [Ne~{\small III}]). 
Our goal in this section is to investigate the stellar masses (and associated uncertainties) and constrain the recent star formation histories. 
We close by comparing the gas-phase properties implied by the photoionization model fits to those derived from the emission line ratios described above.

We describe our fitting methodology and assumptions in Section~\ref{sec:model}. 
To summarize, we use the \texttt{BEAGLE} tool to fit both the NIRCam SED and (if available) the C~{\small III}], H$\gamma$, H$\beta$, and [O~{\small III}]~$\lambda5007$ EWs. 
Our default models consider a two component SFH (TcSFH), consisting of an exponentially delayed component and a recent ($1-20$~Myr) constant star formation component, following the approach introduced in \citet{Endsley2024}. 
The TcSFH model enables flexibility in fitting the recent star formation history, allowing bursts or lulls of star formation in the few Myr before observation. 
Below we will briefly discuss the impact of using several different star formation histories on the inference of stellar mass. 

Our results reveal that the $z>9$ stellar masses inferred from the TcSFH models are all relatively low (left panel of Figure~\ref{fig:mass_muv_sfr}, shown as a function of M$_{\rm{UV}}$), with 16th-50th-84th percentiles of $1.0\times10^8\ M_{\odot}$, $3.0\times10^8\ M_{\odot}$, and $5.2\times10^8\ M_{\odot}$, respectively. 
The stellar masses are at most $\simeq10^9$ M$_\odot$. 
For the subset of systems with stellar masses reported in literature \citep{ArrabalHaro2023a,Castellano2023,Curtis-Lake2023,Tacchella2023a,Curti2025,Kokorev2025}, we compare these values with the stellar masses derived from \texttt{BEAGLE} TcSFH models. 
We find good agreement between the two, with differences less than $0.3$~dex. 

Outshining can play a significant role in driving uncertainties in the stellar masses \citep[e.g.,][]{Roberts-Borsani2020,Laporte2021,Tang2022,Tacchella2023b,Whitler2023b}. 
However, we find this effect is not as pronounced at $z>9$, given the limited cosmic time prior to the redshift of observation. 
In addition to the TcSFH models, we consider Prospector non-parametric SFH models with a continuity prior that allows a significant older component (see Section~\ref{sec:model}). 
We also fit galaxies with CSFH models using the BEAGLE tool. 
In general, we find that the TcSFH and Prospector continuity models give similar stellar masses for our sample, as expected since both setups allow an older stellar component on top of a recent burst of star formation. 
The stellar masses inferred from CSFH models tend to be modestly smaller ($\simeq1.5\times$) than those from TcSFH. 
For the strongest H$\beta$ line emitters ($\rm{EW}>240$~\AA), the variation is at most a factor of two. 
This is a departure from $z\simeq7$, where outshining in young systems could create order of magnitude differences in stellar masses with varying treatments of the SFH \citep[e.g.,][]{Whitler2023b}. 
We will adopt the TcSFH masses in our analysis that follows, but we note that a small uncertainty will be present in the stellar masses of the largest H$\beta$ line emitters. 

In the right panel of Figure~\ref{fig:mass_muv_sfr}, we show the SFRs of our $z>9$ sample as a function of stellar mass (i.e., star forming main sequence, SFMS). 
We define the SFR as that averaged over the most recent $10$~Myr and use the values derived from the TcSFH models. 
The SFRs scale with stellar mass, with 16th-50th-84th percentiles of $1.4$, $3.5$, and $7.2\times10^8\ M_{\odot}\ {\rm yr}^{-1}$. 
Following the approach that is commonly used in literature \citep[e.g.,][]{Pearson2018,Popesso2019,Cole2025,Simmonds2025}, we compare the SFMS at $z>9$ with that derived at $z=6-7$ \citep{Speagle2014,Clarke2024} in the right panel of Figure~\ref{fig:mass_muv_sfr}, which is derived by fitting SEDs assuming a delayed-$\tau$ SFH that is similar to our SFH assumption (Section~\ref{sec:model}). 
We find that the SFRs of $z>9$ galaxies are on average $1.5\times$ higher than that at $z=6-7$ at fixed stellar mass. 
A detailed comparison of evolution in the SFMS would require careful modeling of mass incompleteness at $z\gtrsim9$, which is beyond the scope of this current analysis. 
However, we note that there is a shift toward larger SFR at fixed stellar mass, as expected based on the evolution of baryon accretion rates with redshift \citep{Neistein2008,Dekel2009,McBride2009}. 

The sSFRs constrain the recent SFH in $z>9$ galaxies. 
Here we define the sSFR as that averaged over the $10$~Myr prior to the redshift of observation, and we use the values derived from the TcSFH models. 
The sSFR is constrained by both the emission line EWs and the shape of the NIRCam SED. 
We find that sSFRs of the $z>9$ galaxies tend to be large, with the distribution having 16th-50th-84th percentiles of $2.5$~Gyr$^{-1}$, $14$~Gyr$^{-1}$, and $34$~Gyr$^{-1}$, respectively (left panel of Figure~\ref{fig:mass_muv_sfr}). 
The middle and upper-end of the distribution suggest recent upturns in star formation, whereas the lowest sSFRs correspond to galaxies that have likely experienced a recent lull or downturn in star formation (as discussed below). 

To better quantify the recent SFHs at $z>9$, we calculate the ratio between the SFR averaged over the most recent $3$~Myr (SFR$_{\rm 3Myr}$) and that over the past $3-50$~Myr (SFR$_{\rm 3-50Myr}$) for each galaxy at $z>9$. 
This ratio provides a more direct measure of whether the SFH has recently risen or declined \citep[e.g.,][]{Endsley2025,Kokorev2025}. 
We find that galaxies with the strongest emission lines (H$\beta$ EW $>240$~\AA\ or C~{\small III}] EW $>20$~\AA) have the largest SFR$_{\rm 3Myr}$/SFR$_{\rm 3-50Myr}$ ratios of $>4$ (up to $\simeq15$) inferred from TcSFH models. 
In these cases, the SED needs to be dominated by a young stellar population to reproduce the optical emission line EWs and continuum SED. 
Among the $61$ galaxies at $z>9$, $14$ have SFR$_{\rm 3Myr}$/SFR$_{\rm 3-50Myr}$ ratio larger than $4$ ($>2$ at $84\%$ confidence interval). 
This suggests that $23\%$ of our $z>9$ galaxies may have recently experienced strong upturns in their SFHs. 

We also expect sources to be in an off-mode if star formation is very bursty at $z>9$. 
In these galaxies, we expect weaker emission lines, as O stars disappear from the stellar population. 
At lower redshifts, modest rest-frame optical emission line EWs are common and easily explained by constant star formation histories extending for several $100$~Myr. 
However, at $z>9$ the age of the universe ($\lesssim500$~Myr) is not sufficient to facilitate lower rest-frame optical emission lines if star formation is constant. 
As we showed in Section~\ref{sec:balmer}, the CSFH BEAGLE models require stellar population ages of $\gtrsim500$~Myr to produce H$\beta$ EWs with $<75$~\AA. 
As a result, we find that the sources in our sample with comparably weak rest-frame optical lines tend to be significantly better fit with SFHs having recent SFR downturns, facilitating the diminished emission line strengths. 

Indeed, we find $8$ systems in our $z>9$ sample with weak (or undetected) H$\beta$ emission lines (EW $<75$~\AA). According to the TcSFH models, these sources have low SFR$_{\rm 3Myr}$/SFR$_{\rm 3-50Myr}$ ratios ($<0.2$), indicating significant recent downturns in star formation. 
At $z\simeq6-8$, it has been shown that such systems tend to be found at the faint end of the UV luminosity function \citep[e.g.,][]{Endsley2024,Begley2025}. 
While statistics remain poor at $z>9$, we also find that low inferred SFR$_{\rm 3Myr}$/SFR$_{\rm 3-50Myr}$ ratios are frequently found in faint galaxies. 
For the seven systems with M$_{\rm UV}>-18.5$ in our sample, four ($57\%$) have inferred SFR$_{\rm 3Myr}$/SFR$_{\rm 3-50Myr}$ ratios below $0.2$ (see low sSFR galaxies at faintest magnitudes in the left panel of Figure~\ref{fig:mass_muv_sfr}). 
On the other hand, this fraction decreases dramatically to $9\%$ for galaxies with M$_{\rm UV}<-18.5$. 
These results may suggest that at $z>9$, strong recent downturns in SFR become more common in galaxies with the lowest UV luminosities, as may be expected if the brightest systems tend to be bright in part because they have experienced a recent burst (see also \citealt{Gelli2025}). 
We will discuss the implications of our inferred SFHs for the nature of stellar populations at $z>9$ further in Section~\ref{sec:stellar_pop}.

%%%% Figure: O32 vs. R23 %%%%

\begin{figure}
\includegraphics[width=\linewidth]{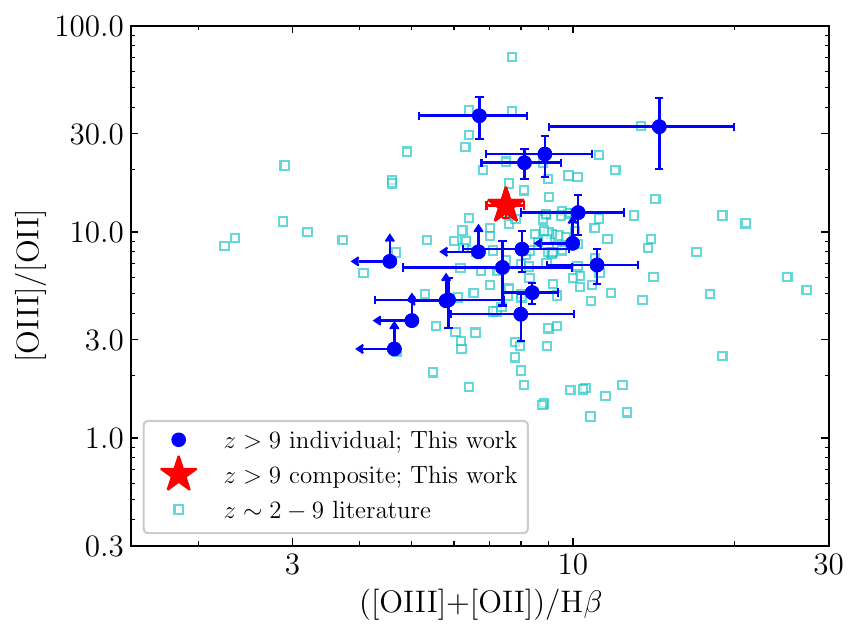}
\caption{O32 versus R23 ratio for for $z>9$ galaxies (blue circles). The average value measured from the composite spectrum of galaxies at $9.0<z<9.6$ is shown by the red star. For comparison, we overplot the data at $z\sim2-9$ in literature \citep{Tang2019,Tang2023,Cameron2023b,Mascia2023,Shapley2023,Boyett2024b,Saxena2024a}.}
\label{fig:o32_r23}
\end{figure}

\subsection{Gas-phase oxygen abundances} \label{sec:OH}

We are able to constrain gas-phase metallicities using the rest-frame optical line ratios reported in Section~\ref{sec:zg9_spectra}. 
The [O~{\small III}]~$\lambda4363$ detection in the composite spectrum enables inference of the direct method metallicity. 
Assuming an electron density ($n_{\rm e}\simeq2\times10^4$~cm$^{-3}$) that is typical of the high-ionization zones of $z>9$ galaxies (derived using [C~{\small III}]~$\lambda1907$/C~{\small III}]~$\lambda1909$ flux ratios in \citealt{Topping2025b}), we infer an electron temperature $T_{\rm e}$(O~{\small III}) of $2.1^{+0.3}_{-0.4}\times10^4$~K for the O$^{++}$ zone using the \texttt{PYTHON} package \texttt{PyNeb} \citep{Luridiana2015} to reproduce the observed line ratios in the composite.
We next derive the O$^+$ zone electron temperature $T_{\rm e}$(O~{\small II}). 
Because the [O~{\small II}]~$\lambda\lambda7320,7330$ auroral lines are not covered by the NIRSpec spectra at $z>9$, we apply the relation $T_{\rm e}$(O~{\small II}) $=0.7\times T_{\rm e}$(O~{\small III}) $+3000$~K \citep{Campbell1986,Garnett1992}. 
Using the measured [O~{\small III}]~$\lambda5007$/H$\beta$ and [O~{\small III}]~$\lambda3728$/H$\beta$ ratios, we infer an average gas-phase oxygen abundance $12+\log{\rm (O/H)}=7.59^{+0.14}_{-0.16}$ ($0.08^{+0.03}_{-0.03}\ Z_{\odot}$, where solar metallicity corresponds to $12+\log{\rm (O/H)}=8.71$; \citealt{Gutkin2016}). 
This indicates moderately metal poor gas is common in the $z>9$ galaxies. 

We also characterize the gas-phase metallicity via strong line ratios: [O~{\small III}]~$\lambda5007$/H$\beta$ (O3) and ([O~{\small III}]+[O~{\small II}])/H$\beta$ (R23) ratios.
For $17$ galaxies with both H$\beta$ and [O~{\small III}]~$\lambda5007$ detections in the NIRSpec spectra, we measure O3 ratios ranging from $3.1$ to $10.2$, with a median value of $5.0$. 
This is consistent with the composite which shows an average O3 of $5.2\pm0.4$. 
Using the relationship between O3 and oxygen abundance derived from $z\simeq2-10$ galaxies in \citet{Sanders2025}, we infer an average oxygen abundance $12+\log{\rm (O/H)}=7.50^{+0.06}_{-0.06}$ ($0.06^{+0.01}_{-0.01}\ Z_{\odot}$) given the composite O3 ratio. 
The R23 ratios suggest a similar picture. 
In Figure~\ref{fig:o32_r23}, we plot the R23 ratios of $z>9$ galaxies. 
We find that the dust-corrected R23 ratios span between $<4.5$ and $15$, with an average of $7.5\pm0.6$ measured from the composite spectrum. 
These are comparable to the R23 ratios seen in star-forming galaxies at lower redshift ($z\sim2-9$; e.g., \citealt{Tang2019,Tang2023,Cameron2023b,Mascia2023,Shapley2023,Boyett2024b,Saxena2024a}). 
Using the R23 $-$ O/H relation in \citet{Sanders2025}, we infer an average $12+\log{\rm (O/H)}=7.30^{+0.07}_{-0.05}$, consistent with the oxygen abundance inferred from O3. 
Both measurements are similar to those reported in \citet{Roberts-Borsani2024} and also comparable to the direct method metallicity derived above. 

The \texttt{BEAGLE} photoionization models also constrain the metallicity through fits to rest-frame optical emission lines. 
In particular, for the composite spectrum at $9.0<z<9.8$, we fit the [O~{\small II}]~$\lambda3728$, [Ne~{\small III}]~$\lambda3869$, H$\gamma$, [O~{\small III}]~$\lambda4363$, H$\beta$, [O~{\small III}]~$\lambda4959$, and [O~{\small III}]~$\lambda5007$ emission line fluxes as well as H$\gamma$ and H$\beta$ EWs.
The flux ratios and EWs are well-reproduced by \texttt{BEAGLE} models. 
The elevated O32 and Ne3O2 ratios at $z>9$ are fit by a large ionization parameter of $\log{U}=-2.11^{+0.08}_{-0.07}$.
The best-fit metallicity inferred by \texttt{BEAGLE} models is $0.11^{+0.02}_{-0.02}\ Z_{\odot}$, corresponding to a gas-phase oxygen abundance $12+\log{\rm (O/H)}=7.76^{+0.05}_{-0.05}$. 
This value is slightly larger than that inferred from the direct method and the O3 and R23 ratios. 
We note that the observed [O~{\small III}]~$\lambda4363$ flux is slightly underpredicted in the best-fitting models (in spite of the strong lines being well-fit), likely contributing to the higher-implied metallicity. 
Slight offsets between metallicities inferred via the direct method and photoionization models are common, potentially reflecting temperature (or other property) variations in the ionized gas that are not captured in the models. Nevertheless, the key result remains that the average gas-phase oxygen abundance in spectroscopic samples at $z>9$ appears to be moderately low, of order $0.04$ to $0.11\ Z_{\odot}$ (Table~\ref{tab:zg9_properties}).

%%%% Figure: abundance patterns %%%%

\begin{figure*}
\includegraphics[width=\linewidth]{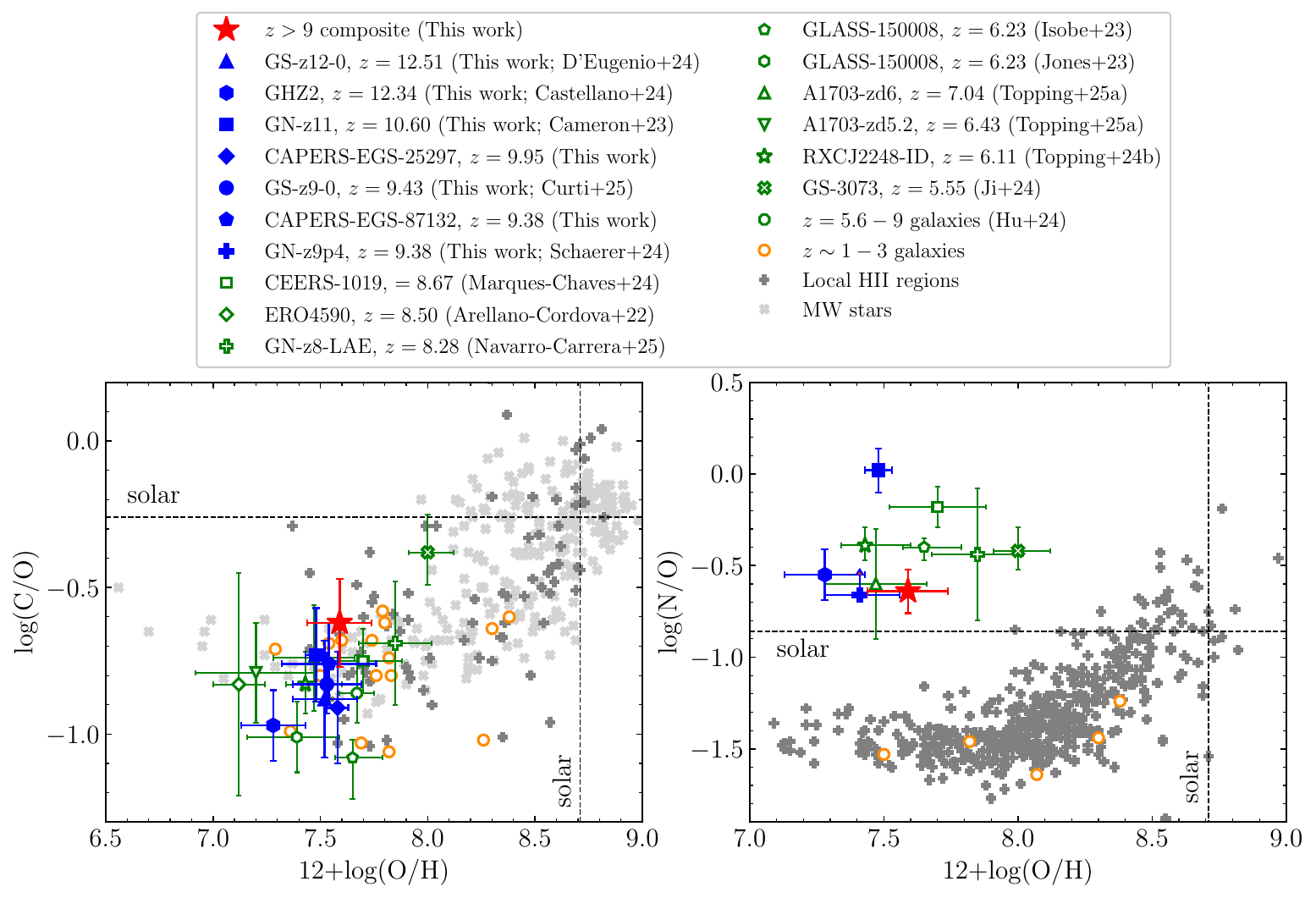}
\caption{C/O ratio (left) and N/O ratio (right) as a function of oxygen abundance O/H. 
We show the abundance patterns of individual $z>9$ galaxies as solid blue symbols (triangle: GS-z12-0; hexagon: GHZ2; square: GN-z11; diamond: CAPERS-EGS-25297; circle: GS-z9-0; pentagon: CAPERS-EGS-87132; plus: GN-z9p4) and the composite spectrum at $z>9$ as solid red stars. For comparison, we overplot the abundance patterns of high redshift galaxies with {\it JWST}/NIRSpec measurements in literature as open green symbols: CEERS-1019 (square, \citealt{Marques-Chaves2024}), ERO4590 (diamond, \citealt{Arellano-Cordova2022}), GN-z8-LAE (plus, \citealt{Navarro-Carrera2025}), GLASS-150008 (pentagon, \citealt{Isobe2023}; hexagon, \citealt{Jones2023}), A1703-zd6 and A1703-zd5.2 (triangle, \citealt{Topping2025a}), RXCJ2248-ID (star, \citealt{Topping2024a}), GS-3073 (cross, \citealt{Ji2024}), and the average at $z=5.6-9$ (octagon, \citealt{Hu2024}). We also present $z\sim1-3$ galaxies (open orange circles, \citealt{Erb2010,Christensen2012,Bayliss2014,James2014,Stark2014,Steidel2016,Amorin2017,Berg2018,Mainali2020,Matthee2021,Iani2023,Citro2024}), local H~{\scriptsize II} regions (solid dark grey plus, \citealt{Esteban2004,Esteban2009,Esteban2014,Garcia-Rojas2004,Garcia-Rojas2005,Garcia-Rojas2007,Lopez-Sanchez2007,Pilyugin2012,Berg2016,Berg2019,ToribioSanCipriano2016,Senchyna2017}), and Milky Way stars (solid light grey cross, \citealt{Gustafsson1999,Akerman2004,Bensby2006,Nissen2014}).}
\label{fig:abundance}
\end{figure*}

\subsection{Carbon-to-oxygen ratios} \label{sec:CO}

At the low metallicities of our galaxies, we expect the gas-phase C/O ratios to be sub-solar, assuming that the C/O $-$ O/H trends seen in local and lower redshift galaxies \citep[e.g.,][]{Garnett1995,Berg2016,Berg2019,Steidel2016} hold at $z>9$. 
However, if the primary or pseudo-secondary production mechanisms of carbon are different in $z>9$ galaxies, we may expect to see variations in the C/O ratios. 
In particular, at extremely low metallicities probed by halo stars and damped Ly$\alpha$ absorbers (DLAs), observations have indicated that C/O ratios begin to increase, reaching close to the solar value \citep[e.g.,][]{Cooke2011}. 
This effect is thought to be related to the yields of the first generation(s) of massive stars. 
{\it JWST} spectroscopy offers the potential to test if a similar C/O `excess' is seen in the lowest metallicity galaxies at very high redshift. 
In this subsection, we will summarize what is known about C/O ratios at $z>9$ given current constraints on the rest-frame UV spectra of the 61 galaxies in our sample. 

We first consider the composite spectrum of galaxies at $9.0<z<14.3$. The stack shows both C~{\small III}] and O~{\small III}] (the latter blended with He~{\small II}), allowing us to constrain the average C/O ratio.
To do so, we first compute the C$^{++}$/O$^{++}$ ratio using the C~{\small III}]/O~{\small III}] flux ratio with \texttt{PyNeb}. 
Here we assume an electron density for highly-ionized gas $n_{\rm e}=2\times10^4$~cm$^{-3}$ \citep{Topping2025b} and an electron temperature $T_{\rm e}=2.1\times10^4$~K inferred from [O~{\small III}]~$\lambda4363$ detection (Section~\ref{sec:OH}). 
Because of the low spectral resolution ($R\sim100$), it is difficult to deblend O~{\small III}] and He~{\small II}. 
To estimate the strength of O~{\small III}], we assume a He~{\small II}/O~{\small III}] flux ratio of $0.25$, which is the typical value inferred from $z>4$ galaxies with He~{\small II} or O~{\small III}] detections in the grating spectra \citep{Bunker2023,Jones2023,Cameron2024,Marques-Chaves2024,Topping2024a,Topping2025a,Curti2025}. 
This results in a C~{\small III}]/O~{\small III}] ratio of $2.5\pm0.8$ and hence C$^{++}$/O$^{++}=0.23\pm0.07$. 
Finally, we convert the C$^{++}$/O$^{++}$ ratio to C/O ratio by applying an ionization correction factor (ICF). 
This step accounts for the fraction of carbon that is triply ionized while oxygen is still primarily doubly ionized due to the higher ionization potential of O$^{++}$ ($55$~eV) than C$^{++}$ ($48$~eV). 
To estimate the ICF, we use the [O~{\small III}]~$\lambda\lambda4959,5007$/[O~{\small II}]~$\lambda3728$ ratio (Section~\ref{sec:opt_lines}) and convert it to the ionization parameter ($\log{U}=-2.06$) and ICF ($1.05$) using the relations in \citet{Berg2019}. 
This gives a value of $\log{\rm (C/O)}=-0.62\pm0.15$ ($0.44\pm0.14$~C/O$_{\odot}$; left panel of Figure~\ref{fig:abundance} and Table~\ref{tab:zg9_properties}). 
We note that the solar abundance ratio is $\log{\rm (C/O)}=-0.26$ from \citealt{Asplund2009}. 
Thus, based on the composite spectrum, we conclude that typical $z>9$ galaxies have sub-solar C/O ratios.

However, it is conceivable that we may find individual examples of $z>9$ galaxies with elevated C/O ratios. 
Using the methods described above, we infer the C/O ratios for the six galaxies with O~{\small III}] and C~{\small III}] detections. 
Our assumptions vary depending on the spectra, so we explain our methods for each source below. 
We note that the C/O ratios of several of these sources have been discussed in the literature previously \citep{Cameron2023a,Castellano2024,DEugenio2024,Curti2025}. 
But to ensure a self-consistent database, we re-derive the C/O ratios with our spectral reductions, even if previous measurements exist. 

One of the first steps is characterizing the electron temperature. 
In the case of two galaxies (GS-z9-0 and CAPERS-EGS-25297), we see [O~{\small II}], [O~{\small III}]~$\lambda4363$, and [O~{\small III}]~$\lambda5007$ detections. 
For these two sources, we derive electron temperatures ($2.0\times10^4$~K and $2.7\times10^4$~K) from the [O~{\small III}]~$\lambda4363$/[O~{\small III}]~$\lambda5007$ ratios with \texttt{PyNeb} and ICF ($1.05$ and $1.30$) from the [O~{\small III}]/[O~{\small II}] ratios \citep{Berg2019}. 
At the higher redshifts of the other four galaxies (GN-z11, GS-z12-0, GHZ2, CAPERS-EGS-87132), the [O~{\small III}] lines are shifted out of the NIRSpec spectra, precluding measurement of the temperature. 
For these systems, we adopt the electron temperature and ICF that was derived from the composite spectrum. 

We next constrain the C~{\small III}]/O~{\small III}] flux ratios. 
The deblended O~{\small III}] emission lines are detected in the medium resolution grating spectra of GS-z9-0 and GN-z11, enabling direct measurement of the flux ratio.
We derive $\log{\rm (C/O)}=-0.83\pm0.10$ ($0.27\pm0.06$~C/O$_{\odot}$) for GS-z9-0 and $\log{\rm (C/O)}=-0.73\pm0.16$ ($0.34\pm0.13$~C/O$_{\odot}$) for GN-z11, consistent with the C/O ratios reported for these sources previously \citep{Cameron2023a,Curti2025}. 
For the other four galaxies (GS-z12-0, GHZ2, CAPERS-EGS-25297, and CAPERS-EGS-87132), we only detect blended He~{\small II}+O~{\small III}] emission in the prism spectra. 
As with our composite, we assume the He~{\small II}/O~{\small III}] ratio is $0.25$. 

Following the steps described above, we derive $\log{\rm (C/O)}=-0.97\pm0.12$ ($0.19\pm0.05$~C/O$_{\odot}$) for GHZ2, consistent with that reported in \citet{Castellano2024}. 
For the two galaxies with newly-identified O~{\small III}] and C~{\small III}], we derive $\log{\rm (C/O)}=-0.91\pm0.19$ ($0.22\pm0.09$~C/O$_{\odot}$; CAPERS-EGS-25297) and $\log{\rm (C/O)}=-0.76\pm0.14$ ($0.32\pm0.10$~C/O$_{\odot}$; CAPERS-EGS-87132). 

We finally consider GS-z12-0, which was first discussed in \citet{DEugenio2024}. 
This source presents a confident detection of C~{\small III}] (S/N $=4$) and a somewhat weaker (S/N $=3$) detection of O~{\small III}] (blended with He~{\small II}). 
Adopting the approach described above, we infer a C/O ratio of $\log{\rm (C/O)}=-0.88\pm0.20$ from the C~{\small III}]/O~{\small III}] measurement.
This implies a sub-solar C/O ratio ($0.24\pm0.11$~C/O$_{\odot}$), similar to what is seen in the other galaxies described above. 
\citet{DEugenio2024} have pointed out that a larger C/O ratio is obtained for this galaxy based on the C~{\small III}]/([O~{\small II}]+[Ne~{\small III}]) ratio. 
However, they find a sub-solar C/O ratio (similar to what we derived above) when using the more standard rest-frame UV-based methods. 
Based on these results, we conclude that there is not enough information to reliably confirm that GS-z12-0 is an outlier in its C/O ratio. 
A deeper grating spectrum separating O~{\small III}] and He~{\small II} should yield a more confident C/O ratio.

There are ten galaxies in our sample that have C~{\small III}] emission without showing detection of O~{\small III}]. 
We compute the $3\sigma$ lower limits on the C/O ratios in these systems, following the same methods described above. 
We find that the spectra imply C/O ratio lower limits ranging from $0.15$~C/O$_{\odot}$ to $0.44$~C/O$_{\odot}$, with a median of $0.22$~C/O$_{\odot}$. 
Deeper spectra will be required to verify the C/O ratios in these systems, but the current lower limits appear consistent with the C/O $-$ O/H trends seen at lower redshifts \citep[e.g.,][]{Garnett1995,Berg2019}.

%%%% Table: physical properties for composite spectra %%%%

\begin{deluxetable}{cc}
\tablecaption{Average physical properties of galaxies in our $z>9$ sample derived for the composite spectra.}
\tablehead{\multicolumn{2}{c}{From \texttt{BEAGLE} photoionization modeling}}
\startdata
Age$_{\rm CSFH}$ (Myr) & $23^{+9}_{-7}$ \\
sSFR (Gyr$^{-1}$) & $43^{+21}_{-12}$ \\
$\log{(\xi_{\rm ion}/{\rm erg}^{-1}\ {\rm Hz})}$ & $25.60^{+0.05}_{-0.06}$ \\
$\log{U}$ & $-2.11^{+0.08}_{-0.07}$ \\
$Z/Z_{\odot}$ & $0.11^{+0.02}_{-0.02}$ \\
$\xi_{\rm d}$ & $0.38^{+0.08}_{-0.14}$ \\
\hline
\hline
\multicolumn{2}{c}{From emission line ratios} \\
\hline
$E(B-V)$ & $0.06^{+0.20}_{-0.06}$ \\
$T_{\rm e}$(O~{\small III}) (K) & $21000^{+3000}_{-4000}$ \\
$12+\log{\rm (O/H)}_{T_{\rm e}}$ & $7.59^{+0.14}_{-0.16}$ \\
$12+\log{\rm (O/H)}_{\rm R23}$ & $7.30^{+0.07}_{-0.05}$ \\
$12+\log{\rm (O/H)}_{\rm O3}$ & $7.50^{+0.06}_{-0.06}$ \\
$\log{\rm (C/O)}$ & $-0.62\pm0.15$ \\
$\log{\rm (N/O)}$ & $-0.64\pm0.12$ \\
\enddata
\tablecomments{C/O and N/O ratios are derived using the O~{\small III}] emission line measured from the composite spectrum at $z>9$, which is deblended from He~{\small II} assuming a He~{\small II}/O~{\small III}] flux ratio of $0.25$ implied from the best-fit \texttt{BEAGLE} model.}
\label{tab:zg9_properties}
\end{deluxetable}

%%%% Figure: composite spectra at 6 < z < 9 %%%%

\begin{figure*}
\includegraphics[width=\linewidth]{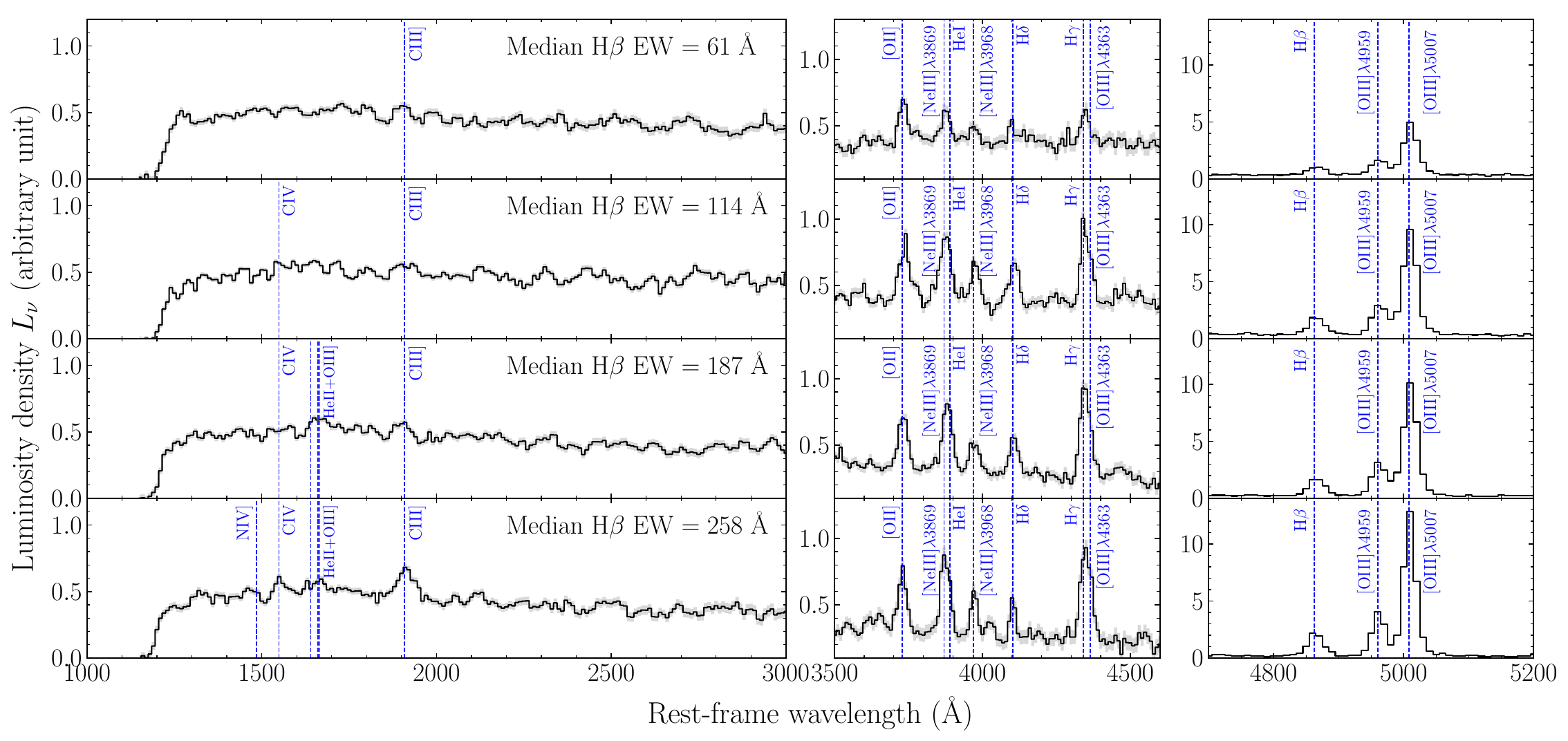}
\caption{Composite prism spectra of galaxies in our $6<z<9$ sample, binned by H$\beta$ EW. From the top to the bottom we show the composites of galaxies with H$\beta$ EW $=10-80$~\AA, $80-150$~\AA, $150-230$~\AA, and $230-500$~\AA. Emission line detections are marked by blue dashed lines.}
\label{fig:z6to9_comp}
\end{figure*}

\subsection{Nitrogen-to-oxygen ratios} \label{sec:NO}

We characterized the N~{\small IV}] and N~{\small III}] emission lines in our $z>9$ sample in Section~\ref{sec:n4_n3}. Previous investigations have computed the N/O ratios for the small number of $z>9$ galaxies with nitrogen line detections, finding evidence for elevated N/O abundance ratios that are similar to those seen in many globular cluster stars (\citealt{Cameron2023a,Senchyna2024,Castellano2024,Schaerer2024,Curti2025}; c.f., \citealt{Flury2025}). 
With a sample size of $61$ galaxies, we can now put the $z>9$ detections in context, both considering a composite spectrum and the large number of non-detections. 
Here we use the emission line measurements to characterize the typical N/O ratios in $z>9$ galaxies. 

We first consider the N/O ratio implied by the composite (top panel of Figure~\ref{fig:zg9_comp}). 
The detection of N~{\small IV}] in the composite hints at nitrogen enhanced conditions. 
In quantifying the N/O ratio, we will assume that the oxygen abundance is dominated by O$^{++}$, following other recent analyses \citep[e.g.,][]{Isobe2023,Castellano2024,Marques-Chaves2024,Topping2024a}. 
Since N~{\small III}] is not detected in the composite spectrum, we derive the N/O ratio from the N~{\small IV}]/O~{\small III}] flux ratio as N/O $\simeq$ N$^{3+}$/O$^{++}$. 
Using \texttt{PyNeb} and assuming $T_{\rm e}=2.1\times10^4$~K (Section~\ref{sec:o3_4363}) and $n_{\rm e}=2\times10^4$~cm$^{-3}$ \citep{Topping2025b}, we derive an average N/O ratio of $\log{\rm (N/O)}=-0.64\pm0.12$ ($1.7\pm0.5$~N/O$_{\odot}$; right panel of Figure~\ref{fig:abundance} and Table~\ref{tab:zg9_properties}), suggesting a super-solar N/O ratio (solar abundance ratio $\log{\rm (N/O)}=-0.86$; \citealt{Asplund2009}). 

This result suggests that typical galaxies at $z>9$ may have nitrogen-enhanced gas conditions. 
To test whether this result is driven by the small number of detections, we generate another composite spectrum without the four nitrogen line emitters (GN-z9p4, GN-z11, UNCOVER-3686, GHZ2). 
Following the same steps described above, we derive an elevated N/O ratio that is consistent with that inferred from the composite including those four systems. 

For completeness, we also assemble the N/O ratios in the four individual $z>9$ galaxies in our sample with N~{\small IV}] detections. 
Two galaxies have spectra that reveal the rest-frame UV nitrogen lines and O~{\small III}] (GN-z11 and GHZ2). 
Both N~{\small IV}] and N~{\small III}] are detected in the NIRSpec spectra of these two systems, and we infer their N/O ratios as N/O $\simeq$ (N$^{3+}$+N$^{++}$)/O$^{++}$. 
We find $\log{\rm (N/O)}=0.02\pm0.12$ ($7.6\pm2.1$~N/O$_{\odot}$) for GN-z11 and $\log{\rm (N/O)}=-0.55\pm0.14$ ($2.0\pm0.6$~N/O$_{\odot}$) for GHZ2 (right panel of Figure~\ref{fig:abundance}), consistent with those inferred in the respective discovery papers \citep{Bunker2023,Cameron2023a,Senchyna2024,Castellano2024}. 
In the other two N~{\small IV}] emitters in our $z>9$ sample (GN-z9p4 and UNCOVER-3686), we do not detect O~{\small III}] emission. 
Using the $3\sigma$ upper limits of O~{\small III}] flux, we derive N/O of $>1.6$~N/O$_{\odot}$ for both GN-z9p4 and UNCOVER-3686. 

Finally, there are four $z>9$ galaxies in our database with O~{\small III}] emission but without nitrogen line detections (GS-z9-0, GS-z12-0, CAPERS-EGS-25297, and CAPERS-EGS-87132). 
Using the $3\sigma$ upper limits of N~{\small IV}] and N~{\small III}] fluxes, we infer N/O ratios of $<0.5-2.9$~N/O$_{\odot}$ for these four systems, with a median $3\sigma$ upper limit of N/O of $<1.8$~N/O$_{\odot}$. 
Deeper spectra will be required to verify if the gas is nitrogen-enriched in these galaxies.

%%%%%%%%%%%% Evolution in Emission Lines at z > 9 %%%%%%%%%%%%

\section{Evolution in Galaxy Spectra at $\lowercase{z}>9$} \label{sec:evolution}

One of the primary goals of this paper is to investigate whether there is evidence for a sudden shift in the burst fraction or star formation efficiency at $z>9$. 
The first step is empirical, establishing whether the spectral properties are notably different than those at lower redshifts.
To facilitate such a comparison, we build a reference sample of $401$ galaxies at $6<z<9$ with NIRSpec observations. 
The selection of spectroscopic sources at $6<z<9$ is conducted similarly to that at $z>9$ and is described in Appendix~\ref{sec:z6to9}. 
The composite spectra of our $6<z<9$ sample is shown in Figure~\ref{fig:z6to9_comp}, separated into bins of H$\beta$ EW (see Appendix~\ref{sec:z6to9}).
In the following subsections, we directly compare the distribution of emission line strengths at $z>9$ to those at $6<z<9$, paying close attention to whether the data reveal the emergence of a population with strong bursts.

\subsection{H$\beta$ and H$\gamma$ emission} \label{sec:balmer_evolution}

The hydrogen recombination lines provide the most direct constraint on the presence of strong bursts. 
If strong bursts become more common at $z>9$, we should see an increase in the presence of galaxies with extremely large H$\beta$ and H$\gamma$ EWs. 
We have demonstrated that $23^{+8}_{-7}\%$ of the $z>9$ galaxies in our sample present extremely strong H$\beta$ (EW $>240$~\AA) or H$\gamma$ emission lines (EW $>80$~\AA) linked to very young stellar populations formed in recent bursts. Using the spectroscopic sample at $6<z<9$, we can test whether this population is becoming more common at earlier epochs. 

\begin{figure}
\includegraphics[width=\linewidth]{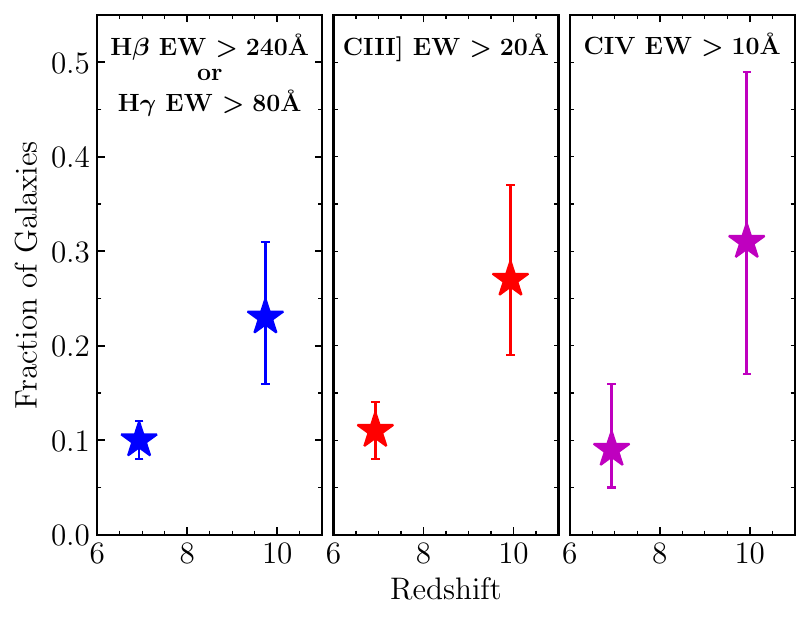}
\caption{Redshift evolution in the fraction of galaxies presenting strong emission lines. The left, middle, and right panels show the fraction of galaxies with strong H$\beta$ (EW $>240$~\AA) or H$\gamma$ (EW $>80$~\AA), C~{\scriptsize III}] (EW $>20$~\AA), or C~{\scriptsize IV} (EW $>10$~\AA) emission lines, respectively.}
\label{fig:eelg_frac}
\end{figure}

Because H$\beta$ emission lines have higher S/N ratios than H$\gamma$ emission lines, we primarily focus on the fraction of galaxies with strong H$\beta$ (EW $>240$~\AA) at $6<z<9$. 
All of the $401$ $6<z<9$ galaxies in our spectroscopic sample have deep enough spectra to reach H$\beta$ EW limit of $<240$~\AA\ at $3\sigma$, and $41$ of them show strong H$\beta$ with EW $>240$~\AA. 
This indicates a fraction of strong H$\beta$ emitters of $10^{+2}_{-2}\%$. 
This fraction is consistent with that inferred from the $6<z<9$ sample in \citet{Roberts-Borsani2024} ($11\%$).
We also quantify the fraction of strong H$\gamma$ emitters (EW $>80$~\AA) at $6<z<9$ using the subset of systems with spectra reaching H$\gamma$ EW limit of $<80$~\AA\ at $3\sigma$ ($141$ objects). 
Among this subsample, we identify $13$ galaxies with H$\gamma$ EW $>80$~\AA, indicating a fraction of $9^{+3}_{-2}\%$ that is similar to the fraction of strong H$\beta$ emitters. 
This reveals that strong Balmer line emitters (H$\beta$ EW $>240$~\AA\ or equivalently H$\gamma$ EW $>80$~\AA) are $\simeq2\times$ more common at $z>9$ than at $6<z<9$ (left panel of Figure~\ref{fig:eelg_frac}), as would be expected if $z>9$ galaxies are significantly more likely to be undergoing strong bursts of star formation. 

\begin{figure*}
\includegraphics[width=\linewidth]{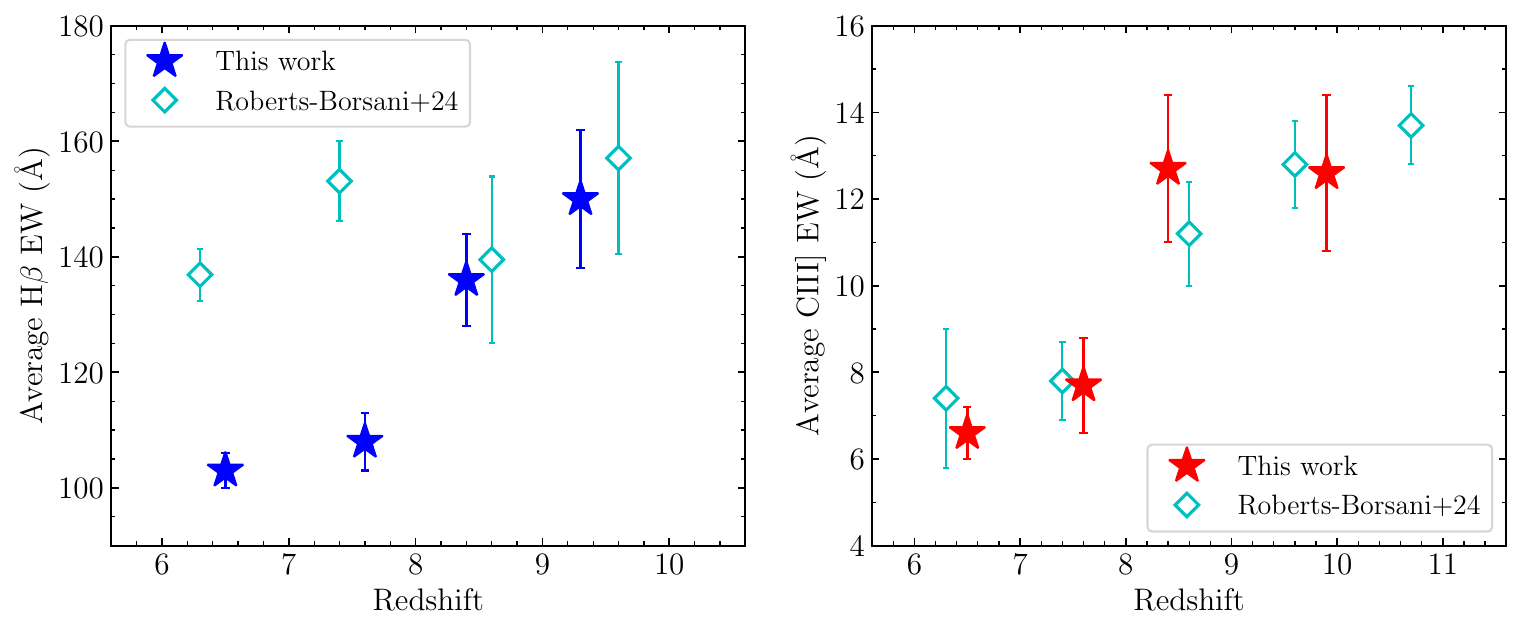}
\caption{Redshift evolution of the average H$\beta$ EW (left panel) and C~{\scriptsize III}] EW (right panel) measured from composite spectra from $6<z<9$ to $z>9$ (blue stars: H$\beta$; red stars: C~{\scriptsize III}]). We overplot data from \citet{Roberts-Borsani2024} as open cyan diamonds as comparison.}
\label{fig:ew_z}
\end{figure*}

We can also investigate the evolution in the mean H$\beta$ EW, using composite spectra centered at $z\simeq6.5$, $7.5$, and $8.5$ (see Appendix~\ref{sec:z6to9}) together with our composite at $z>9$. 
We find $103$~\AA\ ($z\simeq6.5$), $108$~\AA\ ($z\simeq7.5$), $136$~\AA\ ($z\simeq8.5$), while our $z>9$ composite is characterized by H$\beta$ EW $=150$~\AA\ (Table~1). 
Hence our data support mild evolution in the mean H$\beta$ EWs, with slightly larger values at $z>9$ (left panel of Figure~\ref{fig:ew_z}), similar to what was reported in \citet{Roberts-Borsani2024}.

Overall, our database supports evolution in the H$\beta$ EWs at $z>9$, with the most prominent shift at the upper end of the H$\beta$ EW distribution. 
This provides our first line of evidence suggesting that the emission line properties are fundamentally different at $z>9$, with the most extreme emitters forming a larger fraction of the population. 
We note that this is unlikely to be a selection effect since the $z>9$ sample should not be any more biased toward emission line sources than that at $6<z<9$. 
Indeed, at $z>9$, most of the strong lines are redshifted out of the NIRSpec bandpass, reducing the dependence on emission lines for redshift confirmation. 
The rise in the strongest line emitters at $z>9$ is precisely what would be expected if the galaxy population is becoming more dominated by systems that have recently experienced a significant upturn in star formation. 
In the following two subsections, we investigate whether similar signatures are seen in other emission lines. 

%%%% Figure: evolution of strong line emitter fraction %%%%

%%%% Figure: mean EW evolution %%%%

\subsection{C~{\small III}] emission} \label{sec:c3_evolution}

Previous work has demonstrated that the mean C~{\small III}] EW increases with redshift at $z>6$ \citep{Roberts-Borsani2024}. 
Our spectra suggest a similar result. 
We have generated composites from our lower redshift sample, with bins centered at $z\simeq6.5$, $7.5$, $8.5$ (Appendix~\ref{sec:z6to9}). 
The C~{\small III}] EWs of the stacks increase with redshift (right panel of Figure~\ref{fig:ew_z}), from $6.6\pm0.6$~\AA\ at $z\simeq6.5$, to $7.7\pm1.1$~\AA\ at $z\simeq7.5$, to $12.7\pm1.7$~\AA\ at $z\simeq8.5$. 
The C~{\small III}] EW in our $z>9$ composite ($12.6\pm1.8$~\AA) is larger than the range found at $6<z<9$, suggesting continued evolution toward larger C~{\small III}] EWs at higher redshifts.

Perhaps more striking is the fact that a large fraction ($27^{+10}_{-8}\%$) of $z>9$ galaxies have extremely large C~{\small III}] EWs ($>20$~\AA). 
Here we test whether the fraction of strong C~{\small III}] emitters increases sharply with redshift at $z>9$ through comparison to our sample at $6<z<9$.
Among the galaxies with deep enough spectra to reach C~{\small III}] EW limits of $20$~\AA\ at $3\sigma$ ($149$ galaxies), $16$ present C~{\small III}] with EW $>20$~\AA, indicating a large C~{\small III}] EW fraction of $11^{+3}_{-3}\%$ at $6<z<9$. 
This is significantly less than the fraction ($27^{+10}_{-8}\%$) of $z>9$ galaxies that show similarly large C~{\small III}] EWs (middle panel of Figure~\ref{fig:eelg_frac}). 
This suggests that the upper end of C~{\small III}] EW distribution likely undergoes significant evolution at $z>9$, with extremely large C~{\small III}] EWs $2-3\times$ more common than at $6<z<9$. 

To gain intuition for why the C~{\small III}] EW distribution evolves with redshift at $z>9$, we measure the dependence of C~{\small III}] EW on the H$\beta$ EW at $6<z<9$. 
Since the latter quantity tracks the light-weighted stellar population age (and based on Section~\ref{sec:balmer_evolution}, appears to increase at $z>9$), we may expect evolution in the C~{\small III}] EWs if the galaxy population comes to be dominated by stronger bursts or larger star formation efficiency.

\begin{figure}
\includegraphics[width=\linewidth]{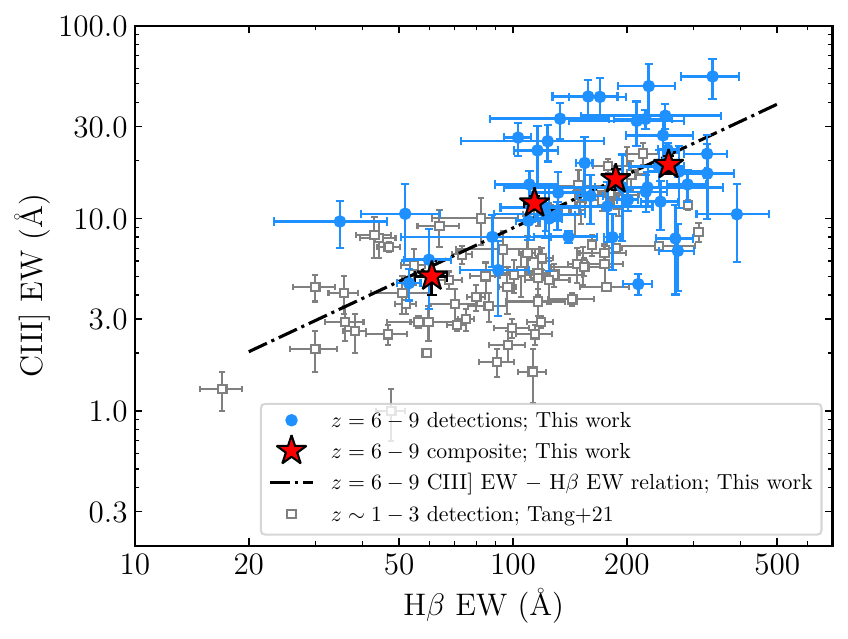}
\caption{Dependence of C~{\scriptsize III}] EW on H$\beta$ EW. We show the average C~{\scriptsize III}] EWs and H$\beta$ EWs measured from the composite spectra at $6<z<9$ (red stars), binning by H$\beta$ EW. The relationship between C~{\scriptsize III}] EW and H$\beta$ EW derived from the composite spectra at $6<z<9$ (Eq. 1) is presented by the black dash-dotted line. We also show the subset with C~{\scriptsize III}] detections with S/N $>3$ at $6<z<9$ (blue circles). As a comparison, we overplot the data at $z\sim1-3$ as grey open squares (see \citealt{Tang2021} and references therein).}
\label{fig:c3ew_hbew}
\end{figure}

As described in Appendix~\ref{sec:z6to9}, we have created stacked spectra in four bins of H$\beta$ EW (Figure~\ref{fig:z6to9_comp}). 
Using the composite spectra, we find that the C~{\small III}] EW increases steadily with H$\beta$ EW (Figure~\ref{fig:c3ew_hbew}), from C~{\small III}] EW $=5\pm1$~\AA\ (at H$\beta$ EW $=61$~\AA) to C~{\small III}] EW $=19\pm2$~\AA\ (at H$\beta$ EW $=258$~\AA). 
The individual galaxies at $6<z<9$ present a similar trend, though with scatter. 
To quantify the connection between C~{\small III}] EW and H$\beta$ EW, we fit the stacked C~{\small III}] EWs and H$\beta$ EWs (in the logarithm) with a linear function: 
\begin{eqnarray}
\log_{10}{({\rm EW}_{\rm CIII]}/{\rm \AA})}&=&0.92\cdot\log_{10}{({\rm EW}_{{\rm H}\beta}/{\rm \AA})} \nonumber \\
& & -0.89
\end{eqnarray}
Based on this relation, we see that the very high C~{\small III}] EW ($>20$~\AA) galaxies which comprise $27\%$ of the population at $z>9$ tend to have H$\beta$ EW $=240$~\AA\ at $6<z<9$, corresponding to light-weighted stellar population ages of $5$~Myr, for constant star formation with BEAGLE. 
We emphasize that this equation gives the average C~{\small III}] EW as a function of H$\beta$ EW. 
The scatter about this relation is driven by variations in gas-phase metallicity and C/O ratios \citep[e.g.,][]{Gutkin2016,Jaskot2016,Plat2019,Tang2021}. 
In the context of the results at $6<z<9$, the observed shift in the C~{\small III}] EW distribution at $z>9$ can be explained by evolution in the light-weighted stellar population ages (which regulate the H$\beta$ EWs) coupled with variations in the gas-phase properties at $z>9$. 
It is also plausible that an increasing fraction of AGN may additionally contribute to the evolution in C~{\small III}] emission.
We will discuss these possibilities in more detail in Section~\ref{sec:discussion}. 

\subsection{C~{\small IV} emission} \label{sec:c4_evolution}

We have found that our composite spectrum at $z>9$ exhibits C~{\small IV} emission (EW $=5.3$~\AA), indicating hard radiation fields capable of triply ionizing carbon are likely typical. 
Moreover, roughly $31\%$ of $z>9$ galaxies have extremely strong C~{\small IV} emission (EW $>10$~\AA). 
Here our goal is to determine if such strong C~{\small IV} emitters are more common at $z>9$ than in similar samples at $6<z<9$, as may be expected if there is a rise in hard ionizing sources at $z>9$.

We first quantify the fraction of $6<z<9$ galaxies presenting C~{\small IV} emission with EW $>10$~\AA\ directly from the spectroscopic sample. 
There are $43$ galaxies with deep enough spectra to reach C~{\small IV} EW limits of $10$~\AA\ at $3\sigma$. 
We identify $4$ with C~{\small IV} detections having EW $>10$~\AA. 
This indicates the fraction of strong C~{\small IV} emitter of $9^{+7}_{-4}\%$ at $6<z<9$, consistent with that of $z>4$ galaxies ($8\%$) presented in \citet{Topping2025a}. 
Notably, this fraction is well below that measured at $z>9$ ($31^{+18}_{-14}\%$; right panel of Figure~\ref{fig:eelg_frac}). 
While statistics remain poor at the highest redshifts, our results suggest that the strongest C~{\small IV} emitters are indeed more common among the $z>9$ population. 

We can gain insight into the physics regulating the emergence of strong C~{\small IV} emitters at $z>9$ by investigating the connection between the C~{\small IV} and H$\beta$ EWs in the composite spectra at $6<z<9$ (Figure~\ref{fig:z6to9_comp}).
In the stack of galaxies with H$\beta$ EW $=10-80$~\AA, we do not identify C~{\small IV} emission, placing a $3\sigma$ upper limit on the average C~{\small IV} EW $<3$~\AA. 
We only start seeing the presence of C~{\small IV} emission in the composites of galaxies with larger H$\beta$ EWs. 
The C~{\small IV} EW increases from $3.2\pm0.9$~\AA\ to $4.9\pm1.3$~\AA\ to $7.1\pm1.4$~\AA\ in the composites of galaxies with H$\beta$ EW $=80-150$~\AA, $150-230$~\AA, and $230-500$~\AA, respectively. 
To quantify the relationship between C~{\small IV} EW and H$\beta$ EW, we fit the logarithm of EWs measured from composite spectra with a linear function: 
\begin{eqnarray}
\log_{10}{({\rm EW}_{\rm CIV}/{\rm \AA})}&=&0.97\cdot\log_{10}{({\rm EW}_{{\rm H}\beta}/{\rm \AA})} \nonumber \\
& & -1.45
\end{eqnarray}

The C~{\small IV} EW $-$ H$\beta$ EW trend suggests a picture that is consistent with what is implied from the demographics of C~{\small IV} detections at lower redshifts. 
The hard ionizing spectra powering strong C~{\small IV} is more typical among metal poor ($\lesssim0.1\ Z_\odot$) galaxies dominated by extremely young stellar populations formed in recent strong bursts of star formation \citep{Senchyna2019,Topping2025a}. 
Based on these results, the increase in the fraction of strong (EW $>10$~\AA) C~{\small IV} emitters at $z>9$ can be explained by the emergence of a metal poor population with extremely large H$\beta$ EW of $>330$~\AA, consistent with the increase in large H$\beta$ and H$\gamma$ EWs presented in Section~\ref{sec:balmer_evolution}. 
Such H$\beta$ EWs correspond to a very young stellar population age of $<3$~Myr assuming CSFH, as expected in galaxies that have experienced a recent strong burst of star formation. 
It is also possible that the evolution in C~{\small IV} emission may reflect the rise of AGN with hard ionizing spectra at $z>9$. 
Indeed, one of the C~{\small IV} emitters (GHZ9) has been claimed as a potential AGN based on X-ray emission \citep{Kovacs2024,Napolitano2025b}. 
We will return to comment on implications of the C~{\small IV} evolution for ionizing sources at $z>9$ in Section~\ref{sec:ionizing_source}. 

%%%% Figure: CIII] EW vs. H-beta EW %%%%

%%%%%%%%%%%% DISCUSSION %%%%%%%%%%%%

\section{Discussion} \label{sec:discussion}

\subsection{Stellar Populations at $z>9$} \label{sec:stellar_pop}

\begin{figure}
\centering
\includegraphics[width=\linewidth]{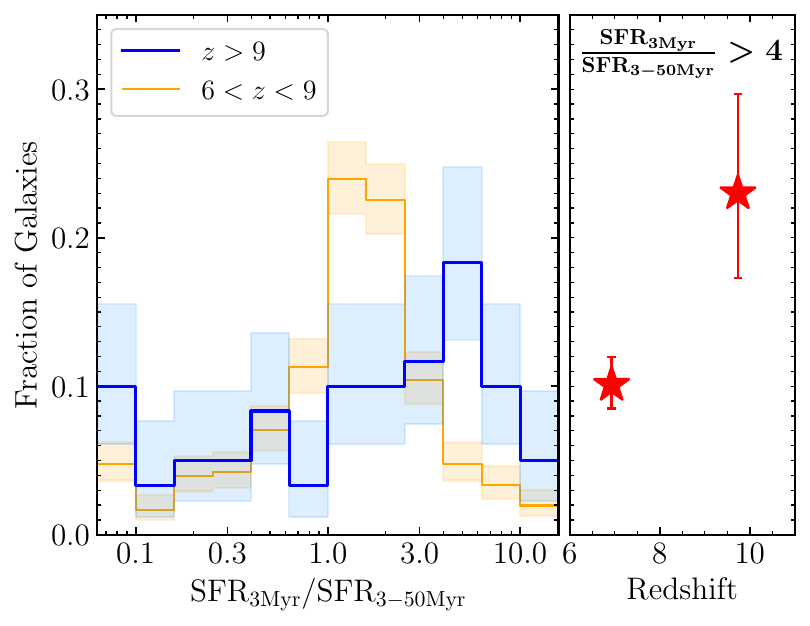}
\caption{Left panel: Distribution of SFR ratios (SFR$_{\rm 3Myr}$/SFR$_{\rm 3-50Myr}$) of galaxies at $z>9$ (blue line) and $6<z<9$ (orange line) in our spectroscopic sample. The SFR ratios are derived from TcSFH models with \texttt{BEAGLE} (Section~\ref{sec:model}), with shaded regions showing $1\sigma$ confidence intervals. Galaxies with large upturns in SFR (SFR$_{\rm 3Myr}$/SFR$_{\rm 3-50Myr}>4$) are more common among $z>9$ galaxies than $6<z<9$ galaxies. Right panel: Evolution of the fraction of galaxies with large upturns in SFR (SFR$_{\rm 3Myr}$/SFR$_{\rm 3-50Myr}>4$) from $6<z<9$ to $z>9$.}
\label{fig:sfr_ratio}
\end{figure}

The spectra of $z\gtrsim9$ galaxies are starting to provide a window on what appears to be a significant transformation in the galaxy population at the highest redshifts probed by {\it JWST}. 
We have shown that H$\beta$, H$\gamma$, and C~{\small III}] emission line EWs undergo rapid evolution between $6<z<9$ and $z>9$, as would be expected if younger stellar populations (formed in recent bursts) come to dominate the spectra at $z>9$. 
Not only is the average emission line EW larger at $z\gtrsim9$ (as previously shown in \citealt{Roberts-Borsani2024}), but a greater fraction of the population is found with extremely large EW line emission. 
In the context of the population synthesis models, these results suggest that nearly $23^{+7}_{-6}\%$ of $z\gtrsim9$ galaxies have undergone strong upturns in recent star formation, with the SFR averaged over the most recent $3$ Myr more than $4$ times that averaged over the previous $3-50$ Myr.
In contrast, only $10^{+2}_{-2}\%$ of the $z\simeq6-9$ population was found in the midst of such strong bursts of star formation. 
To illustrate the implied evolution, we show the inferred SFR$_{\rm 3Myr}$/SFR$_{\rm 3-50Myr}$ ratios in $6<z<9$ and $z>9$ samples in Figure~\ref{fig:sfr_ratio}. 
Overall, the $z>9$ distribution of SFR$_{\rm 3Myr}$/SFR$_{\rm 3-50Myr}$ ratios appears more flattened, with the upper end of the distribution shifted toward larger values. 

One potential explanation for the observed evolution is that the UV scatter ($\sigma_{\rm UV}$) increases with redshift above $z\simeq9$, broadening the range of UV luminosities exhibited by galaxies of a given halo mass. 
In this case, the observed galaxy population at higher redshifts will have a larger contribution from low mass halos which have been upscattered to larger luminosities. 
The impact of $\sigma_{\rm UV}$ evolution has been discussed in the context of the UV luminosity function \citep[e.g.,][]{Mason2023,Mirocha2023,Shen2023,Sun2023,Gelli2024,Kravtsov2024}, providing a plausible explanation for the abundance of the brightest galaxies at $z\simeq14$ (c.f., \citealt{Feldmann2025}). 
However it will also lead to an increase in the prevalence of galaxies with extremely large emission line EWs, as we have shown is likely to be the case at $z\gtrsim9$ (Section~\ref{sec:evolution}, see also \citealt{Roberts-Borsani2024,Kokorev2025}). 
An increase in $\sigma_{\rm UV}$ should also introduce a larger population of galaxies with recent downturns in star formation, broadening the distribution of SFR$_{\rm 3Myr}$/SFR$_{\rm 3-50Myr}$ ratios. 
We have demonstrated some of the weakest emission line sources in our sample are likely to be in such a phase (see Section~\ref{sec:mass_sfh}). 

Other mechanisms may also contribute to an increase in the frequency of galaxies with strong bursts at $z\gtrsim9$. 
These additional explanations mostly require the UV scatter to have a strong dependence on halo mass, with larger $\sigma_{\rm UV}$ in lower mass halos as is often found in hydrodynamical simulations \citep[e.g.,][]{Katz2023,Sun2023,Feldmann2025}. 
In this case, any physical effect which causes higher redshift samples to probe lower mass halos (at a given UV luminosity) will also cause higher redshift samples to sample halos with larger $\sigma_{\rm UV}$. 
One such mechanism is evolution in baryon accretion rates, which are expected to scale as $(1+z)^{2.5}$ \citep{Neistein2008,Dekel2009,McBride2009}, and increasing the SFR at fixed halo mass by a corresponding factor. 
As a result, higher redshift galaxies should tend to sit in lower halo masses at a given M$_{\rm UV}$. 
If $\sigma_{\rm UV}$ is larger in the lower mass population, the higher redshift samples will preferentially probe halos where stronger bursts are present. 
Evolution toward larger SFE at higher redshift \citep[e.g.,][]{Dekel2023,Qin2023,Somerville2025} would also cause the $z\gtrsim9$ samples to sample lower halo masses at fixed M$_{\rm UV}$ than those at lower redshift, further amplifying the evolution in the strong burst fraction. 
Finally, we note that the observed shift in emission line EWs could be additionally impacted by evolution in the initial mass function or an increasing fraction of AGN. 
Meanwhile if there is significant positive evolution in the ionizing photon escape fractions, we may observe a rise in the fraction of weak emission line sources. 
Progress toward decoupling these effects should come with deeper spectroscopy of bright $z>9$ galaxies. 

While the various physical mechanisms described above each broaden the distribution of SFR$_{\rm 3Myr}$/SFR$_{\rm 3-50Myr}$ ratios, they are nevertheless distinct in how they alter the distribution. 
In a future work (Gelli et al., in preparation), we will quantify the impact that the proposed evolution in $\sigma_{\rm UV}$, baryon accretion rates, and SFE (or a shallow relationship between halo mass and SFE as in \citealt{Feldmann2025}) have on the star formation histories. 
This work offers the potential for a new observational constraint on explanations for the slow evolution of the UV luminosity function at $z\simeq9-14$, ensuring that the proposed mechanisms are consistent with the inferred SFR$_{\rm 3Myr}$/SFR$_{\rm 3-50Myr}$ ratios seen in Figure~\ref{fig:sfr_ratio}. 
The utility of such comparisons will be improved as more robust limits are placed on weak emission lines in sources that may have experienced recent downturns of star formation. 
The existing upper limits are often not stringent enough to place meaningful constraints on the recent SFH, particularly at the highest redshifts where we primarily sample the rest-frame UV. 
Approved {\it JWST} programs in Cycle 4 should enable progress through combination of MIRI spectroscopy (probing the rest-frame optical) with new ultra-deep NIRSpec observations (probing the rest-frame UV). 

%%%% Figure: Evolution of SFR ratio %%%%

\subsection{Implications for Ionizing Sources at $z>9$} \label{sec:ionizing_source}

A shift in star formation histories at $z\gtrsim9$ may also help explain the prevalence of the extreme rest-frame UV spectra at the highest redshifts probed by {\it JWST} in the last few years. 
The high ionization emission in GNz11 \citep{Bunker2023} provided the first evidence that hard ionizing agents may be common at $z\gtrsim9$. 
Subsequent investigations have built on this picture, revealing extremely strong C~{\small IV} emission in GHZ2 \citep{Castellano2024}, GHZ9 \citep{Napolitano2025a}, and in MoM-z14 \citep{Naidu2025}. 
The CAPERS program (GO 6368, PI: M. Dickinson) has recently added another strong C~{\small IV} emitter (CAPERS-COSMOS-109917) to the $z>9$ database (Section~\ref{sec:c4_he2}). 

In this paper, we have demonstrated that $31\%$ of $z>9$ galaxies have C~{\small IV} emission with EW $>10$~\AA\ (Figure~\ref{fig:eelg_frac}), more than three times the percentage of the $6<z<9$ population with similarly strong C~{\small IV}. 
While AGN activity may help power C~{\small IV} emission in some cases (perhaps likely in GHZ9 given its X-ray detection, \citealt{Napolitano2025a}), metal poor stellar populations with young ages can power C~{\small IV} emission up to EW $>60$~\AA\ (Plat et al. 2025, in preparation). 
This model-based prediction is supported by the observed strong relationship between C~{\small IV} EW and H$\beta$ EW (Section~\ref{sec:c4_evolution}). 
Based on these results, we suggest that the rise of hard ionizing spectra at $z>9$ is likely a consequence of the increase in the percentage of the $z>9$ population that is experiencing strong bursts of star formation (i.e., those with the largest H$\beta$ EWs and largest C~{\small III}] EWs).

Given the shift in star formation conditions, we may also expect to see evolution in the ionized gas properties at $z>9$. 
There has been considerable focus on the nitrogen-enhanced abundance pattern seen in GNz11 \citep{Bunker2023} and GHZ2 \citep{Castellano2024} (along with similar galaxies at slightly lower redshifts, e.g., \citealt{Isobe2023,Marques-Chaves2024,Topping2024a,Topping2025a}). 
Here we find that the composite spectrum of 61 $z>9$ galaxies shows evidence for N~{\small IV}] emission, with line ratios that suggest mild nitrogen enhancement (N/O $=1.7\pm0.5$~N/O$_{\odot}$) is common in $z>9$ galaxies. 
The physics that causes the nitrogen-enriched abundance pattern remains debated \citep[e.g.,][]{Nagele2023,Cameron2023a,Charbonnel2023,Isobe2023,Isobe2025,Kobayashi2024,Senchyna2024,Flury2025,Ji2025}, but the demographics suggest a link to the high density conditions which form during compact bursts of star formation \citep{Schaerer2024,Topping2024a,Topping2025a}. 
\citet{Topping2025a} have demonstrated that nitrogen-enhancement only becomes common among galaxies with [O~{\small III}]+H$\beta$ EW $>2500$~\AA. 
While such strong bursts are relatively rare at $z\simeq6$, our results suggest they become fairly common at $z>9$, providing a natural explanation for the rise of nitrogen-enhancement in the early galaxy population. 
The nitrogen emitters may provide signposts of dense stellar environments where stellar collisions are likely \citep{Charbonnel2023,Marques-Chaves2024,Senchyna2024}. 
These dense star clusters may provide the ideal environments for the build-up of intermediate mass black holes (IMBHs; e.g., \citealt{Miller2002,PortegiesZwart2002,Kremer2020,Rantala2025}; see \citealt{Greene2020} for a review). 
The {\it JWST} spectroscopic database is revealing that galaxies at $z\gtrsim9$ are more likely to have nitrogen-enhancements than those at $6<z<9$, as we might expect if galaxies are more likely to form stars in denser cluster complexes at $z\simeq9-14$. 
If our initial spectroscopic census is correct, we should continue to find extreme spectra like GNz11 and GHZ2 as $z\gtrsim9$ samples grow in number, allowing us to better study the potential connection between the nitrogen emitters and dense star cluster or IMBH formation.

\subsection{Dust in $z>9$ Galaxies} \label{sec:dust}

The dust content of $z>9$ galaxies has been a topic of interest in the last several years \citep[e.g.,][]{Ferrara2023,Narayanan2025}, largely driven by {\it JWST} measurements of UV continuum slopes. 
NIRCam provided our first statistical exploration of UV slopes at $z\gtrsim9$, revealing extremely blue colors indicative of negligible attenuation \citep[e.g.,][]{Cullen2024,Morales2024,Topping2024b,Austin2025}. 
Prism spectroscopy has suggested a similar picture, with the average UV colors approaching the values expected from the combined emission of stellar and nebular continuum. 
At redshifts above $z\simeq9$, \citet{Saxena2024b} have suggested that the evolution in UV colors may plateau (with UV colors potentially even becoming slightly redder), potentially reflecting the strong contribution of nebular continuum to the underlying SED.

In this paper, we have measured UV continuum slopes of $60$ $z\simeq9-14$ galaxies using the NIRSpec prism spectra, a factor of $2\times$ larger than the sample considered in \citet{Saxena2024b}. 
The average UV slopes we measure are similarly blue as those found in \citet{Saxena2024b}.
However, our spectroscopic sample suggests that the UV slopes may continue to evolve toward bluer colors between $z\simeq9.7$ ($\beta=-2.28^{+0.05}_{-0.05}$) and $z\simeq12.3$ ($\beta=-2.48^{+0.13}_{-0.08}$), continuing the trend seen at slightly lower redshifts. 
While uncertainties remain significant at $z\gtrsim9$, in our prism dataset, we do not find clear evidence for a plateau above $z\simeq9$. 
This is consistent with the trend seen in photometric SEDs of the parent population of continuum-selected galaxies \citep{Cullen2024,Topping2024b,Austin2025}. 
Notably these studies have demonstrated UV slope evolution at $z\gtrsim9$ among fainter (M$_{\rm UV}\simeq-18.5$) galaxies. 
Accordingly, for spectroscopic samples to fully capture evolution in UV slopes at $z\gtrsim9$, it will be necessary to extend existing $z\gtrsim9$ samples to lower luminosities (see Figure~\ref{fig:beta}). 
Larger samples would be able to clarify whether UV slope evolution slows down at the highest redshifts ($z\simeq12$), as might be expected at redshifts where galaxies have negligible dust and minimal scatter in intrinsic slopes. 

It remains an outstanding question whether there is a population of dusty sources at $z\gtrsim9$. 
As described in Section~\ref{sec:uv_slope}, we have identified five red ($\beta\gtrsim-1.5$) galaxies at $z\gtrsim9$ in our current sample. 
In the context of the picture put forward by \citet{Ferrara2023}, these galaxies may correspond to galaxies without sufficient sSFR to eject their dust (see \citealt{Ziparo2023}) or those caught before the dust has been expelled. 
Red colors may also be expected in periods between bursts of star formation \citep{Mirocha2023,Narayanan2025} or in galaxies with high electron densities ($n_{\rm e}\gtrsim10^5$~cm$^{-3}$) and stellar populations dominated by very hot stars \citep{Katz2025}. 
In the latter case, very red slopes arise because the spectrum becomes increasingly nebular-dominated at high stellar temperatures in the far-to-near-UV, with densities large enough to collisionally de-excite the contribution from two photon emission in the far-UV. 
The slope of the nebular continuum SED without two photon emission is red without requiring dust. 
Which physical picture is responsible for the red galaxies at $z\gtrsim9$ is not clear. 

Knowledge of the dust content in the five systems is one of the keys to progress. 
Perhaps not surprisingly, the red galaxies have the largest inferred dust optical depths in our sample. 
Using the TcSFH \texttt{BEAGLE} models with a SMC extinction curve, we find that the $V$-band optical depths are in the range $\tau_V$ of $0.23-0.35$. 
Of course, these inferences depend on knowledge of the intrinsic continuum and the dust law. 
If the intrinsic spectrum is redder than allowed in our TcSFH models (as would be predicted in the high density nebular-dominated SEDs suggested in \citealt{Katz2025}), then the inferred dust attenuation will be overestimated. 
We can place further constraints on the impact of dust via the Balmer decrement. 
As discussed in Section~\ref{sec:balmer}, the two red sources with constraints on H$\beta$ and H$\gamma$ have Balmer decrements that are consistent with modest reddening (i.e. $E(B-V)\simeq0.2$). 
One of the two sources (JADES-GS-20077159) has the smallest H$\gamma$/H$\beta$ ratio in the sample, suggesting that in addition to showing the reddest UV slopes, these sources also have Balmer decrements that suggest the most nebular reddening in the sample. 
More complete coverage of the Balmer lines should yield more confident constraints on the attenuation in the future.

The emission lines in the red galaxies also provide useful insight. 
If the red colors arise during a post burst phase, we may expect to see weaker emission lines as O stars disappear \citep[e.g.,][]{Endsley2025}. 
In contrast, if they are high gas density galaxies with very hot stars, we should see strong emission line spectra from transitions with large critical densities (i.e., [O~{\small III}]~$\lambda\lambda4959,5007$). 
Intriguingly, the reddest object in our sample, JADES-GS-20064312 ($\beta=-1.17$) has no emission line detections in its spectrum. 
To reproduce the weak emission lines, the TcSFH SED fits of this source require the lowest SFR$_{\rm 3Myr}$/SFR$_{\rm 3-50Myr}$ ratios in the sample. 
However, in two of the other galaxies with red UV colors, we find extremely strong emission line spectra (CAPERS-EGS-25297 and GHZ9), either providing signposts of AGN or strong bursts of star formation. 
It is plausible these sources could be genuinely dusty, but we also cannot rule out the high density nebular continuum model posted in \citet{Katz2025}. 
Higher resolution spectra probing the far-UV should enable constraints on the ionized gas density in the future, enabling more detailed modeling of the origin of the red UV continuum colors.

%%%%%%%%%%%% SUMMARY %%%%%%%%%%%%

\section{Summary} \label{sec:summary}

We present an analysis of {\it JWST}/NIRSpec spectra of $61$ galaxies at $z=9-14$, characterizing their spectroscopic properties. 
By comparing against the population at $6<z<9$, we quantify the evolution in the stellar populations, gas properties, and dust content. 
We summarize our key results below. 

1. Using the latest publicly available NIRSpec database, we construct a sample of $61$ spectroscopically confirmed galaxies at $z>9$. Spectroscopic redshifts of $30$ of these systems are newly confirmed. We measure the rest-frame UV to optical emission line properties and UV continuum slopes for subsets of the $61$ galaxies, assembling a sample with size that is at least $2\times$ larger than previous studies of NIRSpec spectra of galaxies at $z>9$ \citep{Roberts-Borsani2024,Saxena2024b}. 

2. We derive the UV slopes of galaxies at $z>9$ using NIRSpec prism spectra. The $z>9$ galaxies are generally blue, with a median UV slope of $\beta=-2.33^{+0.04}_{-0.04}$. We find a flat trend between UV slopes and M$_{\rm UV}$, in contrast to what has been seen at lower redshifts. At fixed M$_{\rm UV}$ ($<-19.8$), galaxies become slightly bluer between $9<z<11$ (median $\beta=-2.28^{+0.05}_{-0.05}$) and $z>11$ (median $\beta=-2.48^{+0.13}_{-0.08}$). The average UV slope of $z>11$ galaxies approaches the intrinsic value expected from stellar and nebular continuum, suggesting minimal obscuration by dust, as found in previous analysis of spectra and multi-band imaging. 

3. We identify five $z>9$ spectroscopically confirmed galaxies with red UV slopes ($\beta\gtrsim-1.5$). \texttt{BEAGLE} models infer the largest dust optical depths for these five red systems among our sample. The only two sources with Balmer line measurements reveal small Balmer decrements, providing independent evidence for modest reddening from dust. These sources may represent a population of galaxies at $z>9$ with significant dust attenuation. Alternatively, the red colors may arise in galaxies dominated by nebular continuum emission with high enough electron densities to collisionally de-excite the two photon continuum \citep{Katz2025}. We demonstrate that the emission lines span a range of properties, suggesting a mixture of star formation histories may be linked to the red colors. 

4. We quantify the evolution in the distributions of H$\beta$, C~{\small III}], and C~{\small IV} EWs from $z\sim6$ to $z>9$. Using the composite spectra, we measure that the average H$\beta$ EW and C~{\small III}] EW increase by a factor of $\simeq2$ from $z\sim6$ to $z>9$. The shift at the upper end of the EW distributions is similar in magnitude. We find that $23-31\%$ of the $z>9$ galaxies show extremely large EWs (H$\beta$ EW $>240$~\AA, C~{\small III}] EW $>20$~\AA, or C~{\small IV} EW $>10$~\AA), which are $2-3\times$ larger than that at $6<z<9$. We quantify the 
evolution in star formation histories (parameterized by SFR$_{\rm 3Myr}$/SFR$_{\rm 3-50Myr}$ ratios) that would be required to explain the emission line EW evolution using \texttt{BEAGLE}. The fits indicate an overall shift toward larger SFR$_{\rm 3Myr}$/SFR$_{\rm 3-50Myr}$ ratios, with the strongest SFR upturns becoming significantly more common between $6<z<9$ and $z>9$. We discuss the connection between the shift in emission line properties and the slow evolution in the UV luminosity function at $z>9$, commenting on how evolution and mass dependence in several quantities ($\sigma_{\rm{UV}}$, SFE) may alter the inferred star formation histories. 

5. We constrain the average ionized gas properties of $z>9$ galaxies. The composite spectra indicate large ionization-sensitive line ratios (O32 $=13.4$, Ne3O2 $=0.88$), as expected if the ISM of $z>9$ galaxies is highly ionized or has very high densities. The [O~{\small III}]~$\lambda4363$ detection enables calculation of the gas-phase metallicity via the direct method, revealing a moderately low gas-phase oxygen abundance of $0.08\ Z_{\odot}$. The C~{\small III}] and O~{\small III}] measurements imply a sub-solar C/O ratio ($0.44$~C/O$_{\odot}$), consistent with the C/O $-$ O/H trends seen at lower redshifts. The composite spectrum also presents the N~{\small IV}] emission line, with line ratios suggesting nitrogen-enhanced abundance (N/O~$=1.7$~N/O$_{\odot}$). At lower redshifts, this abundance pattern is found primarily in galaxies with extremely young stellar populations formed in recent bursts of star formation where high SFR surface densities are likely present. We suggest that this abundance pattern may arise more frequently at $z>9$ as stronger bursts become more common in the galaxy population. The shift in star formation histories may also explain the increased incidence of hard ionizing agents in $z>9$ galaxies, as seen by the prevalence of strong C~{\small IV}-emitters in the existing spectroscopic database.

%%%%%%%%%%%% ACKNOWLEDGEMENT %%%%%%%%%%%%

%% Also note that the akcnowlodgment environment does not support long amounts of text. If you have a lot of people and institutions to acknowledge, do not use this command. Instead, create a new \section{Acknowledgments}.

\section*{acknowledgments}

The authors thank the anonymous referee for insightful comments.
We would like to thank Ad\`ele Plat for useful discussions.
We also thank St\'{e}phane Charlot and Jacopo Chevallard for providing access to the \texttt{BEAGLE} tool used for SED fitting analysis. 
MT acknowledges funding from the \textit{JWST} Arizona/Steward Postdoc in Early galaxies and Reionization (JASPER) Scholar contract at the University of Arizona. 
DPS acknowledges support from the National Science Foundation through the grant AST-2109066. 
CAM, VG, and ZC acknowledge support by the Carlsberg Foundation under grant CF22-1322.
The Cosmic Dawn Center (DAWN) is funded by the Danish National Research Foundation under grant DNRF140.

This work is based on observations made with the NASA/ESA/CSA \textit{James Webb Space Telescope}. 
The data were obtained from the Mikulski Archive for Space Telescopes at the Space Telescope Science Institute, which is operated by the Association of Universities for Research in Astronomy, Inc., under NASA contract NAS 5-03127 for \textit{JWST}. 
These observations are associated with the following public-available programs: GTO 1180, 1181, 1210, 1286, 1287, and GO 3215 (JADES, doi: 10.17909/8tdj-8n28; \citealt{Rieke2023_doi}), ERS 1324 (GLASS, doi: 10.17909/kw3c-n857; \citealt{Treu2023_doi}), ERS 1345 and DDT 2750 (CEERS, doi: 10.17909/z7p0-8481; \citealt{Finkelstein2023_doi}), GO 2561 (UNCOVER), GO 4233 (RUBIES), GO 6368 (CAPERS), GO 1871, GO 3073, GO 4287. 
The authors acknowledge the above teams led by Daniel Eisenstein \& Nora L\"uetzgendorf, K. Isaak, Tommaso Treu, Steven L. Finkelstein, Pablo Arrabal Haro, Ivo Labb\'e \& Rachel Bezanson, Anna de Graaff \& Gabriel Brammer, Mark Dickinson, John Chisholm, and Marco Castellano for developing their observing programs. 
Part of the data products presented herein were retrieved from the Dawn \textit{JWST} Archive (DJA). 
DJA is an initiative of the Cosmic Dawn Center, which is funded by the Danish National Research Foundation under grant DNRF140. 
This work is based in part upon High Performance Computing (HPC) resources supported by the University of Arizona TRIF, UITS, and Research, Innovation, and Impact (RII) and maintained by the UArizona Research Technologies department.


\vspace{5mm}
%\facilities{}


\software{\texttt{NumPy} \citep{Harris2020}, \texttt{Matplotlib} \citep{Hunter2007}, \texttt{SciPy} \citep{Virtanen2020}, \texttt{Astropy} \citep{AstropyCollaboration2013,AstropyCollaboration2018,AstropyCollaboration2022}}, \texttt{BEAGLE} \citep{Chevallard2016}, \texttt{CLOUDY} \citep{Ferland2013}, \texttt{Prospector} \citep{Leja2019,Johnson2021}, \texttt{PyNeb} \citep{Luridiana2015}. 



\appendix
\restartappendixnumbering

\section{Spectroscopic Sample at $6<\lowercase{z}<9$} \label{sec:z6to9}

In this section, we describe the selection of $6<z<9$ sample and emission line measurements. 
Our selection of $6<z<9$ galaxies largely follows our approach to identifying $z>9$ galaxies described in Section~\ref{sec:sample} as well as our methods described in earlier papers \citep{Tang2024c,Topping2025a}. 
We create redshift catalogs from our reductions of NIRSpec observations in a large number of programs in the public archive (see Section~\ref{sec:data}). 
Unlike at $z>9$, strong rest-frame optical emission lines are accessible with NIRSpec throughout the $6<z<9$ redshift range. Redshift confirmation is typically achieved via identifications of multiple emission lines (i.e., H$\beta$, [O~{\small III}], H$\alpha$). 
We identify $393$ galaxies at $6<z<9$ with rest-frame optical emission line detections. 
The precise spectroscopic redshifts of these $393$ galaxies are computed by simultaneously fitting the available strong optical lines with multiple Gaussian profiles and using the fitted line centers. 
To maximize the sample size and avoid bias against sources with weak emission lines, we also search for objects with Ly$\alpha$ breaks but no emission line detections at $6<z<9$. 
We identify $8$ such objects. 
In total we assemble the $6<z<9$ NIRSpec sample containing $401$ galaxies. 

Among the $401$ galaxies in our $6<z<9$ sample, there are $55$, $118$, $73$, and $155$ galaxies in the Abell 2744, EGS, GOODS-N, and GOODS-S fields, respectively. 
$332$ of these $401$ galaxies have low resolution ($R\sim100$) prism spectra. 
Medium resolution ($R=1000$) grating spectra have been obtained in $243$ galaxies, and $49$ more galaxies have high resolution ($R=2700$) grating spectra. 
There are $223$ systems with both prism and grating spectra. 

Using the methods introduced in Section~\ref{sec:spec_measure}, we measure the fluxes and EWs of the available rest-frame UV to optical emission lines in both the prism and grating spectra of galaxies in our $6<z<9$ sample. 
More details about the rest-frame optical spectra of individual systems at $6<z<9$ will be presented in a future paper. 
Here we focus on H$\beta$ emission lines and rest-frame UV lines (i.e., C~{\small III}], C~{\small IV}) in $6<z<9$ galaxies, with the goal of establishing the connection between H$\beta$ EW and UV line EW. 

We detect H$\beta$ emission lines in $374$ of the $401$ galaxies in our $6<z<9$ sample. 
For the $332$ galaxies with prism spectra, we identify H$\beta$ emission detections in $307$ systems. 
From the prism spectra we measure a median H$\beta$ EW of $117$~\AA\ for these $307$ galaxies, with 16th-84th percentiles of $59$~\AA\ and $229$~\AA, respectively. 
For the $292$ galaxies with medium or high resolution grating spectra, we detect H$\beta$ emission lines in $252$ objects. 
For these $252$ systems we measure that the 16th-50th-84th percentiles of H$\beta$ EW are $43$~\AA, $104$~\AA, and $237$~\AA, respectively. 
There are $185$ galaxies at $6<z<9$ with H$\beta$ detections in both prism and grating spectra. 
We find that the H$\beta$ fluxes and EWs measured from both spectra for these $185$ objects are consistent, with a standard deviation of the difference between both EWs just $\simeq0.05$~dex. 
Combining the H$\beta$ EW measurements from prism and grating spectra, the overall H$\beta$ EW distribution of the $374$ galaxies with H$\beta$ detections in our $6<z<9$ sample has 16th-50th-84th percentiles of $53$~\AA, $112$~\AA, and $230$~\AA.

C~{\small III}] emission lines are detected (S/N $>3$) in $48$ galaxies at $6<z<9$. 
We identify C~{\small III}] detections in the prism spectra of $24$ galaxies, with 16th-50th-84th percentiles of C~{\small III}] EW of $10$~\AA, $15$~\AA, and $33$~\AA, respectively. 
For galaxies with medium or high resolution grating spectra, we detect C~{\small III}] emission in $30$ systems, with 16th-50th-84th percentiles of C~{\small III}] EW of $8$~\AA, $14$~\AA, and $32$~\AA. 
Six galaxies have C~{\small III}] detections in both their prism and grating spectra. 
The C~{\small III}] emission line fluxes and EWs measured from both spectra for these six galaxies are broadly consistent, with a standard deviation of the difference between both EWs $\simeq0.13$~dex. 
For the vast majority of galaxies in our $6<z<9$ sample that we do not detect C~{\small III}] emission ($332$ systems), we place $3\sigma$ upper limits to their C~{\small III}] EWs. 
The median $3\sigma$ upper limits of C~{\small III}] EWs derived from prism and grating spectra are $33$~\AA\ and $36$~\AA, respectively. 

We identify C~{\small IV} emission detections in the NIRSpec spectra of $7$ galaxies in our $6<z<9$ sample. 
We measure C~{\small IV} EWs of these seven galaxies ranging from $3.8$~\AA\ to $49$~\AA, with a median C~{\small IV} EW of $12$~\AA. 
C~{\small IV} measurements of six of these seven systems at $6<z<9$ have been reported in literature previously (CEERS-397, JADES-GN-954, JADES-GS-13041, JADES-GN-1899, UNCOVER-10646, CEERS-1019; \citealt{Fujimoto2024,Tang2024b,Tang2024c,Topping2025a,Witstok2025a}). 
We newly-identify C~{\small IV} emission detection in CEERS-1027 (R.A. $=214.88300$, Decl. $=52.84042$). 
The redshift of CEERS-1027 ($z=7.819$) was first confirmed from the CEERS observations using the medium resolution ($R=1000$) NIRSpec grating spectra \citep{Heintz2023,Nakajima2023,Tang2023,Sanders2024}. 
GO 4287 program has obtained high resolution ($R=2700$) rest-frame UV (G140H/F100LP) spectrum for CEERS-1027. 
We detect C~{\small IV} doublet emission lines from the GO 4287 G140H spectrum, with a total EW of $3.8\pm1.0$~\AA.

Finally, to investigate how the average spectroscopic properties change with redshift and H$\beta$ EW, we create composite $6<z<9$ spectra. 
We first stack individual prism spectra binning by redshift. 
We group systems at $6<z<7$ (median $z\simeq6.5$), $7<z<8$ (median $z\simeq7.5$), and $8<z<9$ (median $z\simeq8.5$), with each bin containing $181$, $109$, and $42$ galaxies, respectively.
Composite spectra are stacked following the same procedures described in Section~\ref{sec:stack} for the $z>9$ composite. 
We also creates composites of the prism spectra binning by H$\beta$ EW. 
To do so, we group galaxies with H$\beta$ EW $=10-80$~\AA\ (median $61$~\AA), $80-150$~\AA\ (median $114$~\AA), $150-230$~\AA\ (median $187$~\AA), $230-500$~\AA\ (median $258$~\AA).
The four groups have $99$, $103$, $56$, and $49$ galaxies, respectively. 
The composite spectra of our $6<z<9$ galaxies binning by H$\beta$ EW are shown in Figure~\ref{fig:z6to9_comp}. 
We identify C~{\small III}]~$\lambda1908$ emission lines with S/N of $5-7$ in all the four $6<z<9$ composite spectra grouped by H$\beta$ EW. 
We also detect emission line features with S/N of $3-5$ at the expected wavelength of C~{\small IV}~$\lambda1549$ in the composite spectra of galaxies with H$\beta$ EW $=80-150$~\AA, $150-230$~\AA, and $230-500$~\AA, as well as N~{\small IV}]~$\lambda1485$ in the composite of galaxies with H$\beta$ EW $=230-500$~\AA.
At rest-frame optical wavelengths, we identify a suite of emission lines including [O~{\small II}]~$\lambda3728$, [Ne~{\small III}]~$\lambda3869$, H$\delta$, H$\gamma$, [O~{\small III}]~$\lambda4363$, H$\beta$, [O~{\small III}]~$\lambda4959$, and [O~{\small III}]~$\lambda5007$.
We measure the available emission line properties from the composite spectra with the same methods described in Section~\ref{sec:spec_measure}. 

\section{Table and NIRSpec Spectra of Galaxies at $\lowercase{z}>9$}

\startlongtable
\centerwidetable
\begin{deluxetable*}{cccccccccc}
\tablecaption{NIRSpec spectroscopically confirmed galaxies at $z>9$.}
\tablehead{
ID & PID & R.A. & Decl. & $z_{\rm spec}$ & M$_{\rm UV}$ & $\beta$ & EW$_{\rm CIII]}$ (\AA) & EW$_{\rm CIV}$ (\AA) & Ref.
}
\startdata
CAPERS-EGS-94554 & 6368 & $214.877988$ & $52.856067$ & $9.01$ & $-18.84$ & $-2.45\pm0.50$ & $<47$ & $<58$ & \\
RUBIES-EGS-952693 & 4233 & $214.977359$ & $52.926498$ & $9.03$ & $-20.47$ & $-2.06\pm0.28$ & $<30$ & $<29$ & \\
JADES-GS-20077159 & 1286 & $53.133607$ & $-27.844892$ & $9.05$ & $-20.39$ & $-1.47\pm0.20$ & $11\pm5$ & $<22$ & \\
JADES-GN-619 & 1181 & $189.158251$ & $62.221361$ & $9.07$ & $-19.99$ & $-2.27\pm0.03$ & $<6$ & $<22$ & (16) \\
JADES-GS-20083087 & 1286 & $53.067692$ & $-27.837861$ & $9.07$ & $-19.52$ & $-2.12\pm0.13$ & $<17$ & $<46$ & \\
RUBIES-EGS-929993 & 4233 & $215.125158$ & $52.986532$ & $9.17$ & $-19.14$ & $-1.36\pm0.18$ & $<53$ & $<67$ & \\
JADES-GN-17858 & 1181 & $189.142208$ & $62.284594$ & $9.21$ & $-19.89$ & $-2.06\pm0.14$ & $<27$ & $34\pm13$ & (16) \\
CAPERS-UDS-28597 & 6368 & $34.465142$ & $-5.217735$ & $9.21$ & $-20.58$ & $-2.18\pm0.17$ & $18\pm8$ & $<20$ & \\
CAPERS-UDS-146485 & 6368 & $34.249791$ & $-5.142105$ & $9.26$ & $-20.18$ & $-2.59\pm0.27$ & $24\pm12$ & $<37$ & \\
CAPERS-UDS-142042 & 6368 & $34.249398$ & $-5.130833$ & $9.26$ & $-20.20$ & ... & ... & ... & \\
CAPERS-UDS-22431 & 6368 & $34.460257$ & $-5.185003$ & $9.27$ & $-20.37$ & $-2.56\pm0.11$ & $14\pm4$ & $<9$ & (21) \\
CAPERS-UDS-127376 & 6368 & $34.436059$ & $-5.101416$ & $9.27$ & $-19.81$ & $-2.29\pm0.26$ & $<34$ & $<32$ & \\
CAPERS-COSMOS-118821 & 6368 & $150.090314$ & $2.306922$ & $9.30$ & $-19.66$ & $-2.21\pm0.19$ & $<21$ & $<26$ & \\
JADES-GN-19715 & 1181 & $189.138322$ & $62.289869$ & $9.30$ & $-20.13$ & $-2.88\pm0.29$ & $<27$ & $<27$ & (16) \\
RUBIES-UDS-29385 & 4233 & $34.434963$ & $-5.271119$ & $9.31$ & $-20.58$ & $-2.39\pm0.28$ & $<23$ & $<37$ & \\
UNCOVER-3686 & 2561 & $3.617199$ & $-30.425536$ & $9.32$ & $-21.71$ & $-2.12\pm0.08$ & $<7$ & $<4$ & (8,13) \\
GN-z9p4 & 1181 & $189.016992$ & $62.241585$ & $9.38$ & $-20.69$ & $-2.14\pm0.15$ & $<10$ & $<10$ & (15) \\
CAPERS-EGS-87132 & 6368 & $215.044034$ & $52.994331$ & $9.38$ & $-18.87$ & $-2.22\pm0.14$ & $34\pm7$ & $<24$ & (21) \\
GS-z9-0 & 3215 & $53.112436$ & $-27.774619$ & $9.43$ & $-19.54$ & $-2.60\pm0.06$ & $12\pm1$ & $4\pm1$ & (4,9,17) \\
JADES-GS-20064312 & 1287 & $53.072431$ & $-27.855456$ & $9.44$ & $-17.22$ & $-1.17\pm0.28$ & $<48$ & $<53$ & \\
GLASS-83338 & 3073 & $3.454686$ & $-30.316832$ & $9.53$ & $-19.06$ & $-1.80\pm0.47$ & $<69$ & $<28$ & (19) \\
UNCOVER-22223 & 2561 & $3.568115$ & $-30.383051$ & $9.57$ & $-17.28$ & $-1.99\pm0.15$ & $26\pm7$ & $25\pm5$ & (13) \\
JADES-GN-59720 & 1181 & $189.239795$ & $62.210830$ & $9.63$ & $-19.70$ & $-3.00\pm0.28$ & $<16$ & $<25$ & (16) \\
JADES-GS-20088041 & 1286 & $53.175514$ & $-27.780606$ & $9.68$ & $-20.14$ & $-1.96\pm0.11$ & $41\pm13$ & $<36$ & \\
JADES-GN-80088 & 1181 & $189.239122$ & $62.210934$ & $9.74$ & $-20.15$ & $-2.09\pm0.26$ & $<29$ & $<37$ & (16) \\
JADES-GN-55757 & 1181 & $189.217683$ & $62.199490$ & $9.74$ & $-19.86$ & $-2.51\pm0.17$ & $18\pm3$ & $10\pm4$ & (16,21) \\
CEERS-80026 & 1345 & $214.811848$ & $52.737113$ & $9.75$ & $-20.08$ & $-1.73\pm0.43$ & $<12$ & $<8$ & (1) \\
CAPERS-COSMOS-52185 & 6368 & $150.116572$ & $2.197038$ & $9.80$ & $-21.82$ & $-2.33\pm0.18$ & $16\pm8$ & $<12$ & \\
UNCOVER-13151 & 2561 & $3.592505$ & $-30.401463$ & $9.80$ & $-17.46$ & $-2.46\pm0.43$ & $<27$ & $<22$ & (13) \\
JADES-GS-20076576 & 1286 & $53.055108$ & $-27.845593$ & $9.90$ & $-19.04$ & $-2.53\pm0.32$ & $<14$ & $<26$ & \\
JADES-GN-11508 & 1181 & $189.184460$ & $62.262491$ & $9.93$ & $-19.64$ & $-2.64\pm0.07$ & $26\pm10$ & $<40$ & (16) \\
CAPERS-EGS-25297 & 6368 & $214.817153$ & $52.748305$ & $9.95$ & $-19.96$ & $-1.53\pm0.12$ & $28\pm8$ & $<23$ & (21) \\
UNCOVER-26185 & 2561 & $3.567071$ & $-30.377862$ & $10.06$ & $-18.93$ & $-2.36\pm0.17$ & $36\pm12$ & $<14$ & (6,13) \\
CEERS-64 & 2750 & $214.922774$ & $52.911525$ & $10.07$ & $-19.47$ & $-2.20\pm0.38$ & $<14$ & $<13$ & (2) \\
GLASS-z11-17225 & 3073 & $3.507323$ & $-30.343183$ & $10.07$ & $-19.06$ & $-2.67\pm0.01$ & $<36$ & $<83$ & (19) \\
GHZ9 & 3073 & $3.478735$ & $-30.345495$ & $10.16$ & $-19.80$ & $-1.45\pm0.07$ & $24\pm7$ & $76\pm7$ & (19,20) \\
GHZ8 & 3073 & $3.451447$ & $-30.321795$ & $10.23$ & $-20.37$ & $-2.36\pm0.07$ & $<14$ & $<13$ & (19) \\
CEERS-80041 & 1345 & $214.732534$ & $52.758092$ & $10.23$ & $-20.37$ & $-2.45\pm0.04$ & $<10$ & $<9$ & (1) \\
CAPERS-COSMOS-109917 & 6368 & $150.142945$ & $2.288012$ & $10.28$ & $-19.62$ & $-2.84\pm0.28$ & $20\pm6$ & $36\pm8$ & \\
UNCOVER-37126 & 2561 & $3.590110$ & $-30.359742$ & $10.39$ & $-19.80$ & $-2.67\pm0.11$ & $<8$ & $<7$ & (13) \\
GS-z10-0 & 1210 & $53.158837$ & $-27.773500$ & $10.39$ & $-18.05$ & $-2.38\pm0.04$ & $<31$ & $<22$ & (5,9) \\
CAPERS-UDS-z10 & 6368 & $34.456023$ & $-5.121952$ & $10.44$ & $-19.90$ & $-2.64\pm0.26$ & $19\pm10$ & $<16$ & (18) \\
GHZ7 & 3073 & $3.451347$ & $-30.320738$ & $10.44$ & $-19.84$ & $-2.22\pm0.07$ & $24\pm6$ & $<52$ & (19) \\
JADES-GS-20176151 & 1287 & $53.070746$ & $-27.865487$ & $10.53$ & $-19.08$ & $-2.11\pm0.17$ & $<10$ & $<15$ & \\
GN-z11 & 1181 & $189.106056$ & $62.242052$ & $10.60$ & $-21.89$ & $-2.38\pm0.13$ & $16\pm3$ & $6\pm2$ & (3) \\
GHZ4 & 3073 & $3.513717$ & $-30.351564$ & $10.72$ & $-19.50$ & $-2.26\pm0.16$ & $19\pm11$ & $<55$ & (19) \\
JADES-GS-20177294 & 1287 & $53.078997$ & $-27.863582$ & $10.89$ & $-17.65$ & $-2.34\pm0.55$ & $<63$ & $<62$ & \\
CAPERS-UDS-z11 & 6368 & $34.264455$ & $-5.096191$ & $11.06$ & $-20.41$ & $-2.38\pm0.31$ & $<21$ & $<26$ & (18) \\
JADES-GS-20021387 & 1287 & $53.067147$ & $-27.883190$ & $11.06$ & $-18.33$ & $-1.85\pm0.15$ & $<23$ & $<29$ & \\
JADES-GS-20015720 & 1287 & $53.117610$ & $-27.888327$ & $11.26$ & $-19.88$ & $-2.67\pm0.12$ & $15\pm5$ & $9\pm4$ & \\
CEERS-10 & 2750 & $214.906630$ & $52.945507$ & $11.39$ & $-20.18$ & $-2.02\pm0.32$ & $<19$ & $<21$ & (2) \\
CEERS-1 & 2750 & $214.943138$ & $52.942444$ & $11.55$ & $-20.07$ & $-2.71\pm0.29$ & $<20$ & $<12$ & (2) \\
GS-z11-0 & 1210 & $53.164768$ & $-27.774627$ & $11.55$ & $-19.46$ & $-2.20\pm0.27$ & $<18$ & $<12$ & (5,9,14) \\
GHZ2 & 3073 & $3.498971$ & $-30.324734$ & $12.34$ & $-20.93$ & $-2.58\pm0.14$ & $21\pm4$ & $47\pm3$ & (11) \\
UNCOVER-38766 & 2561 & $3.513562$ & $-30.356798$ & $12.39$ & $-18.91$ & $-2.35\pm0.03$ & $<38$ & $<14$ & (7,13) \\
GS-z12-0 & 3215 & $53.166346$ & $-27.821558$ & $12.51$ & $-19.03$ & $-2.19\pm0.11$ & $29\pm7$ & $<40$ & (5,9,12) \\
UNCOVER-13077 & 2561 & $3.570886$ & $-30.401556$ & $13.05$ & $-19.22$ & $-2.35\pm0.23$ & $<57$ & $<32$ & (7) \\
JADES-GS-z13-1-LA & 1287 & $53.064771$ & $-27.890294$ & $13.06$ & $-18.27$ & $-2.87\pm0.27$ & $<19$ & $<70$ & (22) \\
GS-z13-0 & 3215 & $53.149886$ & $-27.776504$ & $13.22$ & $-18.78$ & $-2.64\pm0.39$ & $<45$ & $<17$ & (5,9,14) \\
GS-z14-1 & 1287 & $53.074245$ & $-27.885980$ & $14.04$ & $-18.79$ & $-2.85\pm0.29$ & $<23$ & $<23$ & (10) \\
GS-z14-0 & 1287 & $53.082917$ & $-27.855628$ & $14.22$ & $-20.99$ & $-2.19\pm0.14$ & $7\pm2$ & $<9$ & (10) \\
\enddata
\tablecomments{References: (1) \citet{ArrabalHaro2023a}; (2) \citet{ArrabalHaro2023b}; (3) \citet{Bunker2023}; (4) \citet{Cameron2023b}; (5) \citet{Curtis-Lake2023}; (6) \citet{Goulding2023}; (7) \citet{Wang2023}; (8) \citet{Boyett2024a}; (9) \citet{Bunker2024}; (10) \citet{Carniani2024}; (11) \citet{Castellano2024}; (12) \citet{DEugenio2024}; (13) \citet{Fujimoto2024}; (14) \citet{Hainline2024b}; (15) \citet{Schaerer2024}; (16) \citet{Tang2024c}; (17) \citet{Curti2025}; (18) \citet{Kokorev2025}; (19) \citet{Napolitano2025a}; (20) \citet{Napolitano2025b}; (21) \citet{Pollock2025}; (22) \citet{Witstok2025b}.}
\label{tab:zg9}
\end{deluxetable*}

%%%% Figure: new spectra at z>9 %%%%

\begin{figure*}
\includegraphics[width=0.495\linewidth]{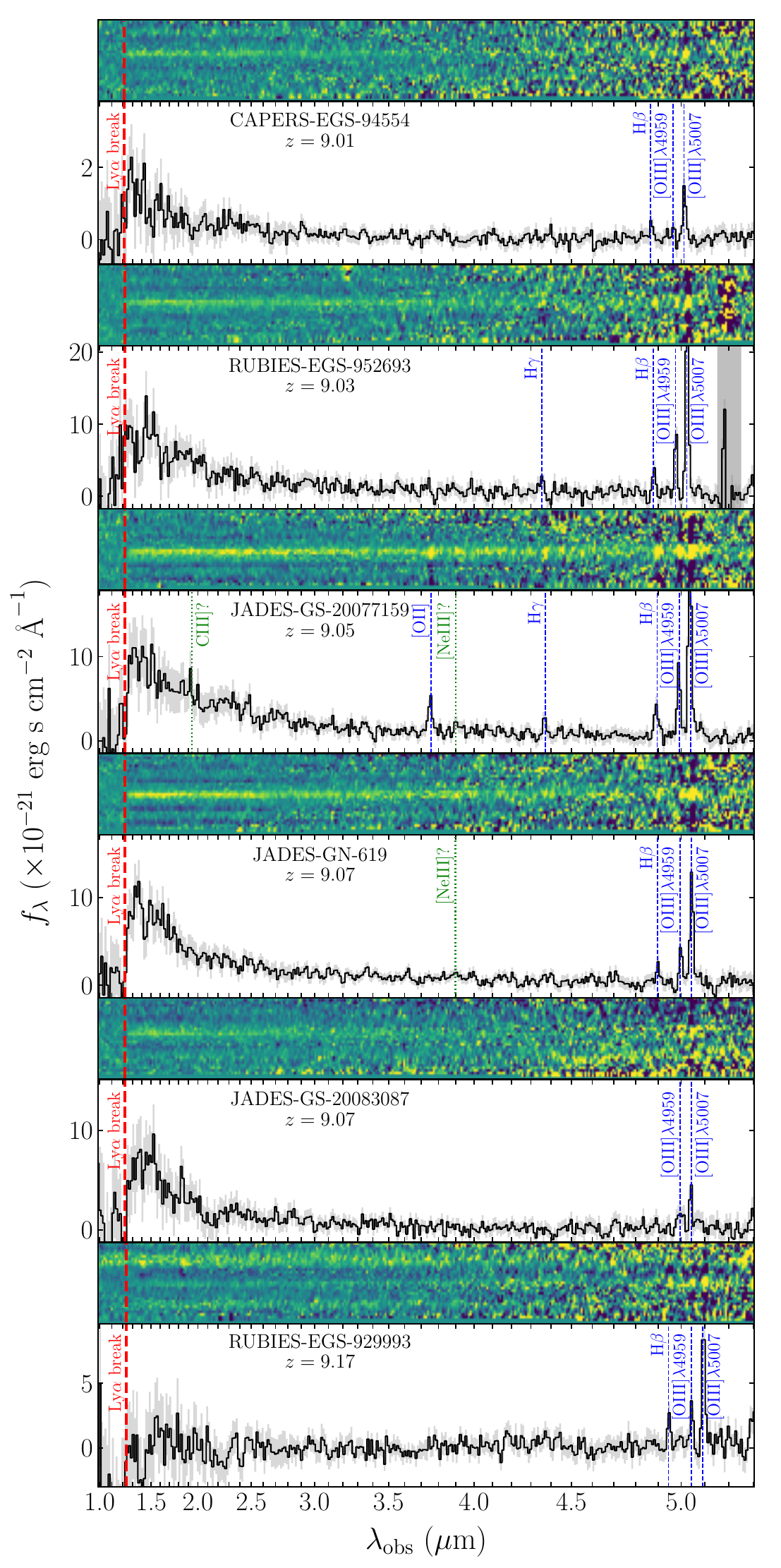}
\includegraphics[width=0.495\linewidth]{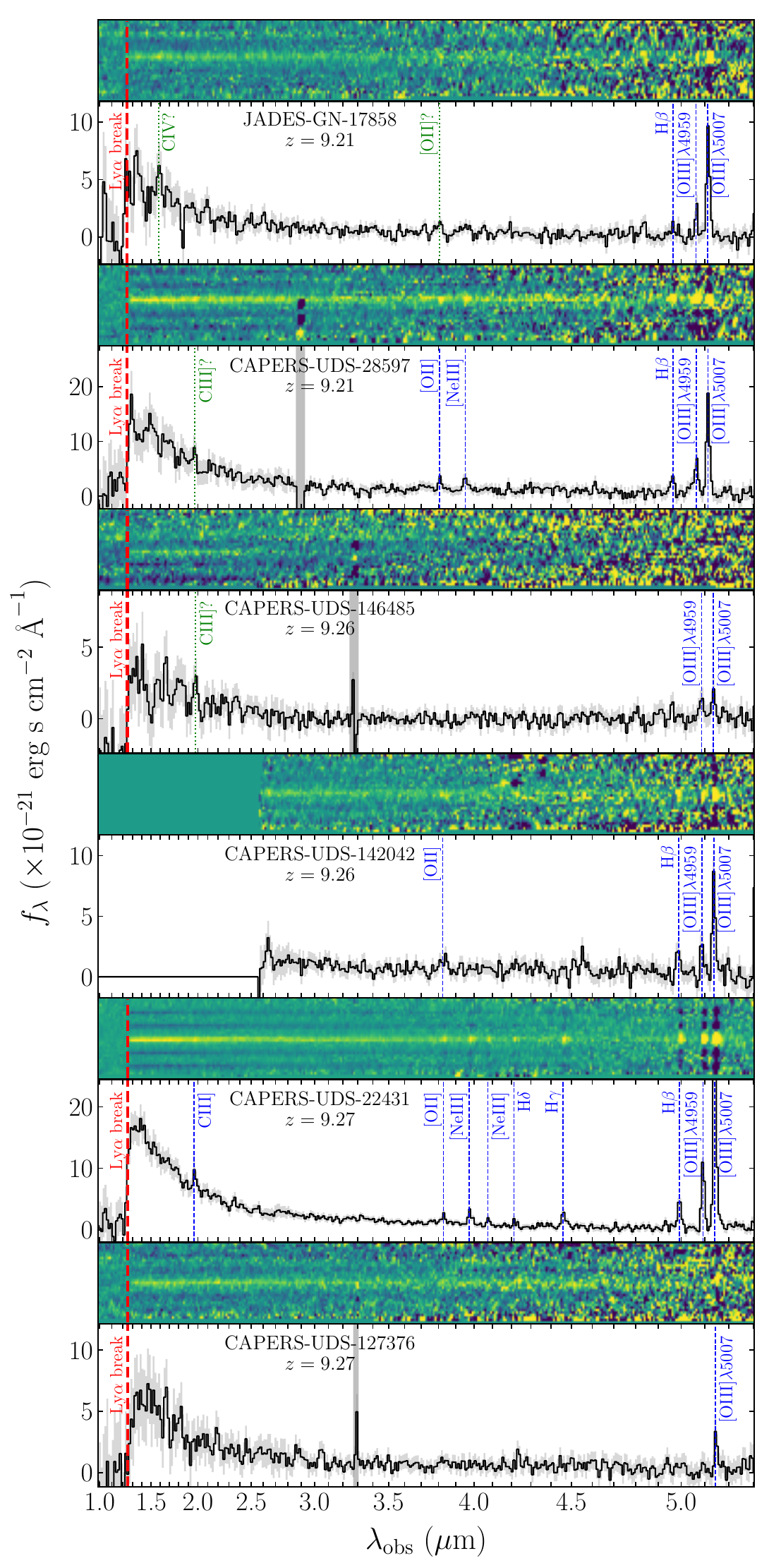}
\caption{NIRSpec prism spectra of the $30$ newly identified galaxies at $z>9$. For each object, the top panel shows the 2D spectrum and the bottom panels shows the 1D spectrum. We mark detected emission lines (S/N $>3$) with blue dashed lines and tentative detections (S/N $=2-3$) with green dotted lines. Spectra contaminated by noise pixels are masked by grey shaded regions.}
\label{fig:zg9_spec}
\end{figure*}

\begin{figure*}
\includegraphics[width=0.495\linewidth]{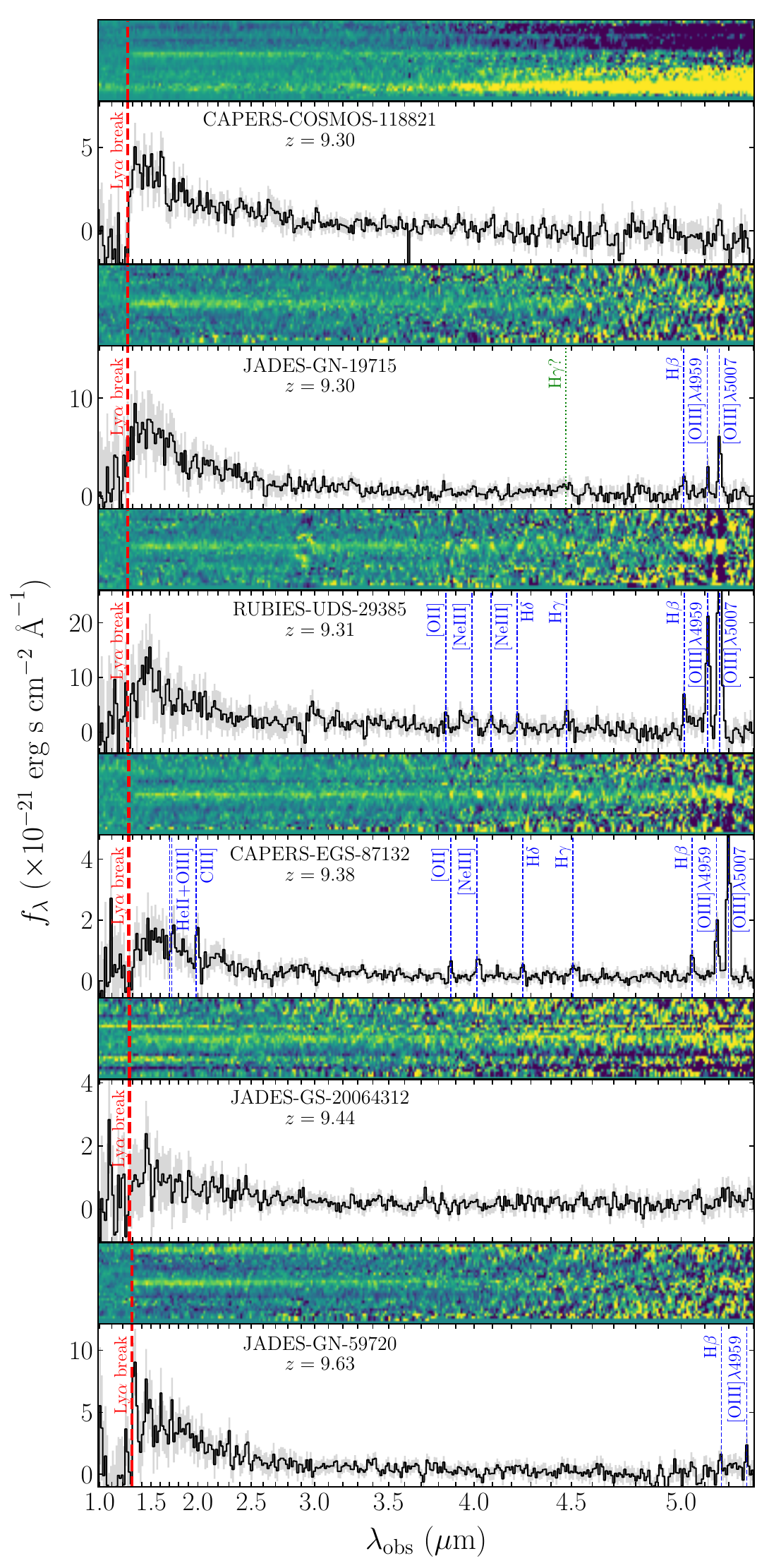}
\includegraphics[width=0.495\linewidth]{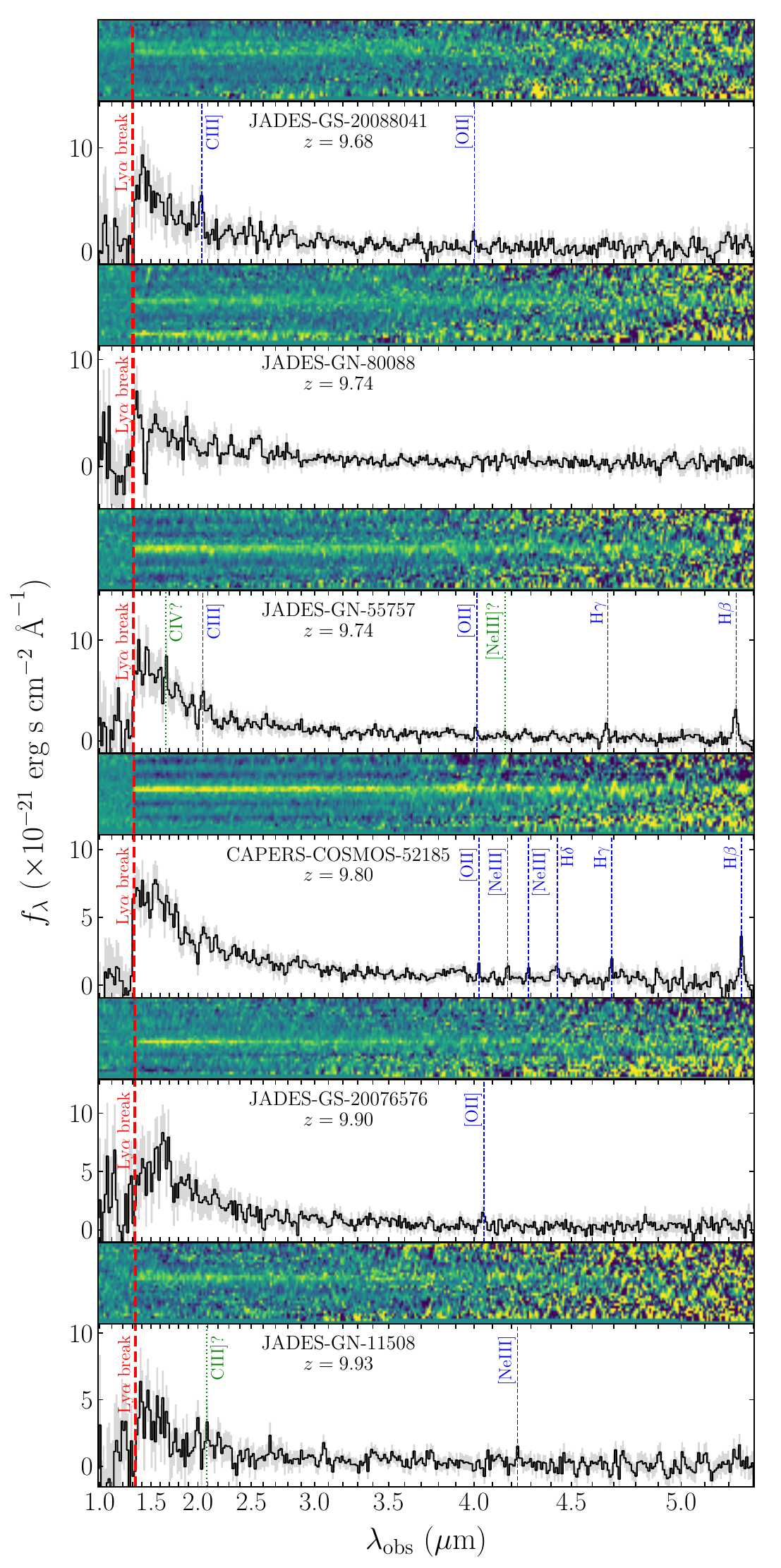}
\setcounter{figure}{0}
\caption{Continued.}
\end{figure*}

\begin{figure*}
\includegraphics[width=0.495\linewidth]{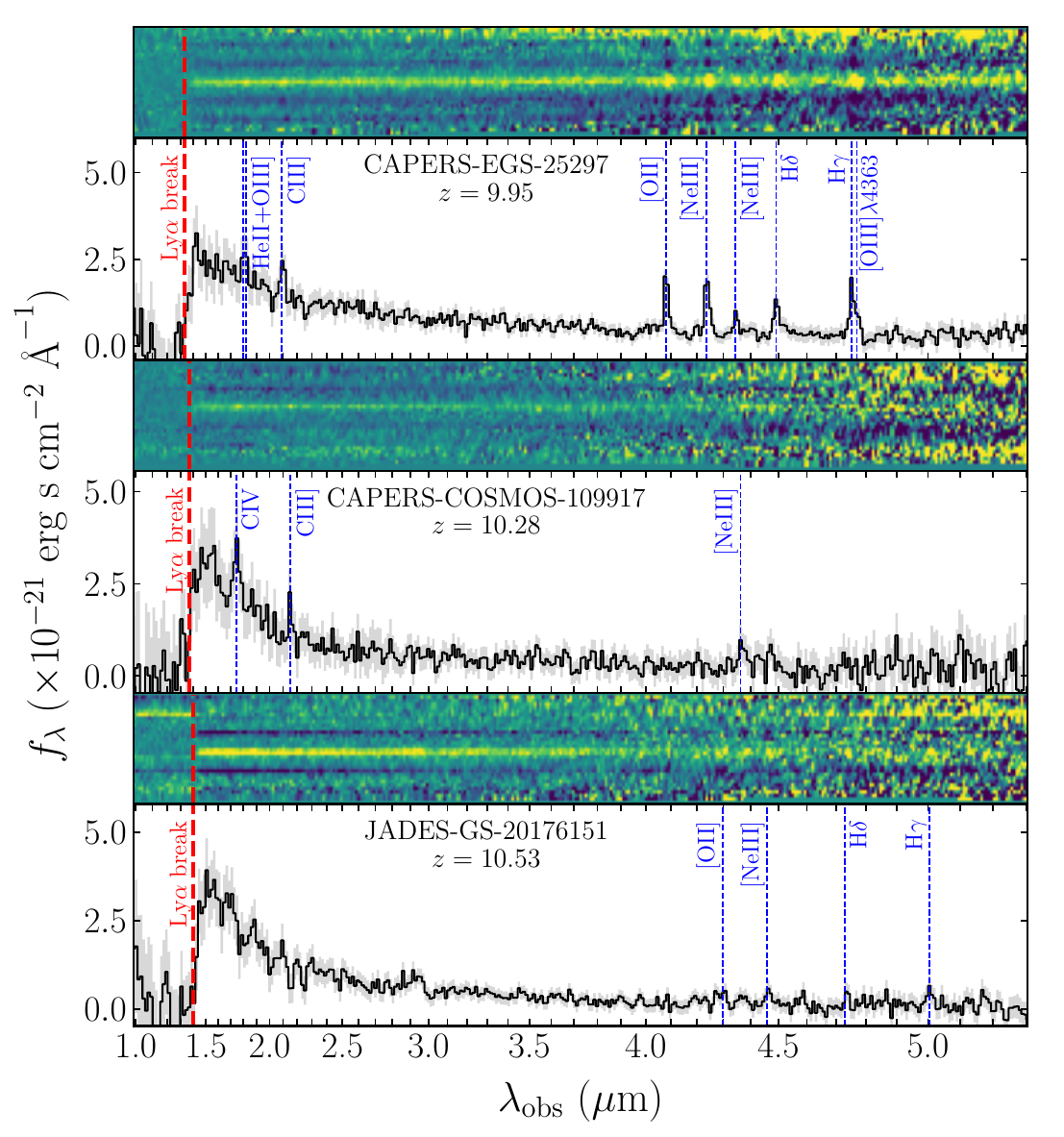}
\includegraphics[width=0.495\linewidth]{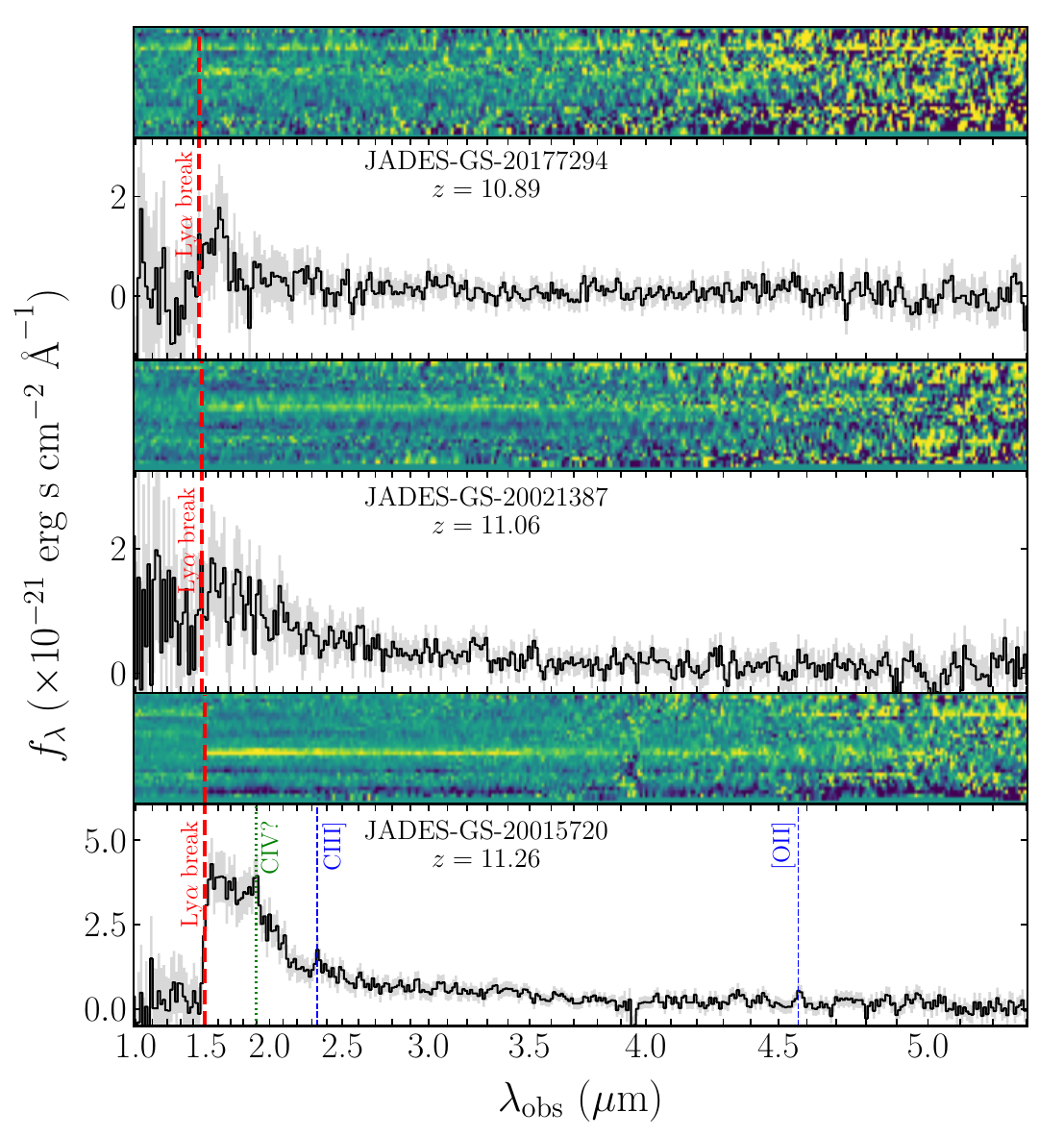}
\setcounter{figure}{0}
\caption{Continued.}
\end{figure*}


\bibliography{NIRSpec_zg9}{}
\bibliographystyle{aasjournal}



\end{document}